\documentclass[
     a4paper,
     12pt,
     twoside,
     openright, 
     ]{report}

\usepackage[
     top=4cm,
     bottom=4cm,
     left=3cm,
     right=3cm
     ]{geometry}
\usepackage{xcolor}
     \definecolor{green_maf}{RGB}{28, 112, 46}
     \definecolor{green_maf_faded}{RGB}{80, 117, 88}
     \definecolor{darkBlue}{HTML}{003366}
     \definecolor{OxfordBlue}{RGB}{0,33,71}
     \definecolor{detail}{RGB}{110,110,110}
\usepackage[
     colorlinks=true,
     linkcolor=darkBlue,
     citecolor=darkBlue,
     filecolor=darkBlue,
     urlcolor=darkBlue,
     linktocpage=true
     ]{hyperref}
\usepackage{setspace} 
\usepackage[normalem]{ulem} 
\usepackage{amsmath} 
\usepackage{nccmath} 
\usepackage{amssymb} 
\usepackage{amsthm} 
\usepackage{upgreek} 
\usepackage{simpler-wick} 
\usepackage{datetime} \settimeformat{ampmtime}
\usepackage{minitoc} 
     \dominitoc[n]      
      \mtcsetoffset{minitoc}{-10pt} 
\usepackage[toc,page]{appendix} 
\usepackage[new]{old-arrows} 
\usepackage{dsfont} 
\usepackage{bm} 
\usepackage{pbox} 
\usepackage{tikz} 
\usepackage{colortbl} 
\usepackage{mathrsfs} 
\usepackage{mathtools} 
     \mathtoolsset{showonlyrefs} 
\usepackage[utf8]{inputenc} 
\usepackage{fancyhdr} 
\usepackage{float} 

\newif\ifcomments
     \commentsfalse 
\newif\ifdetails
     \detailsfalse 
\newif\ifmafdetails
     \mafdetailsfalse 

\usepackage{titlesec}
\titleformat{\chapter}[display]{\normalfont\bfseries\filcenter}{\vspace{-90pt}\Huge\thechapter}{0pt}{\LARGE}

\newcommand{\detail}[1]{
     \ifdetails{\color{detail}  #1  }
     \fi}

\usetikzlibrary{shapes.geometric}
\def\weight #1 at (#2,#3){\node[fill=black, inner sep=1pt, label=40:{\small #1},circle] at (#2,#3) {}; }

\newcommand{\sep}{\vspace{15pt}}
\newcommand{\tr}[0]{\text{tr}} 
\newcommand{\no}[1]{\textbf{:}#1\textbf{:}} 
\newcommand{\dd}{{\rm d}} 
\newcommand{\widesim}{\raisebox{-4pt}{$\widetilde{\qquad}$}} 
\newcommand{\boxope}[1]{{\,}\boxed{#1}{\,}} 
\DeclarePairedDelimiter\ket{\lvert}{\rangle} 
\DeclarePairedDelimiterX\braket[2]{\langle}{\rangle}{#1 \delimsize\vert #2} 

\newcommand{\voc}[1]{\emph{#1}}

\begin{document}

\pagenumbering{gobble} 

\onehalfspacing
\begin{titlepage}

\begin{center}
{\bf \huge G-structures and Superstrings \\
\vspace{10pt}
from the Worldsheet}

\vspace{30pt}

{\Large Marc-Antoine Fiset}

\vspace{20pt}

Trinity College \\
University of Oxford

\vfill

\includegraphics[width=40mm]{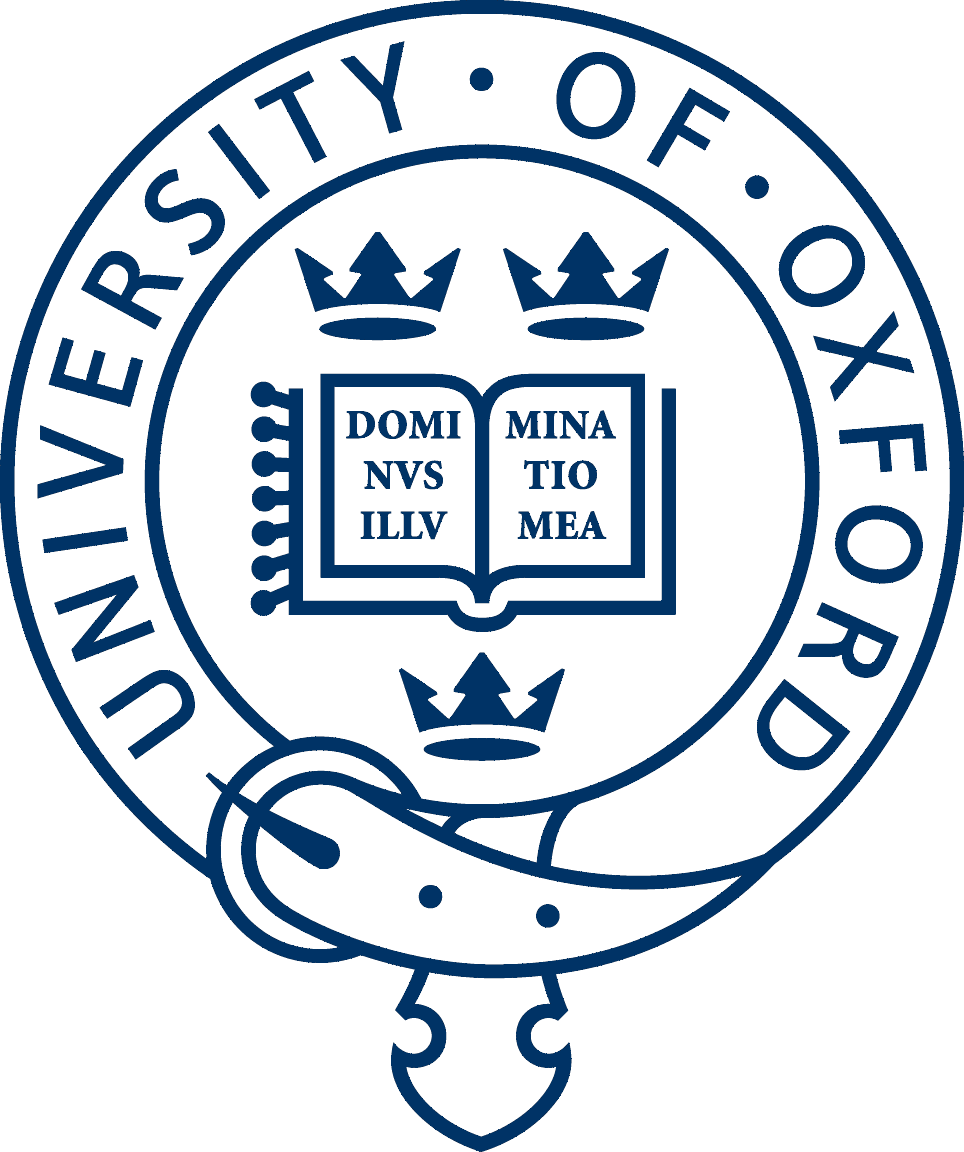}

\vfill

A thesis submitted for the degree of \\
\textit{Doctor of Philosophy} \\
Trinity 2019
\end{center}

%
%
%
%
%
%
%
%
%
%

%
%
%
%
%

\end{titlepage}
\newpage
~\newpage

\newgeometry{top=4cm, bottom=4cm, left=4cm, right=3cm}
\onehalfspacing

\begin{flushright} \textit{
To Justine, my wife and best friend,\\who brightened my stay in Oxford.
}\end{flushright}
\newpage
~\newpage

\begin{center}
{\LARGE G-structures and Superstrings \\
\vspace{10pt}
from the Worldsheet}

\vspace{15pt}

{\large Marc-Antoine Fiset}

\vspace{10pt}
\end{center}

\noindent $\mathcal{G}$-structures, where $\mathcal{G}$ is a Lie group, are a uniform characterisation of many differential geometric structures of interest in supersymmetric compactifications of string theories. Calabi--Yau $n$-folds for instance have torsion-free $SU(n)$-structure, while more general structures with non-zero torsion are required for heterotic flux compactifications. Exceptional geometries in dimensions $7$ and $8$ with $\mathcal{G}=G_2$ and $Spin(7)$ also feature prominently in this thesis.

We discuss multiple connections between such geometries and the worldsheet theory describing strings on them, especially regarding their chiral symmetry algebras, originally due to Odake and Shatashvili--Vafa in the cases where $\mathcal{G}$ is $SU(n)$, $G_2$ and $Spin(7)$. In the first part of the thesis, we describe these connections within the formalism of operator algebras. We also realise the superconformal algebra for $G_2$ by combining Odake and free algebras following closely the recent mathematical construction of twisted connected sum $G_2$ manifolds. By considering automorphisms of this realisation, we speculate on mirror symmetry in this context.

In the second part of the thesis, $\mathcal{G}$-structures are studied semi-classically from the worldsheet point of view using $(1,0)$ supersymmetric non-linear $\sigma$-models whose target $\mathcal{M}$ has reduced structure. Non-trivial flux and instanton-like connections on vector bundles over $\mathcal{M}$ are also allowed in order to deal with general applications to superstring compactifications, in particular in the heterotic case. We introduce a generalisation of the so-called special holonomy W-symmetry of Howe and Papadopoulos to $\sigma$-models with Fermi and mass sectors. We also investigate potential anomalies and show that cohomologically non-trivial terms in the  quantum effective action are invariant under a corrected version of this symmetry. Consistency with supergravity at first order in $\alpha'$ is manifest and discussed.
We finally relate marginal deformations of $(1,0)$ $G_2$ and $Spin(7)$ $\sigma$-models  with the cohomology of worldsheet BRST operators. We work at lowest order in $\alpha'$ and study general heterotic backgrounds with bundle and non-vanishing torsion.

\chapter*{Statement of originality}

\vspace{-13pt}
\noindent Chapter \ref{chap:G-symmetries} and \ref{chap:MarginalDeformations} in this thesis are based on articles written in collaboration:
\begin{itemize}
\item[]
X. de la Ossa and M.-A. Fiset (2018)
\textit{$\mathcal{G}$-structure symmetries and anomalies in $(1,0)$ non-linear $\sigma$-models},
Journal of High Energy Physics \textbf{01}, 62
\texttt{[1809.01138]}
\item[]
M.-A. Fiset, C. Quigley and E. E. Svanes (2018)
\textit{Marginal deformations of heterotic $G_2$ sigma models},
Journal of High Energy Physics \textbf{02}, 52 \texttt{[1710.06865]}
\end{itemize}

\noindent Chapter \ref{chap:TCS} is based on the following single-authored work:

\begin{itemize}
\item[]
M.-A. Fiset (2018)
\textit{Superconformal algebras for twisted connected sums and $G_2$ mirror symmetry}, 
Journal of High Energy Physics \textbf{12}, 11 \\
\texttt{[1809.06376]}
\end{itemize}

\chapter*{Acknowledgements}

\vspace{-13pt}
I first wish to express my gratitude to Xenia de la Ossa, who welcomed me as a graduate student and also more personally, with warmth and generosity. She shaped my path in research. She shared enthusiastically clever insights and her knowledge of geometry and physics. Her incredible support never faltered.

My success owes a great deal to other mentors and collaborators as well, in particular to Andreas Braun, Callum Quigley and Eirik E. Svanes. I learned precious lessons alongside them about research in theoretical high energy physics. They stimulated the work presented in this thesis.

Christopher Beem, Sakura Sch\"afer-Nameki, Philip Candelas, Carlo Mene\-ghelli, Tomasz Łukowski, Mathew Bullimore and Anthony Ashmore are other members of the Mathematical Physics group who had a particularly positive impact on my progression. I thank them for useful comments, guidance and general wisdom. I learned a lot from my peers, in particular Pietro Benetti Genolini, whose diligence and passion for the field are contagious. I also wish to acknowledge my other officemates, past and present: Alex, Omar, Carolina, Juan Carlos, Sebastjan, Pyry, Matteo, Giulia and Atul.
I thank Michele Del Zotto, Katrin Wendland, Sebastian Goette, Chris Hull, Jock McOrist, Magdalena Larfors, Ruben Minasian, Jan de Boer, José Figueroa-O'Farrill, Thomas Creutzig, Ilarion Melnikov and Dominic Joyce for discussions.

My research was financed by a Reidler scholarship from the Mathematical Institute and by a FRQNT scholarship from the Government of Quebec.

I am finally indebted to my family. For trusting me, for her determination and for her continuous love, I thank my wife Justine. Her role in the success of our adventure in the United Kingdom is enormous. As did my parents, Sylvie and Yvon, she believed in me and encouraged me to achieve my goals. I will remain grateful forever to them all.

\dominitoc 

\tableofcontents

\chapter{Introduction}

\doublespacing

\pagenumbering{arabic} 

\begin{figure}
\begin{center}
\begin{tikzpicture}[scale=1]
\node at (-3.5,0) {Worldsheet $(\Sigma^2,\gamma)$};
\node at (-3.5,-0.5) {$(1,0)$ non-linear $\sigma$-model};
\node at (-3.5,-1) {\&};
\node at (-3.5,-1.5) {extra symmetries};

\draw[thick, ->] (-1.5,0) -- (1.5, 0);

\node at (5.0,1) {$\mathcal{V}^n, A$};
\draw[thick, ->] (4.7,0.75) -- (4.7,0.25);
\node at (4,0) {Target space $(\mathcal{M}^d; G, B)$};
\node at (7.5,0) {$\times \quad \mathbb{R}^{1,9-d}$};
\node at (4,-0.75) {with};
\node at (4,-1.5) {$\mathcal{G}$-structure \& supersymmetry equations};

\draw[thick, ->] (-3.5,-2) -- (-3.5, -5);

\node at (-3.5,-5.5) {2D SCFT $(\text{Vir}{}^{\,\mathcal{N}=1} , \overline{\text{Vir}}{}^{\,\mathcal{N}=0})$};
\node at (-3.5,-6) {\&};
\node at (-3.5,-6.5) {extra currents};

\draw[dashed, ->] (4,-2) -- (-0.5, -6);

\end{tikzpicture}
\caption{Schematic overview of the context of this thesis}
\label{fig:triangle}
\end{center}
\end{figure}
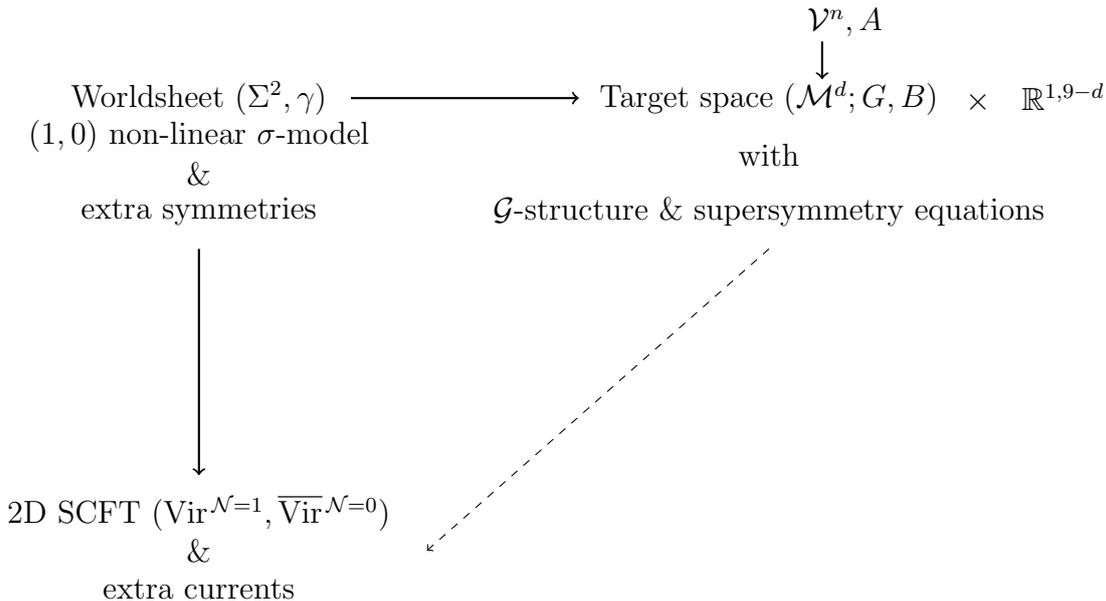

$\mathcal{G}$-structures, superstrings---or rather two-dimensional superconformal field theory and the worldsheet $\sigma$-model: three vertices of a triangle of relationships (figure~\ref{fig:triangle}), which best summarises what this thesis is about.

The top horizontal edge represents the embedding of a real two-dimensional string \emph{worldsheet} $\Sigma$, with Lorentzian metric $\gamma$, in ten-dimensional target spacetime.             The former will remain topologically a punctured sphere in this thesis: we focus on closed strings and ignore string loop effects.            The latter is generically curved, although one first starts with Minkowski spacetime in introductions to string theory \cite{Polchinski:1998rq, Blumenhagen:2013fgp, Green:1987sp}. We take the compromise ansatz where spacetime splits metrically as
\begin{equation} \label{eq:splitSpacetime}
\mathcal{M}^d\times\mathbb{R}^{1,9-d} \,,
\end{equation}
where $\mathcal{M}$ is a closed manifold of real dimension $d$.            This set-up would be convenient to study dimensional reduction for instance. Dynamics can be assigned via an explicit action for fields over the worldsheet; Polyakov's action in the simplest scenario, or for curved spaces, a generalisation like the \textit{non-linear $\sigma$-model} described in chapter~\ref{chap:G-symmetries}.
We will always focus on the \textit{internal sector}, for which the codomain of this $\sigma$-model is $\mathcal{M}$. In other words, we mostly omit the  $\mathbb{R}^{1,9-d}$ factor in \eqref{eq:splitSpacetime}.\footnote{Apart from the $\mathbb{R}^{1,9-d}$ sector, there are many more ingredients which we will silence in this thesis: the ghost sector, the GSO projection, modular invariance, etc. We shall also ignore the dilaton. These concepts are well covered in string theory textbooks.}          In addition to the metric tensor $G$, the target space $\mathcal{M}$ supports a $B$ field with NS--NS flux $H$ and eventually other background geometric data, as well as a choice of ``$\mathcal{G}$-structure'', to which we will return shortly.

The vertical edge of the triangle (figure~\ref{fig:triangle}) essentially represents quantization.            Our grasp of the quantum $\sigma$-model depends on the level of simplicity of target space. When $\mathcal{M}^d = \mathbb{R}^d$ (or $\mathbb{T}^d$) is flat, canonical quantization is possible and our control over the quantum theory is maximal.
All books on string theory cited above take this approach in their early chapters.
More intricate $\sigma$-models have spheres, Lie groups or coset spaces as targets and a lot can be said about them analytically \cite{Pisarski:1979gw, Brezin:1980ms, Wess:1971yu, Witten:1983ar}. Generically however, the quantum theory can only be directly accessed with a $\sigma$-model in perturbation theory. Our reach is then limited by how reliably we can ignore corrections proportional to Riemannian and flux curvature on target space, as measured in units of the string length $\sqrt{\alpha'}$, which plays for the worldsheet theory the same loop-counting role as $\hbar$ in ordinary quantum field theory. This is the ``large-radius'' limit of the $\sigma$-model.

This old-fashioned route to quantization will be explored in section~\ref{sec:Anomalies}, yielding interesting new results at order $\alpha'$. It enjoys tight relationships with geometry of target space, but it suffers from technical limitations. For instance, global aspects of the dynamics are hard to capture because local coordinate patches are essential for perturbative calculations in the worldsheet non-linear $\sigma$-model. It is thus instructive to compare this approach to the one based on exact symmetries of the internal theory, which we now recall. This approach is represented by the lowermost triangle tip on figure~\ref{fig:triangle}.          

For critical strings, the classical invariance under local Weyl rescalings of the metric $\gamma$ by a positive function must not be anomalous. In conformal gauge, where $\gamma$ looks flat, this and diffeomorphism invariance give rise to \textit{conformal symmetry}. In two dimensions, this means the Hilbert space organises into representations of two copies of the Virasoro algebra, one for the left-moving degrees of freedom and one for the right-moving ones. Worldsheet \emph{supersymmetry} must also complement this conformal symmetry by construction (in the Ramond--Neveu--Schwarz approach). The minimal number of Majorana--Weyl supercharges in two dimensions offers the greatest flexibility and will be our reference case. We will denote this baseline by ``$(1,0)$'', where $0$ refers to the Virasoro algebra, and $1=\mathcal{N}$ to its supersymmetric version.

The most natural application of $(1,0)$ is to \textit{heterotic superstrings}. In this case, a fermionic sector must be incorporated in the $\sigma$-model to describe sections of a rank-$n$ vector bundle $\mathcal{V}$ over $\mathcal{M}$ acted upon by a gauge connection $A$. By performing the so-called standard embedding of $A$ in the spin connection on $T\mathcal{M}$, one can also obtain general $(1,1)$ theories, useful in particular to formulate {type II} superstrings. In a sense then, $(1,0)$ theories are building blocks for superstring theories and they are well worth our attention.

The $\mathcal{N}=1$ supersymmetric version of the Virasoro algebra imposes very light geometric constraints on target space. Any background $(\mathcal{M}, \mathcal{V}; G, B, A)$ is indeed consistent with global $\mathcal{N}=1$ worldsheet supersymmetry. Full conformal invariance on the other hand yields differential conditions generalising Einstein's relativity. At low orders in $\alpha'$, these conditions bear out supergravity equations of motion (see e.g.\ the excellent reviews \cite{Hull:1986hn, Callan:1989nz, McOrist:2010ae}).

Central to this thesis are \emph{extra worldsheet symmetries in addition to the $\mathcal{N}=1$ super-Virasoro}, the blank canvas against which extra layers can be appreciated. Crucial also are their \textit{relations to specific target space geometries}. We will mainly examine extensions arising when demanding spacetime supersymmetry in a dimensionally-reduced effective theory on $\mathbb{R}^{1,9-d}$.

The most important example is the enhancement from worldsheet $\mathcal{N}=1$ to $\mathcal{N}=2$ super-Virasoro, especially when it happens for both left and right chiralities, yielding $(2,2)$ superconformal field theories (SCFT). The associated target space geometry is then typically\footnote{More general bi-Hiermitian geometries are also allowed \cite{Gates:1984nk}.} Calabi--Yau. Exploiting only the symmetries, one can infer many useful features of the quantum theory in this case: chiral rings, spectral flow, topological twists, exactly marginal deformations, relation between target space cohomology and ground states, holomorphicity, etc. \cite{Lerche:1989uy, Witten:1988xj} Some of these properties also generalise to $(2,0)$ SCFTs \cite{Adams:2005tc, McOrist:2010ae, Sharpe:2015vza}.

The Calabi--Yau/$\mathcal{N}=2$ scenario is but one example in a much larger class of extensions, less appreciated because less phenomenologically relevant, but nevertheless of considerable theoretical and mathematical interest. To various degrees of certainty, many of the features above were established in the realm of $(1,1)$ SCFTs. This may surprise but goes back to Shatashvili and Vafa \cite{Shatashvili:1994zw}, who convincingly argued that the key role traditionally played by the $U(1)$ R-symmetry may be assumed by more exotic worldsheet symmetries extending $\mathcal{N}=1$ super-Virasoro. The chiral algebras they studied---which we simply name after these authors---have a major importance in this thesis. They are connected to the exceptional Lie groups $G_2$ and $Spin(7)$ through $\mathcal{G}$-structures, to which we now return.

In the context of string compactifications on $\mathcal{M}^d$, minimal spacetime supersymmetry in the $\mathbb{R}^{1,9-d}$ factor in \eqref{eq:splitSpacetime} requires that $\mathcal{M}$ admits a real nowhere vanishing spinor, as well as further differential constraints on $(\mathcal{M}, \mathcal{V}; G, B, A)$, the \textit{Killing spinor} or \textit{supersymmetry equations}. The very existence of the spinor on $\mathcal{M}$ can be phrased in the mathematical language of $\mathcal{G}$-structures and the supersymmetry equations then refine the types of $\mathcal{G}$-structures of interest. (They also place constraints on the vector bundle.) A {$\mathcal{G}$-structure} is defined as follows. Let $\mathcal{F}$ be the frame bundle of $\mathcal{M}^d$, a principal $GL(d,\mathbb{R})$-bundle. Let $\mathcal{G}\subseteq GL(d,\mathbb{R})$ be a Lie subgroup. A \emph{$\mathcal{G}$-structure} on $\mathcal{M}$ is the reduction to a principal subbundle of $\mathcal{F}$ with fibre $\mathcal{G}$ \cite[p.\,36]{Joyce2007}.

$\mathcal{G}$-structures are closely related to connections on $T\mathcal{M}$. For instance, a connection $\nabla$ with holonomy group contained in $\mathcal{G}$ automatically guarantees the existence of a $\mathcal{G}$-structure, compatible in a suitable sense \cite[p.\,37]{Joyce2007}. If the connection is torsion-free, the $\mathcal{G}$-structure is also called \emph{torsion-free}.\footnote{There is also an intrinsic definition of torsion for $\mathcal{G}$-structures \cite[p.\,38]{Joyce2007}, but $\mathcal{G}$-structures and connections are essentially equivalent, and we prefer to work in terms of the latter.} Spaces whose Levi--Civita connection have holonomy $\mathcal{G}$ are simple examples of manifolds with a torsion-free $\mathcal{G}$-structure.
Calabi--Yau manifolds correspond to the case $\mathcal{G}=SU(n)$. For $\mathcal{G}=G_2$, we are typically interested in $d=7$ dimensional manifolds $\mathcal{M}$. In many ways, torsion-free $G_2$-holonomy manifolds are the closest analogues to Calabi--Yau manifolds in dimension $7$; for example they admit Ricci-flat metrics. More generally, heterotic strings compactified on $d=7$ dimensional manifolds require $G_2$-structures with non-vanishing torsion. Similarly compactifications on $d=8$ dimensional manifolds shed the spotlight on $Spin(7)$-structure manifolds with non-zero torsion. M-theory compactifications are another application of $G_2$-holonomy, as such manifolds are necessary for $\mathcal{N}=1$ supersymmetry in four dimensions \cite{Papadopoulos:1995da}.

There are simpler characterisations of $G_2$- and $Spin(7)$-structures than the definition above. We will describe them in due course (chapter~\ref{chap:OA}), along with the Shatashvili--Vafa superconformal algebras they are related to.

Let us finally highlight one more incentive to tighten the loose link established by the intricate process described above between target space differential geometry and worldsheet extended symmetry algebras (dashed oblique edge in figure~\ref{fig:triangle}). In cases where $\mathcal{M}$ is a torus or Calabi--Yau, it is known that this relation fails to be bijective: more than one target space geometries lead to the same conformal field theory (up to automorphism). This is known as \textit{T-duality} and \textit{mirror symmetry} respectively. Shatashvili and Vafa, in the paper \cite{Shatashvili:1994zw} cited above, conjecture a generalisation of this principle applicable to their exotic algebras. Perhaps mirror symmetry could hold for $G_2$ and $Spin(7)$ manifolds? Perhaps, they suggest, the deficiency of the conformal field theory to decipher aspects of the target manifold is precisely explained by the existence of multiple (mirror) geometries corresponding to the same conformal field theory (at least up to automorphisms). Mirror symmetry for certain $G_2$ manifolds is the subject of chapter~\ref{chap:TCS}. Other potential mathematical outputs of a tighter relation between geometry and CFT include topological invariants such as analogues of elliptic genera and eventually Gopakumar--Vafa and Gromov--Witten invariants.

\sep

Two main technical frameworks are exploited in this thesis to help clarify the triangle of relationships described above: CFT techniques, in particular the notion of \emph{operator algebra}, and $(1,0)$ non-linear $\sigma$-models. These formalisms divide the main body in two largely independent parts (although, in the last chapter, we briefly bend this rule). Geometry of target space permeates the whole thesis.

In the next chapter, we define and provide examples of operator algebras. We introduce in particular the Shatashvili--Vafa algebras for $G_2$ and $Spin(7)$ string compactifications. We also describe the notion of \textit{algebra realisation} and illustrate realisations in free theories of the operator algebras of main interest. This chapter gives motivation for the rest as well as some background.

Chapter~\ref{chap:TCS}, based on \cite{Fiset:2018huv}, is essentially an example of a realisation of the $G_2$ Shatashvili--Vafa algebra in the particular case where the $G_2$ manifold can be obtained by a ``twisted connected sum'' construction. We review this recent mathematical developement \cite{MR2024648, MR3109862,Corti:2012kd} and infer properties of the geometry/CFT relationship on figure~\ref{fig:triangle} to construct the associated operator algebra. We also comment on generalised mirror symmetry.

We shift gears in chapter~\ref{chap:G-symmetries} to formulate $(1,0)$ non-linear $\sigma$-models with a fermionic gauge bundle sector. This chapter assumes little background and can be read independently. We clarify chiral symmetries, of which superconformal transformations and \emph{extended $\mathcal{G}$-structure symmetries} are examples. The latter is the main result of this chapter, first communicated in \cite{delaOssa:2018azc}.
A special case is the $\sigma$-model symmetry enhancement from $\mathcal{N}=1$ to $\mathcal{N}=2$ superconformal symmetry. It correlates with a structure group reduction of the target space: $SO(d) \rightarrow U(d/2)$. We explain in chapter~\ref{chap:G-symmetries} how this is part of an enlightening overarching principle, giving a worldsheet symmetry for any reduction $SO(d)\rightarrow\mathcal{G}$ of the structure group of target space. It holds regardless of the application, in type II or heterotic string compactifications for example.

Order-$\alpha'$ anomalies at one-loop in perturbation theory are also probed in this chapter. We examine both superconformal transformations and $\mathcal{G}$-structure symmetries via the worldsheet quantum effective action. This is an essential step taken to bridge the gap between Lagrangian $\sigma$-models and the exact quantum description based on conserved currents. Corrections in $\alpha'$ expected in heterotic string theory nicely arise from this analysis. At order $\alpha'$, the B-field must be assigned anomalous Green--Schwarz gauge and Lorentz transformations and the associated gauge-invariant $3$-form flux must be corrected with Chern--Simons forms locally as
\begin{equation} \label{eq:H}
H=\dd B + \frac{\alpha'}{4}(\text{CS}_3(A)-\text{CS}_3(\Theta)) \,,
\end{equation}
where $A$ is the gauge field and $\Theta$ is a connection on $T\mathcal{M}$.

A closer look at free operator algebras in the last chapter finally instructs us on how to perturb $G_2$ and $Spin(7)$ heterotic models preserving all worldsheet symmetries. Using the $(1,0)$ $\sigma$-model description, we obtain constraints satisfied by these infinitesimal deformations, called \emph{marginal deformations} or \emph{moduli}. We do not rely on the standard embedding nor on any artificial assumptions on the NS--NS flux or the torsion in target space. We work at tree-level in the $\alpha'$ expansion and, in this limit, our results for $G_2$ agree with those obtained in \cite{Clarke:2016qtg, delaOssa:2016ivz, delaOssa:2017pqy} using the supergravity perspective. Although our leading order result is already known from these works, a \emph{BRST operator} takes central stage in the process, which suggests protected quasi-topological sectors. The $Spin(7)$ result is entirely new and deserves, to this very day, a comparison with supergravity. The $G_2$ computation was first communicated in \cite{Fiset:2017auc}.

\part{Operator algebras} \label{part:I}

\chapter{Operator algebras and geometry} \label{chap:OA}

\onehalfspacing
\minitoc
\vspace{25pt}

\doublespacing

In this chapter, we agree on terminology and conventions about chiral symmetry algebras in two-dimensional conformal field theory (CFT). There are multiple possible approaches to this subject. While ours is decisively from physics and may not fully quench the reader's mathematical thirst, we state clearly our assumptions and the fundamentals of so-called \textit{operator algebras}, which we use. We then embark on a tour of some key examples of operator algebras. Most serve as building blocks in the construction described in chapter~\ref{chap:TCS}.
Contact with these algebras from perturbative quantum field theory is also the motivation behind chapter~\ref{chap:G-symmetries}. The free field realisations presented here and the hidden Virasoro operators $T_-$ in sections~\ref{sec:SVSpin(7)} and \ref{sec:SVG2} are finally crucial elements to constrain marginal deformations in chapter~\ref{chap:MarginalDeformations}.

This chapter is primarily meant as a review, but the presentation is somewhat original. Section~\ref{sec:exercise} in particular introduces an unexpectedly clean organisational principle underlying the operator algebras of interest to us.

Notations and key definitions from this chapter are collected in table~\ref{tab:OAnotations} for the reader's convenience.

\newpage

{\singlespacing

\begin{table}[H]
\begin{center}
\renewcommand\arraystretch{1.5}
\begin{tabular}{ c | c | c }
Notation & Definition & Reference \\\hline
 & Operator algebra & p.\pageref{p:OA}\\
$\no{A B}$ & Normal ordered product & p.\pageref{p:NO} \\
$\{A(z) B(w)\}$ & Full OPE & p.\pageref{p:OPE}\\
$\wick{\c A(z) \c B(w)}$ & Singular OPE & p.\pageref{p:OPE} \\
$\no{A(z) B(w)}$ & Regular OPE & p.\pageref{p:OPE} \\
$\langle \cdots \rangle$ & Generating set, Ideal & p.\pageref{p:generate} \\
$\Big/$& Quotient by an ideal & p.\pageref{p:ideal} \\
$T$, $c$ & Virasoro operator, central charge & p.\pageref{p:Virasoro} \\
$h(A)$ & Weight & p.\pageref{p:Virasoro} \\
 & Primary & p.\pageref{p:Virasoro} \\
 & Superprimary & p.\pageref{p:superprimary} \\
$\longhookrightarrow$ & Realisation & p.\pageref{p:realisation} \\
& Automorphism & p.\pageref{p:Autom} \\
\hline
$(\text{Free})^d$ & Free field algebra & p.\pageref{p:Free} \\
$\text{Vir}^{0}_c$ or $\text{Vir}_c$ & Virasoro algebra & p.\pageref{p:Virasoro} \\
$\text{Vir}^{1}_c$ & $\mathcal{N}=1$ Virasoro algebra & p.\pageref{p:Virasoro1} \\
$\text{Vir}^{2}_c$ & $\mathcal{N}=2$ Virasoro algebra & p.\pageref{p:Virasoro2} \\
$\text{Od}^{n}$ & Odake $n$ algebra & p.\pageref{p:Od} \\
$\text{SV}^{Spin(7)}$ & Shatashvili--Vafa $Spin(7)$ algebra & p.\pageref{p:Spin7} \\
$\text{SV}^{G_2}$ & Shatashvili--Vafa $G_2$ algebra & p.\pageref{p:G2} \\
$\mathcal{W}^1_c(\tfrac{3}{2};2)$ &  & p.\pageref{p:W} \\
\end{tabular}
\caption{Notations and definitions about operator algebras}
\label{tab:OAnotations}
\end{center}
\end{table}

}

\newpage



\section{Fundamentals} \label{sec:OABasics}

It is a standard practise in quantum field theory to expand local operators about each other when they are inserted nearby on the space where the theory is formulated. On a Riemann surface such as $\Sigma=\mathbb{CP}^1$, where we restrict ourselves\footnote{Wick rotation to an Euclidean worldsheet is assumed to have been performed in this chapter.}, there is a natural notion of holomorphic operators. For these, the operator expansion (OPE) leads to a particularly useful algebraic structure. The prime example of a holomorphic operator, in a conformally-invariant theory, is the $zz$-component\footnote{Here $z$ and $\bar{z}$ are holomorphic and anti-holomorphic local coordinates on $\Sigma$, which we assume are isothermal. The existence of isothermal coordinates is a standard fact about two-dimensional spaces: in these coordinates, by definition, the metric looks flat up to an overall positive factor, which is omitted here.} $T(z)=T_{z z}(z,\bar{z})$ of the energy-momentum tensor $T_{\mu\nu}(z,\bar{z})$. Indeed $\partial_{\bar{z}}T(z) = 0$\,, thanks to the vanishing trace condition. The expansion of this operator about itself is the simplest example of an \emph{operator algebra}:
\begin{equation} \label{eq:VirasoroOA}
T(z) T(w) \sim \frac{c/2}{(z-w)^4}+\frac{2T(w)}{(z-w)^2}+\frac{\partial_wT(w)}{z-w}+\ldots\,, \qquad c\in \mathbb{R} \,.
\end{equation}
As is well-known, this relation reproduces by design commutation relations for the modes $\{L_n\}_{n\in \mathbb{Z}}$ of $T(z)$ in a Laurent expansion in $z$. Abusing terminology, we can thus call \eqref{eq:VirasoroOA} the \emph{Virasoro algebra}, after the corresponding infinite-dimensional Lie algebra structure on the space spanned by these modes. More distinguishably, putting the emphasis on $T$ rather than $L_n$, we could refer to \eqref{eq:VirasoroOA} as the \emph{Virasoro operator algebra}. An advantage of focusing on the operator formulation is to avoid dealing with infinitely many generators: the relevant information is contained in the poles in \eqref{eq:VirasoroOA}. 

If there are additional symmetries, they can sometimes yield holomorphic currents as well.
The Virasoro operator algebra then gets extended by further operators and further OPE relations. Generic terms typically attached in physics to structures arising in this way include ``chiral,'' ``current,'' and ``conformal  algebras,'' as well as ``W-algebras'' \cite{Bouwknegt:1992wg}. Clearly there is a need for a standardised definition of such entities and their properties.

A mathematically rigorous approach, from the angle of Laurent modes, is due to Borcherds \cite{MR843307}, Frenkel, Lepowsky and Meurman \cite{MR996026} and pivots on the notion of \textit{vertex operator algebras} (VOA). For detailed presentation, see \cite{Kac1998VOA,FrenkelBZ2001}. We present instead the approach of Thielemans \cite{Thielemans:1994er}.

Starting from elementary physical principles (e.g. Ward identities), Thielemans reaches the concept of operator algebra, capturing the properties generally expected or assumed in two-dimensional CFT (meromorphicity of correlators, crossing symmetry, etc.). He also proves, in favourable circumstances, \cite{Thielemans:1991uw} the equivalence\footnote{We thank T. Creutzig for this clarification.} between operator algebras and VOAs. Another advantage of Thielemans' language is that he implemented it in a \textit{Mathematica}\textsuperscript{®} package, which is widely used currently in the ``conformal bootstrap'' community. This package was essential to obtain the results in \cite{Fiset:2018huv} presented in chapter~\ref{chap:TCS}.

It was recently pointed out to the author\footnote{Private conversation with C. Beem.} that a further attempt to make rigorous operator algebras from two-dimensional CFT is the concept of \textit{Lie conformal algebra}, in which OPEs translate into so-called \label{p:lambdabracket} \emph{$\lambda$-brackets}. This concept is likely closely related to operator algebras and in turn to VOAs, although we have not attempted to verify this equivalence.

\sep

We proceed with the main definition. An \label{p:OA} \emph{operator algebra}\footnote{We shall eventually call them simply \textit{algebras}.} is composed of the following data.
\begin{enumerate}
\item A vector space\footnote{The vector space here is not to be confused with the vector bundle $\mathcal{V}\rightarrow \mathcal{M}$ discussed in the introduction.} $\mathcal{V}$, whose elements are typically called \emph{operators}, \emph{fields}, or \emph{currents} and denoted $\{\mathds{1},A,B,\ldots\}$. The distinguished element $\mathds{1}$ is called the \emph{identity}. We think of elements of $\mathcal{V}$ as operators acting on the Hilbert space of the quantum field theory.

\item A $\mathbb{Z}_2$-grading $\mathcal{V}=\mathcal{V}^b\oplus \mathcal{V}^f$. The \emph{statistics} of $A$ is defined as follows.
\begin{equation}
|A|=\left\lbrace
\begin{matrix}
0 & A\in \mathcal{V}^b & \text{($A$ is \emph{bosonic}.)} \\
1 & A\in \mathcal{V}^f & \text{($A$ is \emph{fermionic}.)} \\
\end{matrix} \right.
\end{equation}
Also, $|\mathds{1}|=0$, i.e. the identity is bosonic.

\item An even linear map $\partial : \mathcal{V} \rightarrow \mathcal{V}$ (whose role in conformal field theory is typically played by the holomorphic derivative $\partial_{z}$ on $\mathbb{CP}^1$). A prime will also sometimes be used instead of the notation `$\partial$'.
\item Sequence of bilinear box maps $\boxope{- \,, -}_n: \mathcal{V}\otimes \mathcal{V} \rightarrow \mathcal{V}$, $n\in\mathbb{Z}$, compatible with the grading and satisfying the following axioms:
\begin{enumerate}
\item[(0)] $\forall$ $A, B\in \mathcal{V}$, $\exists ~ n_{\text{max}}(A,B)\in \mathbb{Z}$ such that
\begin{equation}
\boxope{AB}_{n}=0 \qquad \forall n\geq n_{\text{max}}(A,B) \,;
\end{equation}
\item[(1)] (\emph{unity}):
\begin{equation} \label{eq:unityAxiom}
\boxope{\mathds{1}A}_n=\delta_{0,n}A \qquad \forall A\in \mathcal{V} \,;
\end{equation}
\item[(2)] (\emph{commutativity}): \begin{equation}  \label{eq:commutativityAxiom}
\boxope{BA}_n
=(-1)^{|A||B|}\sum_{m\geq n}\frac{(-1)^m}{(m-n)!}\partial^{(m-n)}\boxope{AB}_m \qquad \forall n\in \mathbb{Z}\,;
\end{equation}
\item[(3)] (\emph{associativity})\footnote{The binomial coefficient must be generalised to negative uppermost entries \cite{Thielemans:1994er}. Let $x\in \mathbb{R}$ and $n\in \mathbb{Z}^{\geq 0}$. The \emph{binomial coefficien}t is
\begin{equation}
\binom{x}{n} = \frac{(x)_n}{n!} \,,
\qquad \text{where the \emph{Pochhammer symbol} is}  \qquad
(x)_n =\left\lbrace \begin{matrix}
1 & n=0 \\
\prod_{i=0}^{n-1}(x-i) & n\in \mathbb{Z}^{\geq 0}
\end{matrix} \right. \,.
\end{equation}
}:\begin{equation} \label{eq:Jacobilike}
\boxope{A\boxope{BC}_m}_n = (-1)^{|A||B|} \, \boxope{B\boxope{AC}_n}_m
+ \sum_{l\geq 1} \binom{n-1}{l-1} \boxope{\boxope{AB}_l C}_{m+n-l}\,.
\end{equation}
\end{enumerate}
\end{enumerate}

The physical motivation for this definition should be mostly evident, apart perhaps for the box maps, which are nothing but the poles appearing in OPEs. Axiom (0) states that there should be a finite maximal pole between any operators (e.g. $n=4$ for the Virasoro algebra \eqref{eq:VirasoroOA}).

Let us press on with properties (1)--(3) to understand better the connection with standard CFT lore. From these axioms, the following identities follow \cite{Thielemans:1994er}:
\begin{equation}
\boxope{A\mathds{1}}_{n> 0} = 0 \,,
\qquad \text{and} \qquad
\boxope{A\mathds{1}}_{n\leq 0} = \frac{1}{(-n)!}\partial^{(-n)}A \,.
\end{equation}
The proof is particularly easy:
\begin{equation}
\boxope{A\mathds{1}}_n
= \sum_{m\geq n}\frac{(-1)^m}{(m-n)!}\partial^{(m-n)}\boxope{\mathds{1}A}_m
= \sum_{m\geq n}\delta_{0,m}\frac{(-1)^m}{(m-n)!}\partial^{(m-n)}A \,.
\end{equation}
We also have
\begin{align}
\boxope{\partial A B}_{n+1} = -n \boxope{ A B}_n ~\,\qquad~\qquad \forall n \in \mathbb{Z} \quad &\text{(Differentiation rule I)}\,, \label{eq:boxDelAB} \\
\boxope{ A \partial B}_{n+1} = n \boxope{A B}_n + \partial \boxope{AB}_{n+1}
 ~ \forall n \in \mathbb{Z} \quad &\text{(Differentiation rule II)}\,.\label{eq:boxADelB}
\end{align}

A consequence of \eqref{eq:boxDelAB} is that all boxes with negative label can be expressed in terms of the box with label $0$ and powers of $\partial$, hence a special terminology is convenient. The \emph{normal ordered product} \label{p:NO} of $A, B \in \mathcal{V}$ is
\begin{equation}
\no{ A B } = \boxope{AB}_0 \,. 
\end{equation}

The normal ordered product obeys the Leibniz rule $\partial \no{A B} = \no{\partial A B} + \no{A \partial B}$\,. More generally, we have
\begin{equation}\label{eq:boxLeibniz}
\partial \boxope{AB}_n = \boxope{\partial A B}_n + \boxope{A \partial B}_n
\qquad \forall n \in \mathbb{Z} \qquad \text{(Leibniz rule)}\,.
\end{equation}
\textit{Proof.} By the derivative rules~\eqref{eq:boxADelB} and \eqref{eq:boxDelAB},
\begin{align}
\partial \boxope{AB}_{n+1} &= \boxope{A\partial B}_{n+1}-n\boxope{AB}_n \,, \\
&=\boxope{A\partial B}_{n+1} + \boxope{\partial A B}_{n+1} \,.\qquad\qquad\square
\end{align}

Now notice that repeated applications of \eqref{eq:boxDelAB} yield
\begin{equation} \label{eq:boxNegative}
\boxope{AB}_{n\leq 0} = \frac{1}{(-n)!} \no{(\partial^{(-n)}A)B} \qquad \text{(Regular terms)}\,.
\end{equation}
This is precisely as expected from the interpretation of $\boxope{AB}_n$ as poles in the expansion of $A(z)$ about $B(w)$. We may thus now introduce the \textit{operator product expansion} \label{p:OPE} (OPE), be it only as a mnemonic gadget to record the sequence of box maps and its properties. Of course, it can also be interpreted and used as an expansion of complex meromorphic functions, as is usually assumed in two-dimensional CFT.

Let $A, B \in \mathcal{V}$. The $AB$ OPE has a left hand side denoted $\{A(z) B(w)\}$, or simply $A(z) B(w)$ and a right hand side which is a formal power series with the boxes as coefficients. The two sides are joined with the symbol ``$\sim$''. We write
\begin{equation}
A (z) B (w)
\sim \sum_{n\in \mathbb{Z}} \frac{\boxope{AB}_n(w)}{(z-w)^n} \,.
\end{equation}

The \emph{singular OPE} is a part of the right hand side (a finite sum),
\begin{equation}
\wick{\c A (z) \c B (w)}
= \sum_{n\in \mathbb{Z}^{>0}} \frac{\boxope{AB}_n(w)}{(z-w)^n} \,,
\end{equation}
and the \emph{regular OPE} is the remaining part,
\begin{equation}
\no{A(z) B(w)}
= \sum_{n\in \mathbb{Z}^{\leq 0}} \frac{\boxope{AB}_n(w)}{(z-w)^n} = \sum_{n=0}^\infty (z-w)^n ~\frac{1}{n!} \no{(\partial^{n}A)B}(w) \,,
\end{equation}
where we used \eqref{eq:boxNegative} for the second equality. The notation $\no{A(z) B(w)}$ for the regular OPE is due to all of its terms being normal ordered. However it is important to distinguish from $\no{AB}(w)$ which is only the leading term in the regular OPE. Our definition of normal ordering is consistent with the definition via point splitting, namely
\begin{equation}
\no{AB}(w) = \text{lim}_{z\rightarrow w} \left( A(z) B(w) - \wick{\c A(z) \c B(w)} \right) \,.
\end{equation}
For free CFTs, it also coincides with the prescription to place all annihilation modes to the right \cite[p.173]{DiFrancesco:1997nk}.

Note that the derivative formula~\eqref{eq:boxDelAB} is consistent with a naive holomorphic derivative with respect to $z$ on both sides of the $AB$ OPE. Similarly \eqref{eq:boxADelB} is obtained by a naive derivative with respect to $w$. Also one can take linear combinations of OPEs by bilinearity of $\boxope{- \,, -}_n$. \detail{We illustrate such manipulations with OPEs with an example.

\noindent\emph{Example. Free fermion I}\hrulefill

Let $\psi\in\mathcal{V}$ with $|\psi|=1$ ($\psi$ is fermionic). Consider the $\psi\psi$ OPE
\begin{equation}
\psi(z)\psi(w) \sim \frac{\mathds{1}(w)}{z-w} + \text{reg}(\psi(z)\psi(w)) \,.
\end{equation}
The $\partial \psi \psi$ OPE can be obtained by a naive holomorphic derivative with respect to $z$ on both sides:
\begin{align}
\partial\psi(z)\psi(w)
&\sim \frac{\partial}{\partial z} \left(\frac{\mathds{1}(w)}{z-w} + \no{\psi\psi}(w) + (z-w)\no{\partial\psi\psi}(w) + \ldots\right) \\
&=-\frac{\mathds{1}(w)}{(z-w)^2} + \no{\partial \psi\psi}(w) + \ldots \,.
\end{align}
Similarly, we can obtain the $\psi \partial \psi$ OPE by a naive holomorphic derivative with respect to $w$ on both sides:
\begin{align}
\psi(z)\partial\psi(w)
&\sim \frac{\partial}{\partial w} \left(\frac{\mathds{1}(w)}{z-w} + \no{\psi\psi}(w) + (z-w)\no{\partial\psi\psi}(w) + \ldots\right) \\
&=\frac{\mathds{1}(w)}{(z-w)^2} + \partial \no{\psi\psi}(w) - \no{\partial\psi\psi}(w) + (z-w)\partial \no{\partial\psi\psi}(w) + \ldots \,. \label{eq:egPsiDelPsi}
\end{align}
The term at order $(z-w)^0$ is equal to $\no{\psi \partial \psi}$ by Leibniz rule \eqref{eq:boxLeibniz}, as it should be expected from \eqref{eq:boxNegative}. We have also used above
\begin{equation}
\partial \mathds{1} = \boxope{\mathds{11}}_{-1} = 0 \,.
\end{equation}
\noindent\hrulefill}

Moving on, the unity axiom \eqref{eq:unityAxiom} in OPE form is
\begin{equation}
\mathds{1}(z)A(w) \sim A(w) \,.
\end{equation}
This is consistent with the idea that $\mathds{1}$ acts trivially. The commutativity axiom \eqref{eq:commutativityAxiom} is equivalent to allowing that, on the left hand side of the OPE (i.e. inside correlators), the operators commute up to sign,
\begin{equation}
A(z) \cdot B(w) = (-1)^{|A||B|}B(w) \cdot A(z) \,,
\end{equation}
while on the right hand side, the original formal power series can be formally re-expanded to be put in canonical form. \detail{This is best understood from an example.

\noindent \emph{Example. Free fermion II}\hrulefill

The $\psi \partial\psi$ OPE that we have obtained above by differentiation can also be obtained by commutativity from the $\partial \psi \psi$ OPE. We use that $\psi$ and $\partial \psi$ commute up to a sign on the left hand side of the OPE and then perform simple manipulations.
\begin{align}
\partial\psi(z)\psi(w) &\sim -\frac{\mathds{1}(w)}{(z-w)^2} + \no{\partial \psi\psi}(w) + \ldots \\
-\psi(w)\partial\psi(z) &\sim -\frac{\mathds{1}(w)}{(z-w)^2} + \no{\partial \psi\psi}(w) + \ldots \\
\psi(w)\partial\psi(z) &\sim \frac{\mathds{1}(w)}{(z-w)^2} - \no{\partial \psi\psi}(w) - \ldots \\
\psi(z)\partial\psi(w) &\sim \frac{\mathds{1}(z)}{(w-z)^2} - \no{\partial \psi\psi}(z) - \ldots \\
 &= \frac{\mathds{1}(w)}{(w-z)^2} - \no{\partial \psi\psi}(w) + \ldots
\end{align}
In the last step we have formally performed Taylor expansions about $w$ in $z-w$ using the fact that $z=w+(z-w)$. The term at order $(z-w)^0$ can be written\footnote{This is proven as follows. $\no{\psi \psi}=0$ by \eqref{eq:consistencySelfOPE} below, so $0=\partial \no{\psi \psi} = \no{\partial \psi \psi} + \no{\psi \partial \psi}$.} as $\no{\psi \partial \psi} (w)$, matching the result \eqref{eq:egPsiDelPsi} found previously.

\noindent\hrulefill}

Finally the associativity axiom \eqref{eq:Jacobilike} (for $n\geq 1$) in OPE language can be written as a statement about the singular OPE between a field $A$ and the full OPE of other fields $B$ and $C$ \cite[lemma 2.3.1]{Thielemans:1994er}. Practically a simpler statement is very useful, namely the singular OPE of $A$ with the normal ordered product $\no{BC}$\, \cite[section 6.B]{DiFrancesco:1997nk}. This is a generalised version of \textit{Wick's theorem} valid in interacting CFTs. Let $A,B,C\in\mathcal{V}$. We have
\begin{align}
\wick{\c A(z) \c {\no{BC}}}(w) = \frac{1}{2\pi i} \oint_{\mathbb{S}^1_{w}}\frac{dx}{x-w} \left(\{ \wick{\c A(z) \c B(x)}~ C(w) \} + (-1)^{|A||B|}\{B(x) ~ \wick{\c A(z) \c C(w)}\}\right)\,, \label{eq:Wick}
\end{align}
where the brackets $\{ \}$ emphasise that the \textit{full} OPE must be used (not only the singular terms). \detail{Some formulas from complex analysis are useful to have in mind when using this result in practical calculations. Let $f(z)$ be analytic on $\mathbb{C}$.

\begin{align}
&\oint f(z) d z = 0   \qquad    \text{(Cauchy's theorem)}\\
&\oint_{\mathbb{S}^1_w} \frac{dz}{2\pi i} \frac{f(z)}{(z-w)^n}=\frac{f^{(n-1)}(w)}{(n-1)!} \quad n\in\mathbb{Z}^{\geq 1}      \qquad    \text{(Cauchy's integral formula)}\\
&\oint_{\mathbb{S}^1_w} \frac{dx}{2\pi i} \frac{1}{(x-w)^n}\frac{1}{(z-x)^m}
=\frac{(n+m-2)!}{(n-1)!(m-1)!}\frac{1}{(z-w)^{n+m-1}} \quad n, m\in\mathbb{Z}^{\geq 1}
\end{align}

\noindent\textit{Example. }
$\wick{\c T \c{\no{GG}}}$ \hrulefill

Let $T$ be a Virasoro operator, i.e.
\begin{equation}
\wick{\c T(z) \c T(w)} = \frac{c / 2 ~ \mathds{1}(w)}{(z-w)^4} + \frac{2T(w)}{(z-w)^2} + \frac{\partial T(w)}{z-w} \,.
\end{equation}
Let $G$ be a fermionic primary of weight $3/2$ with respect to $T$, i.e.
\begin{equation}
\wick{\c T(z) \c G(w)} = \frac{3/2 G(w)}{(z-w)^2} + \frac{\partial G(w)}{z-w} \,.
\end{equation}
Finally, let $G$ have the following self-OPE
\begin{equation}
\wick{\c G(z) \c G(w)} = \frac{2c/3~\mathds{1}(w)}{(z-w)^3} + \frac{2T(w)}{z-w} \,.
\end{equation}
We will compute the $\wick{\c T \c{\no{GG}}}$ OPE using \eqref{eq:Wick}. We will also show in an example below that $\no{GG} = \partial T$. It will be a nice consistency check that our result matches what can be obtained by differentiation of the $TT$ OPE with respect to $w$.

We will need the $\partial G G$ and $G\partial G$ OPEs at intermediate steps. By differentiation with respect to $z$, we get
\begin{equation}
\wick{\c {\partial G}(z) \c G(w)} = -\frac{2c~\mathds{1}(w)}{(z-w)^4} - \frac{2T(w)}{(z-w)^2} \,
\end{equation}
and
\begin{equation}
\wick{\c G(z) \c {\partial G} (w)} = \frac{2c ~ \mathds{1}(w)}{(z-w)^4}+\frac{2T(w)}{(z-w)^2}+\frac{2\partial T (w)}{z-w} \,.
\end{equation}

Focusing on the first term in \eqref{eq:Wick}, we have
\begin{equation*}
\frac{1}{2\pi i} \oint_{\mathbb{S}^1_w}\frac{dx}{x-w} \{ \wick{\c T(z) \c G(x)} ~ G(w) \}
= \frac{1}{2\pi i} \oint_{\mathbb{S}^1_w}\frac{dx}{x-w} \left\lbrace \left( \frac{3/2 G(x)}{(z-x)^2} + \frac{\partial G(x)}{z-x} \right) ~ G(w) \right\rbrace
\end{equation*}
{\scriptsize
\begin{equation*}
= \frac{1}{2\pi i} \oint_{\mathbb{S}^1_w}\frac{dx}{x-w} \left[ \frac{3/2}{(z-x)^2}\left( \frac{2c/3~\mathds{1}(w)}{(x-w)^3} + \frac{2T(w)}{x-w} + \no{GG}(w) + \ldots \right) + \frac{1}{z-x} \left( -\frac{2c~\mathds{1}(w)}{(x-w)^4} - \frac{2T(w)}{(x-w)^2} + \no{\partial G G}(w) + \ldots \right) \right]
\end{equation*}
\begin{equation*}
= \frac{1}{2\pi i} \oint_{\mathbb{S}^1_w} dx \left[ \frac{1}{(z-x)^2}\left( \frac{c~\mathds{1}(w)}{(x-w)^4} + \frac{3T(w)}{(x-w)^2} + \frac{3/2 \no{GG}(w)}{x-w} + \ldots \right) + \frac{1}{z-x} \left( -\frac{2c~\mathds{1}(w)}{(x-w)^5} - \frac{2T(w)}{(x-w)^3} + \frac{\no{\partial G G}(w)}{x-w} + \ldots \right) \right]
\end{equation*}
}
Next, we use the formulas from complex analysis. The structure of the regular terms that we have represented with ``$\ldots$'' is such that, by Cauchy's theorem, they do not survive the integral. Continuing, the above is equal to
\begin{equation*}
\frac{4c~\mathds{1}(w)}{(z-w)^5} + \frac{6T(w)}{(z-w)^3} + \frac{3/2 \no{GG}(w)}{(z-w)^2} -\frac{2c~\mathds{1}(w)}{(z-w)^5} - \frac{2T(w)}{(z-w)^3} + \frac{\no{\partial G G}(w)}{z-w}
\end{equation*}
\begin{equation*}
=\frac{2c~\mathds{1}(w)}{(z-w)^5} + \frac{4T(w)}{(z-w)^3} + \frac{3/2 \no{GG}(w)}{(z-w)^2} + \frac{\no{\partial G G}(w)}{z-w} \,.
\end{equation*}

We must next do a similar manipulation starting with the second term in the formula in \eqref{eq:Wick}. This is actually much easier.
\begin{equation*}
\frac{1}{2\pi i} \oint_{\mathbb{S}^1_w}\frac{dx}{x-w} \{G(x) ~ \wick{\c T(z) \c G(w)}  \}
= \frac{1}{2\pi i} \oint_{\mathbb{S}^1_w}\frac{dx}{x-w} \left\lbrace G(x)  ~ \left( \frac{3/2 G(w)}{(z-w)^2} + \frac{\partial G(w)}{z-w} \right) \right\rbrace
\end{equation*}
{\scriptsize
\begin{equation*}
=\frac{1}{2\pi i} \oint_{\mathbb{S}^1_w}\frac{dx}{x-w}
\left[
\frac{3/2}{(z-w)^2} \left(
\frac{2c/3~\mathds{1}(w)}{(x-w)^3} + \frac{2T(w)}{x-w} + \no{GG}(w) + \ldots
\right)
+
\frac{1}{z-w} \left(
\frac{2c ~ \mathds{1}(w)}{(x-w)^4}+\frac{2T(w)}{(x-w)^2}+\frac{2\partial T (w)}{x-w} + \no{G\partial G}(w)+ \ldots
\right)
\right]
\end{equation*}
\begin{equation*}
=\frac{1}{2\pi i} \oint_{\mathbb{S}^1_w} dx
\left[
\frac{1}{(z-w)^2} \left(
\frac{c~\mathds{1}(w)}{(x-w)^4} + \frac{3 T(w)}{(x-w)^2} + \frac{3/2\no{GG}(w)}{x-w} + \ldots
\right)
+
\frac{1}{z-w} \left(
\frac{2c ~ \mathds{1}(w)}{(x-w)^5}+\frac{2T(w)}{(x-w)^3}+\frac{2\partial T (w)}{(x-w)^2} + \frac{\no{G\partial G}(w)}{x-w} + \ldots
\right)
\right]
\end{equation*}
}
Now, the factors of $(z-w)$ can be pulled out of the integral and we are left with integrals to be dealt with using Cauchy's formula. Of all the terms, only those with the factor $(x-w)^{-1}$ survive. Continuing, the above is equal to
\begin{equation*}
\frac{3/2\no{GG}(w)}{(z-w)^2}
+
\frac{\no{G\partial G}(w)}{z-w} \,.
\end{equation*}
Summing the two contributions, we have the final result:
\begin{equation}
\wick{\c T(z) \c {\no{GG}}(w)} = 
\frac{2c~\mathds{1}(w)}{(z-w)^5} + \frac{4T(w)}{(z-w)^3} + \frac{3 \no{GG}(w)}{(z-w)^2} + \frac{\partial\no{G G}(w)}{z-w} \,,
\end{equation}
where we have used the Leibniz rule of \eqref{eq:boxLeibniz} to combine the terms at order 1 pole. Identifying $\no{GG} = \partial T$, it is easy to check that this is consistent with the result of differentiating the $TT$ OPE with respect to $w$.

\noindent\hrulefill
}

The associativity axiom is the most constraining behind a consistent operator algebra. In physics terms, it guarantees that the order in which contractions are made within correlators does not affect the final result (crossing symmetry). In simple operator algebras, the Jacobi-like identities \eqref{eq:Jacobilike} are satisfied on the nose. Generically however, one must allow so-called \textit{null} fields $\{N^1,N^2,\ldots\}$ \cite{Thielemans:1994er}. Physically they are characterised by their vanishing correlators with all other fields. In particular, null fields could arise in any consistency condition for the operator algebra without impacting measurable quantities. The infinite set of fields obtained from these null fields by taking any combinations of derivatives, normal ordered products and OPEs with any other fields is called the \textit{ideal} \label{p:ideal} generated by $\{N^1,N^2,\ldots\}$. We denote it by $\left\langle N^1, N^2,\ldots\right\rangle$. We will denote statements holding up to fields in the ideal with a quotient, $\Big/ \langle N^1, N^2,\ldots \rangle$. The technicality of ideals is unavoidable for us because it concerns the operator algebras related to $G_2$ and $SU(3)$-structure \cite{Odake:1988bh, Figueroa-OFarrill:1996tnk}, as described in sections~\ref{sec:SVG2}--\ref{sec:Odake}.

Most operator algebras we will encounter are \textit{finitely generated}. A set of elements $\{g_1, g_2,\ldots\}\subset \mathcal{V}$ is said to \label{p:generate} \emph{generate} $\mathcal{V}$ if and only if all other elements of $\mathcal{V}$ can be obtained as linear combinations of normal ordered products of derivatives of the generators. In such a case, only the singular OPEs between the generators are sufficient to fully specify the algebra. We shall denote a finitely generated operator algebra by listing its generators: $\mathcal{V}=\langle g_1, g_2,\ldots \rangle$.

\subsection{Virasoro $\mathcal{N}=0$ and $\mathcal{N}=1$} \label{sec:Virasoro}

We now consider examples of operator algebras, starting slowly with the non-supersymmetric \label{p:Virasoro} \voc{Virasoro algebra} $\text{Vir}^0_c$ (or just $\text{Vir}_c$), the smoking gun of chiral conformal invariance at the quantum level. It is generated by the single element $T\in\mathcal{V}$ (\emph{Virasoro operator}) with $|T|=0$ and the singular self-OPE as in \eqref{eq:VirasoroOA}:
\begin{equation} \label{eq:OPE_TT}
\wick{\c T(z) \c T(w)} = \frac{c / 2 ~ \mathds{1}(w)}{(z-w)^4} + \frac{2T(w)}{(z-w)^2} + \frac{\partial T(w)}{z-w} \,,
\end{equation}
for some $c\in\mathbb{R}$ called the \textit{central charge}. From now on, we will omit to write the identity field $\mathds{1}$. The importance of this algebra for us is that it sits inside others to be introduced later; sometimes in unexpected ways.

The presence of $T\in \mathcal{V}$ allows to talk about weights and primaries. Let $A\in\mathcal{V}$ be any element. If 
\begin{equation}
\boxope{TA}_{2} = h(A) A \,,
\end{equation} 
for a certain number $h(A)$ (typically real), we shall say that \emph{$A$ has well-defined weight $h(A)$ with respect to $T$}. If moreover
\begin{equation}
\wick{\c T(z) \c A(w)} = \frac{h(A) A(w)}{(z-w)^2} + \frac{\partial A(w)}{z-w} \,,
\end{equation}
we shall call $A$ a \emph{weight-$h(A)$ primary with respect to $T$}.

%

Our interest will often be in the \textit{$\mathcal{N}=1$ supersymmetric Virasoro algebra} \label{p:Virasoro1} $\text{Vir}^1_c$, obtained by adjoining to $\text{Vir}_c$ a fermionic primary $G$ of conformal weight~$3/2$:
\begin{align}
\wick{\c T(z) \c G(w)} &=\frac{3/2}{(z-w)^2}G(w)+\frac{\partial G(w)}{z-w}\,,  \label{eq:OPE_TG} \\
\wick{\c G(z) \c G(w)} &=\frac{2c/3}{(z-w)^3}+\frac{2T(w)}{z-w}\,. \label{eq:OPE_GG}
\end{align}

From \eqref{eq:OPE_TG}, we could work out the $GT$ OPE and, by definition, we would conclude from this and \eqref{eq:OPE_GG} that the ordered pair $(\tfrac{1}{2}G,T)$ forms an \voc{$\mathcal{N}=1$ multiplet}---i.e. $\tfrac{1}{2}G$ and $T$ are \voc{supersymmetric partners}.

The presence of $\text{Vir}^1_c$ inside an operator algebra $\mathcal{V}$ is the signature of chiral supersymmety as well as conformal invariance and, as such, we will call such algebras \textit{superconformal}. Analogously to the non-supersymmetric case, there is a notion of \label{p:superprimary} \textit{superprimary} multiplet $(A,B)$ of weight $h(A)$ defined by
\begin{align}
\wick{\c T(z) \c A(w)} &= \frac{h(A) A(w)}{(z-w)^2} + \frac{\partial A(w)}{z-w} \,, \\
\wick{\c T(z) \c B(w)} &= \frac{(h(A)+1/2) B(w)}{(z-w)^2} + \frac{\partial B(w)}{z-w} \,, \\
\wick{\c G(z) \c A(w)} &= \frac{B(w)}{z-w} \,, \\
\wick{\c G(z) \c B(w)} &= \frac{2h(A) A(w)}{(z-w)^2} + \frac{\partial A(w)}{z-w} \,.
\end{align}

Further superconformal algebras $\text{Vir}^{\mathcal{N}}_c$ with $\mathcal{N}=2$ and $\mathcal{N}=4$ will also be encountered in section~\ref{sec:Odake}.

\vfill

\section{Free fermions and bosons} \label{sec:Free}

Free chiral fermions and bosons are deceptively simple examples of operator algebras. They have a rather rich structure and we shall use them all over the rest of this chapter to gain geometric intuition about other algebras.
We can see them as describing, in conformal field theory, the local geometry of curved target spaces (the large radius limit). By the same token, they enter the description of string dynamics on flat manifolds such as $\mathbb{R}^d$ or real tori $\mathbb{T}^d$. 

Let $d\in \mathbb{Z}^{\geq 1}$ and consider $d$ fermionic fields $\psi^i$ (Majorana--Weyl fermions) and $d$ bosonic fields ${\mathscr{I}}^i$, $i\in \{ 1,\ldots, d \}$. We can think of the ${\mathscr{I}}^i$s as holomorphic derivatives of free bosonic fields: ${\mathscr{I}}^i=i \partial x^i$. The only non-vanishing singular OPEs are defined as
\begin{equation} \label{eq:freeOPEs}
\wick{\c \psi^i (z) \c \psi^j (w)} = \frac{\delta^{ij}}{z-w} \,, \qquad
\wick{\c {\mathscr{I}}^i (z) \c {\mathscr{I}}^j (w)} = \frac{\delta^{ij}}{(z-w)^2} \,.
\end{equation}
For $d=1$, we denote this algebra by \label{p:Free} $\text{Free}=\langle \psi , {\mathscr{I}} \rangle$. For $d\geq 1$, we write $(\text{Free})^d$. 

We can make contact with Virasoro algebras by defining
\begin{equation} \label{eq:freeTfTb}
T_f = \sum_{i=1}^d \frac{1}{2} \no{\partial \psi^i \psi^i} \,, \qquad
T_b = \sum_{i=1}^d \frac{1}{2} \no{{\mathscr{I}}^i {\mathscr{I}}^i} \,.
\end{equation}
It can indeed be checked with the generalised Wick theorem \eqref{eq:Wick} that $T_f$ is a Virasoro operator. It has the singular OPE as in \eqref{eq:OPE_TT} for central charge $c_f=d/2$. Similarly $T_b$ generates a Virasoro algebra with central charge $c_b=d$. Since $\psi^i$s and ${\mathscr{I}}^i$s have vanishing mutual singular OPEs---they \voc{commute}---so do $T_f$ and $T_b$. Therefore
\begin{equation} \label{eq:freeT}
T=T_f+T_b
\end{equation}
is also a Virasoro operator, now with central charge $c_f+c_b=3d/2$.

Relative to their Virasoro operators $T_f$ and $T_b$ (and relative to $T$), $\psi^i$ and ${\mathscr{I}}^i$ transform like primaries of weight $1/2$ and $1$ respectively. One can further extend to $\text{Vir}^1_{3d/2}$ by defining the supersymmetry current
\begin{equation} \label{eq:freeG}
G=\sum_{i=1}^d \no{{\mathscr{I}}^i \psi^i} \,,
\end{equation}
and it can be checked that $(\psi^i,{\mathscr{I}}^i)$ assemble into a weight-$1/2$ superprimary with respect to $\text{Vir}^1_{3d/2}=\langle T, G \rangle$.

Together \eqref{eq:freeTfTb}--\eqref{eq:freeG} define a map
\begin{equation} \label{eq:VirInFree}
\text{Vir}^1_{3d/2} \longhookrightarrow (\text{Free})^d \,,
\end{equation}
which is our first example of \label{p:realisation} ``realisation'' of an operator algebra.
Let $\mathcal{V}$ and $\mathcal{W}$ be two finitely generated operator algebras. We shall say that $\mathcal{V}$ is \emph{realised} in $\mathcal{W}$ if there is a map $\mathcal{V}\longhookrightarrow\mathcal{W}$, defined by its action on generators, such that the singular OPEs of $\mathcal{V}$ are consistent with the singular OPEs of $\mathcal{W}$. For example in \eqref{eq:VirInFree}, the singular OPEs characterising $\text{Vir}^1_{3d/2}$ can be obtained (``realised'') from the singular OPEs of free fields via the explicit definitions \eqref{eq:freeTfTb}--\eqref{eq:freeG}.
We summarise such statements with a hooked arrow in order both to convey that $\text{Vir}^1_{3d/2}$ is included inside $(\text{Free})^d$ and to highlight the map by which this inclusion is borne out. This notation is used abundantly in chapter~\ref{chap:TCS}.

\subsection{Virasoro operators from quartic fermion products} \label{sec:exercise}

The interpretation of equation \eqref{eq:VirInFree} in the physical context of non-linear $\sigma$-models is that flat target spaces such as $\mathbb{R}^d$ or $\mathbb{T}^d$, corresponding to $(\text{Free})^d$, are in particular Ricci-flat, the latter being necessary\footnote{Ricci-flatness is a gross simplification: it is only a leading-order approximation in the $\alpha'$ (large radius) expansion. Moreover even this leading result is affected by the presence of background fluxes in the target space geometry, an issue in which we will be interested in part~\ref{part:II} of the thesis.} for $\mathcal{N}=1$ superconformal invariance. Geometrically this is of course very innocuous, but the algebraic counterpart \eqref{eq:VirInFree} reveals the interesting structure $\text{Vir}_{3d/2}^1$ inside $(\text{Free})^d$. In the rest of this chapter, we identify and describe other subalgebras of $(\text{Free})^d$ which shine the spotlights on other geometric facets of $\mathbb{R}^d$ or $\mathbb{T}^d$, more intricate than Ricci-flatness. This will be achieved through an exercise which we now describe.

Consider $\mathbb{R}^d$ or $\mathbb{T}^d$ with Cartesian coordinates $x^i$ and perform the identification
\begin{equation}
\psi^i \longleftrightarrow \dd x^i
\end{equation}
between fermions and the coordinate basis of $1$-forms. The free fermionic algebra is such that we can generalise to forms of higher degrees, identifying the normal ordered product with wedging:
\begin{equation}
\no{ ~ } \longleftrightarrow \wedge \,.
\end{equation}
For instance, we have $\no{\psi^i \psi^j}=0$ for $i=j$, as expected from $\dd x^i \wedge \dd x^i = 0$. This is a consequence of the general result
\begin{equation} \label{eq:consistencySelfOPE}
\boxope{AA}_n = \frac{1}{2}\sum_{m\geq 1} \frac{(-1)^{m+1}}{m!}\partial^{m}\boxope{AA}_{m+n}\,,
\qquad \forall n \in \mathbb{Z},~n+|A|\text{ odd}\,,
\end{equation}
which is easily proven from the axioms of operator algebras.
All the expected antisymmetry properties of differential forms can be checked directly from the OPEs \eqref{eq:freeOPEs}.

Normal ordering being a pairwise operation, we must specify a convention for successive normally ordered products. We choose \textit{right-nesting}.
As an example, let $\Psi \in \Omega^4(\mathbb{R}^d,\mathbb{R})$ be a $4$-form on $\mathbb{R}^d$ with constant coefficients. We construct the corresponding weight-$2$ operator as
\begin{equation} \label{eq:PformField}
\widehat{\Psi}^{(d)} = \frac{1}{4!}\Psi_{ijkl} \, \no{\psi^i\no{\psi^j\no{\psi^k\psi^l}}} \,.
\end{equation}
Forms with a hat denote the corresponding field theory operator constructed in this way.  To save space, we will use the shorthand $\no{\psi^{ijkl}}$ for the right-nested normal ordered product $\no{\psi^i\no{\psi^j\no{\psi^k\psi^l}}}$\,. The superscript $(d)$ helps us track the dimension we are working in. 

Notice that a quartic fermionic combination like $\widehat{\Psi}^{(d)}$ has weight $4\cdot (1/2) = 2$, precisely like Virasoro operators. Let $\Psi^{(d)}$ be a \emph{constant} $4$-form on $\mathbb{R}^d$ or $\mathbb{T}^d$. In the following sections, we seek values of $\mu \Psi_{ijkl}$ and $\nu$ such that the linear combination
\begin{equation} \label{eq:TminusAnsatz}
\mu \widehat{\Psi}^{(d)} + \nu T_f
\end{equation}
satisfies a Virasoro algebra for some central charge. This exercise may sound artificial, but it will prove interesting in many ways.
It is motivated by the free field realisations of the $Spin(7)$ and $G_2$ algebras discovered by Shatashvili and Vafa that we discuss next.
The first lesson revealed by the exercise is that these two algebras have, in addition to the conformal symmetry generated by $T$, a ``hidden'' conformal symmetry. These hidden sectors influence significantly the spectrum of the full algebras. This is very important for instance for the application in chapter~\ref{chap:MarginalDeformations}. In free field realisations, the hidden Virasoro operators have the form \eqref{eq:TminusAnsatz}.

The exercise is also a nice way to introduce so-called Odake algebras in their free field realisations. We will do this for the Odake algebra associated to $SU(3)$-structure target spaces in section~\ref{sec:Odake}. Again, surprisingly, a hidden Virasoro operator of the form \eqref{eq:TminusAnsatz} exists.

\section{$Spin(7)$ Shatashvili--Vafa} \label{sec:SVSpin(7)}

We take $d=8$ in the free algebra $(\text{Free})^d$ discussed in section~\ref{sec:Free}. Rather than seeking a Virasoro operator \eqref{eq:TminusAnsatz} for an arbitrary $4$-form $\Psi^{(8)}$, we refine the ansatz and consider directly the quartic combination
\begin{align} \label{eq:Psi8}
\widehat{\Psi}^{(8)} &=
\no{\psi^{1234}} 
+ \no{\psi^{1256}} 
+ \no{\psi^{1278}} 
+ \no{\psi^{1357}} 
- \no{\psi^{1368}} 
- \no{\psi^{1458}} 
- \no{\psi^{1467}} \\ &\qquad
- \no{\psi^{2358}}
- \no{\psi^{2367}}
- \no{\psi^{2457}}
+ \no{\psi^{2468}}
+ \no{\psi^{3456}}
+ \no{\psi^{3478}}
+ \no{\psi^{5678}} \,. \nonumber
\end{align}

Endowing $\mathbb{R}^8$ with the $4$-form $\Psi^{(8)}$ associated to this $\widehat{\Psi}^{(8)}$ yields the simplest example of a manifold with a \voc{Spin(7)-structure}. More generally, a $Spin(7)$-structure on an eight-dimensional manifold $\mathcal{M}$ can be understood as the existence on $\mathcal{M}$ of a nowhere vanishing $4$-form for which there exists, at any given point $p\in\mathcal{M}$, an isomorphism $T^*_p\mathcal{M}\rightarrow \mathbb{R}^8$ upon which the $4$-form maps to the reference $\Psi^{(8)}$ on $\mathbb{R}^8$. Manifolds whose Riemannian holonomy group is contained in $Spin(7)$ are examples of $Spin(7)$-structure manifolds, but the latter concept is more general. More details can be found in \cite[p.\,239]{Joyce2007}, whose conventions we follow.

With the choice \eqref{eq:Psi8} and aside from $T_f$ itself (see \eqref{eq:freeTfTb}), explicit calculations reveal that there are only two possible non-trivial Virasoro operators of the form \eqref{eq:TminusAnsatz}:
\begin{align} \label{eq:Tplusminus8}
T_- &= \frac{1}{8}\left(-\widehat{\Psi}^{(8)} +  T_f \right)\,, \qquad c_- = \frac{1}{2}\,, \\
T_{+f} &= \frac{1}{8}\left(\widehat{\Psi}^{(8)} + 7 T_f\right)\,, \qquad c_{+f} = \frac{7}{2}\,.
\end{align}
Notice that
\begin{equation}
T_-+T_{+f}=T_{f} \,,
\end{equation}
so, adding $T_b$ on both sides, we get
\begin{equation}
T_- + (\underbrace{T_{+f}+T_b}_{T_+}) = \underbrace{T_f + T_b}_{T} \,.
\end{equation}

$T_-$ especially draws our attention because its central charge is inferior to $1$. In 2D CFT, only a discrete set of central charges below $1$ are compatible with unitarity. They correspond to the sequence of \voc{minimal models}. We are lucky here that $c_- = 1/2$ is one of them: $\langle T_- \rangle$ is isomorphic to the \voc{Ising model}. As the name suggests, it is related to lattice models of ferromagnetism in statistical mechanics. 


We have now witnessed a remarkable fact thanks to our simple exercise with free fermions: $Spin(7)$-structure manifolds, essentially defined by the intricate $4$-form~\eqref{eq:Psi8}, are intimately connected, through strings and conformal field theory, to Ising lattice models in condensed matter physics. That such complicated geometries are related to physical systems as simple as minimal models is one of the wonders of this subject. A similar statement holds for the group $G_2$, as we will see in section~\ref{sec:SVG2}. We regard this as a deep mystery and as a good incentive to understand better the link between geometry and CFT.

One may argue that this is a figment of the free field algebra and that perhaps this connection only holds for simple $Spin(7)$-manifolds like $\mathbb{R}^8$ or $\mathbb{T}^8$. The hypothesis of Shatashvili and Vafa \cite{Shatashvili:1994zw} is however that an abstract operator algebra---independent of any free field realisation---should still describe string theory on any (curved) manifold with holonomy group $Spin(7)$. Our considerations in this thesis can be taken as indirect evidence of the relevance of this algebra, which we denote $\text{SV}^{Spin(7)}$, even for more general compactifications on $Spin(7)$-structure manifolds with vector bundle. Convincing arguments appeared recently in \cite{Melnikov:2017yvz} explaining that $\text{SV}^{Spin(7)}$ is indeed required by spacetime supersymmetry in heterotic compactifications. For now, we review how $\text{SV}^{Spin(7)}$ obtains from our results so far.

Up to a different $Spin(7)$ convention for $\widehat{\Psi}^{(8)}$ than ours, stripping $T_-$ in \eqref{eq:Tplusminus8} off its overall factor defines the operator $X$ of \cite{Shatashvili:1994zw}:
\begin{equation}
X=-\widehat{\Psi}^{(8)} + T_f \,.
\end{equation}
The pairwise OPEs between $X$ and $T=T_f+T_b$ only produce $X$ and $T$ themselves and their derivatives. However the $GX$ OPE produces a new combination of free fields which cannot be directly written in terms of $X$ and $T$. It is the supersymmetric partner of $X$, denoted by $M$ in \cite{Shatashvili:1994zw}:
\begin{equation}
M=\boxope{GX}_1 \,.
\end{equation}
A priori, taking further OPEs like $MM$ and $MX$, etc., can be expected to produce forever more new combinations of free fields. This however does not happen here: the fields $T, G, X$ and $M$ are sufficient to close a subalgebra of $(\text{Free})^8$. Very precisely, this means that all pairwise OPEs between $T, G, X$ and $M$ produce only singular terms involving operators obtained from $\{T, G, X, M\}$ by taking normal ordered products, derivatives and linear combinations. We have thus described a free field realisation (see p.\pageref{eq:VirInFree}),
\begin{equation} \label{eq:FF_Spin(7)}
\langle T,G,X,M\rangle \longhookrightarrow (\text{Free})^8 \,,
\end{equation}
of an operator algebra
\begin{equation}
\text{SV}^{Spin(7)}=\langle T,G,X,M\rangle \,,
\end{equation}
which also makes sense on its own: the \label{p:Spin7} \emph{$Spin(7)$ Shatashvili--Vafa operator algebra}. When dealing with arbitrary curved manifolds with a $Spin(7)$-structure, one forgets about the free field realisation \eqref{eq:FF_Spin(7)} and focuses abstractly directly on $\text{SV}^{Spin(7)}$. The explicit OPEs defining the $\text{SV}^{Spin(7)}$ algebra are given in \cite{Shatashvili:1994zw}.

As pointed out in \cite{Figueroa-OFarrill:1996tnk}, the linear combinations\footnote{The overall factor is largely irrelevant for us.}
\begin{equation} \label{eq:WUforSpin7}
W=9(\tfrac{1}{3}T-X) \qquad \text{and}\qquad U=9(\tfrac{1}{6}G'-M)
\end{equation}
form a superprimary field $(W,U)$. Meanwhile it was proven in \cite{Figueroa-OFarrill:1990mzn} that there exists a unique family of extensions of $\mathcal{N}=1$ Virasoro by such a weight-$2$ superprimary, parametrised by the central charge and the normalisation of the weight-$2$ superprimary. We denote this family \label{p:W} $\mathcal{W}^1_{c}(\tfrac{3}{2};2)$. By uniqueness then, the $Spin(7)$ algebra corresponds to $c=8\cdot 3/2$ in this family:
\begin{equation}
\text{SV}^{Spin(7)} = \mathcal{W}^{1}_{12}(\tfrac{3}{2};2) \,.
\end{equation}
 
\section{$G_2$ Shatashvili--Vafa} \label{sec:SVG2}

We now obtain the $G_2$ Shatashvili--Vafa operator algebra through its free field realisation, just like we did in the $Spin(7)$ case, repeating the exercise from subsection~\ref{sec:exercise}, now taking $d=7$. We need a $4$-form on $\mathbb{R}^7$ and there is a very natural candidate if we are interested in the group $G_2$.

An oriented \textit{$G_2$-structure} on a seven-dimensional manifold $\mathcal{M}$ is equivalent to the existence of a nowhere vanishing \voc{positive} $4$-form $\Psi^{(7)}$. The positivity condition states that $\Psi^{(7)}$ should be isomorphic pointwise to a reference $4$-form on $\mathbb{R}^7$ such as
\begin{equation}
\dd x^{1357}
+ \dd x^{2345}
+ \dd x^{2367}
+ \dd x^{4567}
-\dd x^{1247}
- \dd x^{1256}
- \dd x^{1346} \,.
\end{equation}
The positive $4$-form canonically allows to define a Riemannian metric $G_{\Psi}$ on $\mathcal{M}$ from which a Hodge dual $3$-form $\Phi^{(7)}=*\Psi^{(7)}$ is derived. Often this is described in reverse: a $G_2$-structure is also equivalent to the existence of a nowhere vanishing positive $3$-form, in terms of which a metric and thus a Hodge dual $4$-form are defined. These approaches are entirely equivalent.

There is a nice trick to derive $\widehat{\Psi}^{(7)}$ and $\widehat{\Phi}^{(7)}$ from the $Spin(7)$ $4$-form operator $\widehat{\Psi}^{(8)}$ in section~\ref{sec:SVSpin(7)}. We are going to perform in reverse a construction in \cite{Shatashvili:1994zw}. Notice from \eqref{eq:Psi8} that we can write \detail{\cite[p.\,242]{Joyce2007}}
\begin{equation}
\widehat{\Psi}^{(8)} = \widehat{\Psi}^{(7)} + \no{\psi^1 \widehat{\Phi}^{(7)}} \,,
\end{equation}
which corresponds in geometry to reducing the $Spin(7)$-structure $\Psi^{(8)}$ on $\mathbb{R}^8_{x^1x^2\ldots x^8}$ to a $G_2$-structure $(\Psi^{(7)},\Phi^{(7)})$ on $\mathbb{R}^7_{x^2\ldots x^8}$. Relabelling $x^i \rightarrow x^{i-1}$ so that $i$ runs from $1$ to $7$, we write the resulting operators as\footnote{This agrees with the $G_2$ conventions of Joyce \cite[p.\,227]{Joyce2007}.}
\begin{align}
\widehat{\Psi}^{(7)} &=
\no{\psi^{1357}}
+ \no{\psi^{2345}}
+ \no{\psi^{2367}}
+ \no{\psi^{4567}}
-\no{\psi^{1247}}
- \no{\psi^{1256}}
- \no{\psi^{1346}} , \quad . \label{eq:Psi7}
\\
\widehat{\Phi}^{(7)} &=
 \no{\psi^{123}} 
+ \no{\psi^{145}} 
+ \no{\psi^{167}} 
+ \no{\psi^{246}} 
- \no{\psi^{257}} 
- \no{\psi^{347}} 
- \no{\psi^{356}} \,. \label{eq:Phi(7)}
\end{align}

We now repeat the exercise of finding Virasoro operators for the ansatz~\eqref{eq:TminusAnsatz} with the $4$-form \eqref{eq:Psi7}. Again, there are only two non-trivial possibilities:
\begin{align} \label{eq:Tplusminus7}
T_- = \frac{1}{5}\left(-\widehat{\Psi}^{(7)} +  T_f \right)\,, \qquad c_- = \frac{7}{10}\,, \\
T_{+f} = \frac{1}{5}\left(\widehat{\Psi}^{(7)} + 4 T_f\right)\,, \qquad c_{+f} = \frac{14}{5}\,,
\end{align}
and again $T_-+T_{+f}=T_f$, so that, adding the free bosonic energy-momentum tensor, we have
\begin{equation}
T_- + (\underbrace{T_{+f}+T_b}_{T_+}) = \underbrace{T_f + T_b}_{T} \,.
\end{equation}
Removing the overall coefficient of $T_-$, we define
\begin{equation}
\widetilde{X}=-X=-\widehat{\Psi}^{(7)}+T_f \,,
\end{equation}
where $X$ is as in \cite{Shatashvili:1994zw}. The opposite sign choice is preferable for homogeneity, but we will also use the conventional $X$, in particular in section~\ref{chap:TCS}. Defining a supersymmetric partner for $\widetilde{X}$ by
\begin{equation}
\widetilde{M}=-M=\boxope{G\widetilde{X}}_1
\end{equation}
is again sufficient to close an operator algebra $\langle T, G, \widetilde{X}, \widetilde{M}\rangle$. Defining
\begin{equation} \label{eq:WUforG2}
W=\tfrac{27}{2}(\tfrac{1}{3}T-\widetilde{X})\,,\qquad U=\tfrac{27}{2}(\tfrac{1}{6}G'-\widetilde{M}) \,,
\end{equation}
one finds that $(W,U)$ form a weight-$2$ superprimary. By uniqueness of the family of extensions of $\text{Vir}^1$ by such a superprimary (see section~\ref{sec:SVSpin(7)}), we thus conclude \cite{Gepner:2001px}
\begin{equation} \label{eq:FSinSVG2}
\langle T, G, \widetilde{X}, \widetilde{M}\rangle = \mathcal{W}^{1}_{21/2}(\tfrac{3}{2};2) \,.
\end{equation}

This is not yet the full $G_2$ operator algebra: the subalgebra \eqref{eq:FSinSVG2} was not in fact noticed in the original article \cite{Shatashvili:1994zw}. However it is remarkable that the \emph{same} family $\mathcal{W}^{1}_c(\tfrac{3}{2};2)$ has to do with string dynamics on manifolds endowed with \textit{either} of $Spin(7)$- or $G_2$-structures, which otherwise appear as exceptional and isolated mathematical objects. It is as if a hidden uniformity existed for these exceptional geometries once stringy corrections are taken into account. Note also the intriguing similarity between \eqref{eq:WUforSpin7} and \eqref{eq:WUforG2}. In a future publication, the author investigates the relevance and further consequences of the $\mathcal{W}^{1}_c(\tfrac{3}{2};2)$ algebra in string theory compactifications  \cite{FisetToAppear2019}.

A key feature of this regularity is the presence of Virasoro minimal models inside $\mathcal{W}^{1}_c(\tfrac{3}{2};2)$ \cite{Gepner:2001px}. We have already encountered one instance of this in $\text{SV}^{Spin(7)}$, where $T_-$ was an Ising model Virasoro operator. In the case of $G_2$, the central charge $c_-=7/10$ for $T_-$ in \eqref{eq:Tplusminus7} is characteristic of the so-called \voc{tri-critical Ising model}. This has important consequences for the analysis of the spectrum of $G_2$ CFTs and for the computation of moduli in chapter~\ref{chap:MarginalDeformations}.

So far we have derived a free field realisation of $\mathcal{W}^1_{21/2}(\tfrac{3}{2};2)$. The full $\text{SV}^{G_2}$ operator algebra is obtained by a further extension with the $3$-form operator $\Phi=\widehat{\Phi}^{(7)}$ (see \eqref{eq:Phi(7)}) and its supersymmetric partner $K=\boxope{G\Phi}_1$. It can be checked that all possible pairs of OPEs between $\{T, G, \widetilde{X}, \widetilde{M}, \Phi, K\}$ produce only linear combinations of derivatives of these fields, so that they close an algebra. We call it the \label{p:G2} \voc{Shatashvili--Vafa $G_2$ operator algebra}. As a W-algebra, it was first noticed in \cite{Blumenhagen:1991nm, Blumenhagen:1992vr}. It is part of a $2$-parameter family \cite{Noyvert:2002mc} of algebras of type $\mathcal{W}^{1}_{c}(\tfrac{3}{2};\tfrac{3}{2}, 2)$ in the nomenclature inspired by \cite{Bouwknegt:1992wg}. One parameter is the central charge. The other is a ``self-coupling'' which vanishes in the case of $\text{SV}^{G_2}$.

The construction presented so far gives the following sequence of inclusions:
\begin{equation} \label{eq:FF_SVG2}
\mathcal{W}^1_{21/2}(\tfrac{3}{2};2) \longhookrightarrow \text{SV}^{G_2} = \langle T, G, X, M, \Phi, K\rangle \longhookrightarrow (\text{Free})^{7} \,.
\end{equation}
All the OPE relations of the $G_2$ Shatashvili--Vafa algebra (see appendix~\ref{app:SV}) can be obtained directly from the free field realisation \eqref{eq:FF_SVG2}. One would then like to forget about the definition of the generators $\langle T, G, X, M, \Phi, K\rangle$ in terms of free fields and focus abstractly directly on the algebra they generate. This is possible and every comments we have made so far about $\text{SV}^{G_2}$ remain true. There is however an aspect of the $G_2$ algebra that cannot be captured by its free field realisation. Abstractly the OPE relations that define it  do not automatically define an associative algebra (i.e. axiom~\eqref{eq:Jacobilike} in section~\ref{sec:OABasics} fails). However they do form an algebra modulo the ideal generated by the null field \cite{Figueroa-OFarrill:1996tnk}

\begin{equation}\label{eq:NullFieldinG2algebra}
N=4\no{GX}-2\no{\Phi K}-4\partial M-\partial^2 G \,.
\end{equation}
In the free field realisation, in can be checked that $N$ vanishes identically, so that there are no conditions to impose and associativity is manifest. In general, we have to deal with the null field. It will be very important for instance in the gluing construction in chapter~\ref{chap:TCS}.

We note that the subalgebra $\mathcal{W}^1_{21/2}(\tfrac{3}{2};2) \hookrightarrow \text{SV}^{G_2}$ does not intersect with the ideal $\langle N \rangle$. This could make this subalgebra more convenient in certain situations.

\section{$U(n)$ and $SU(n)$-structures} \label{sec:Odake}

Consider pursuing one step further the strategy leading from $d=8$ to $d=7$ free realisations of Shatashvili--Vafa algebras (see\ section~\ref{sec:SVG2}). Let us write (consistently with \cite[p.\,230]{Joyce2007})
\begin{align}
\widehat{\Psi}^{(7)}&=\widehat{\Psi}^{(6)}-\no{\psi^1 \text{Im}(\widehat{\Omega})} \,, \label{eq:Psi(7)Joyce}\\
\widehat{\Phi}^{(7)}&=\text{Re}(\widehat{\Omega})+\no{\psi^1 \widehat{\omega}} \,.
\end{align}
Starting from $\widehat{\Psi}^{(7)}$ and $\widehat{\Phi}^{(7)}$ in \eqref{eq:Psi7}--\eqref{eq:Phi(7)}, this uniquely determines the operators on the right hand side.\detail{ It is important that we have picked out $\psi^1$ here because this is the direction fixed by $SU(3)$ in our conventions [Sebastjan Cizel]. }In order to have indices running up to $6$, we cyclically relabel $\psi^i \rightarrow \psi^{i-1}$ and take
\begin{align}
\label{eq:Psi_d=6}
\widehat{\Psi}^{(6)} &= 
\no{\psi^{1234}}
+\no{\psi^{1256}}
+\no{\psi^{3456}} \,,
\\
\widehat{\omega} &=
\no{\psi^{12}} 
+\no{\psi^{34}} 
+\no{\psi^{56}}  \,,
\\
\widehat{\Omega} &= 
\no{\psi^{135}} 
- \no{\psi^{146}} 
- \no{\psi^{236}} 
- \no{\psi^{245}} \nonumber
\\
&\qquad\qquad\qquad
+i(
\no{\psi^{136}}
+ \no{\psi^{145}}
+ \no{\psi^{235}}
- \no{\psi^{246}}
)\,.~\label{eq:OmegaHatFree}
\end{align}

The geometric interpretation of these formulas is that the $G_2$-structure $(\Psi^{(7)},\Phi^{(7)})$ on $\mathbb{R}^7$ decomposes into an\textit{ $SU(3)$-structure} on $\mathbb{R}^6$ defined by the \textit{Hermitian $2$-form} $\omega$ and \textit{holomorphic $3$-form} $\Omega$. The $2$-form alone defines an $U(3)$-structure. An example of a $U(n)$-structure is provided by a K\"ahler structure, which is in fact equivalent to a torsion-free $U(n)$-structure \cite[p.\,39]{Joyce2007}. The $3$-form $\Omega$ is responsible for a further reduction to a $SU(3)$-structure.

Note that, in the differential geometric version of \eqref{eq:Psi(7)Joyce}, the $4$-form is
\begin{equation}
\Psi^{(6)} = \frac{1}{2}\omega \wedge \omega\,,
\end{equation}
while in field theory, identifying $\no{~} \leftrightarrow \wedge$, we find a mismatch:
\begin{equation} \label{eq:Psi(6)&omega}
\widehat{\Psi}^{(6)} = \frac{1}{2}\no{\widehat{\omega} \widehat{\omega}} + T_f \,.
\end{equation}
This is a humbling reminder of the important \emph{differences} between operator algebras and differential geometry. Here it arises when taking the normal ordered product of normally ordered products. In spite of this, some connections are unmissable, in particular between some operators and differential forms. We will have a lot more to say about this in chapter~\ref{chap:G-symmetries}.

It turns out that $T$, $G$, $\widehat{\omega}$, $\widehat{\Omega}$ and their supersymmetric partners generate a consistent operator algebra discovered in \cite{Odake:1988bh}: the \textit{Odake $n=3$ algebra}. \detail{(didn't really check)} We will shortly describe it in full generality, along with its cousins corresponding to $SU(n)$-structure manifolds.
Before this, let us try again our exercise from section~\ref{sec:exercise} and seek a Virasoro operator of the form $\mu \widehat{\Psi}^{(6)}+\nu T_f$ inside $(\text{Free})^6$. There are once more only two non-trivial solutions:
\begin{align}
T_- &= \frac{1}{3}\left(-\widehat{\Psi} +  T_f \right)\,, \qquad c_- = 1\,, \\
T_{+f} &= \frac{1}{3}\left(\widehat{\Psi} + 2 T_f\right)\,, \qquad c_{+f} = 2\,.
\end{align}
By virtue of \eqref{eq:Psi(6)&omega}, we can also write
\begin{equation}
T_- = -\frac{1}{6}\no{\widehat{\omega}\widehat{\omega}} \,.
\end{equation}

Once more the similarities with the free field realisations of Shatashvili--Vafa algebras are remarkable and deserve more scrutiny in the future. The author hopes to return to this point in an upcoming publication. We note that the subalgebra $T_-$ was first noticed in \cite{Odake:1988bh}. It was never much exploited to the author's knowledge. An exception is \cite{Distler:1995mi}, where it was used to decompose differential forms on the Calabi--Yau manifold in similar ways to what we will describe for $G_2$ and $Spin(7)$ in section~\ref{sec:BRST}.

Let us now describe more precisely Odake's algebras. They are extensions of the \emph{$\mathcal{N}=2$ supersymmetric Virasoro algebra} $\text{Vir}^{2}_c$, so we start there.

\subsection{Virasoro $\mathcal{N}=2$}

\label{p:Virasoro2}

In addition to the $\mathcal{N}=1$ Virasoro generators $G=G^{0}$ and $T$ (see section~\ref{sec:Virasoro}), the generators of $\text{Vir}^{2}_c$ \cite{Zumino:1979et} include the real multiplet $(J,G^{3})$. It is composed of a bosonic $U(1)$ current $J$ and its supersymmetric partner, which is a new supersymmetry current $G^{3}$. The role of $J$ was played by $\widehat{\omega}$ above. This is an illustration of the fact that $\text{Vir}^2_c$ is associated to $\sigma$-models whose target spaces have a $U(n)$-structure. In addition to the $\mathcal{N}=1$ OPEs we have the following relations:


\begin{align}
\wick{\c T(z) \c J(w)}&=\frac{J(w)}{(z-w)^2}+\frac{J'(w)}{z-w} \,,
\\
\wick{\c T(z)  \c G^{3}(w)}&=\frac{3/2}{(z-w)^2}G^{3}(w)+\frac{(G^3)'(w)}{z-w} \,,
\\
\wick{\c G^{0}(z) \c J(w)}&=\frac{G^{3}(w)}{z-w} \,,
\\
\wick{\c G^{0}(z) \c G^{3}(w)}&=\frac{2J(w)}{(z-w)^2}+\frac{J'(w)}{z-w} \,,
\\
\wick{\c G^{3}(z) \c G^{3}(w)}&=\frac{2c/3}{(z-w)^3}+\frac{2T(w)}{z-w} \,,
\\
\wick{\c J(z) \c J(w)}&=-\frac{c/3}{(z-w)^2} \,,
\\
\wick{\c G^{3}(z) \c J(w)}&=-\frac{G^{0}(w)}{z-w} \,.
\end{align}

Our conventions are non-standard but better suited to use with real target space geometries such as ``twisted connected sum'' $G_2$ manifolds in chapter~\ref{chap:TCS} because they do not involve imaginary numbers. The most widespread choice of basis---see e.g.  \cite{Odake:1988bh,Quigley:2014rya,Alim:2012gq,Greene:1996cy}---is related to ours as follows. The $U(1)$ current is made imaginary by defining
\begin{equation}
I=-iJ \,,
\end{equation}
and the supercurrents are combined as
\begin{equation}
G^{\pm}=\frac{1}{\sqrt{2}}\left(G^{0}\pm iG^{3}\right).
\end{equation}
Note that $G^+$ and $G^-$ are complex conjugate to each other: $G^-=(G^+)^*$.

With these definitions, the OPEs that do not involve $T$ are as follows:
\begin{align}
\wick{\c I(z) \c I(w)}&=\frac{c/3}{(z-w)^2} \,,
\\
\wick{\c I(z) \c G^{\pm}(w)}&=\pm\frac{G^{\pm}(w)}{z-w} \,,
\\
\wick{\c G^+(z) \c G^-(w)}&=\frac{2c/3}{(z-w)^3}+\frac{2I(w)}{(z-w)^2}+\frac{(I'+2T)(w)}{z-w} \,,
\\
\wick{\c G^\pm(z) \c G^\pm(w)}&=0 \,.
\end{align}

\subsection{Odake $n$} \label{app:Odake}

\label{p:Od}

The\textit{ Odake operator algebra} $\text{Od}^n$ \cite{Odake:1988bh} with central charge $c=3n$, $n\in\mathbb{N}$ is associated to $\sigma$-models mapping in manifolds endowed with certain $SU(n)$\emph{-structures} $(\omega,\Omega)$. Calabi--Yau manifolds have torsion-free $SU(n)$\emph{-structure} for example. It is obtained by extension of $\text{Vir}^2_c$ with four new real generators or equivalently two complex generators:
\begin{equation}
\Omega=A+iB\,, \qquad \text{and}\qquad \Upsilon=\frac{C+iD}{\sqrt{2}}\,.
\end{equation}
Their complex conjugates are denoted with the star; e.g.\ $\Omega^*=A-iB$. The role of $\Omega$ was played by \eqref{eq:OmegaHatFree} at the beginning of this section. It is intuitively due to the existence, on the target space, of a holomorphic $n$-form. We start by providing the OPEs with the $\mathcal{N}=2$ generators.

The field $\Omega$ is a weight-$n/2$ primary with respect to $T$ and it has $U(1)$ charge $n$:
\begin{align}
\wick{\c T(z) \c \Omega(w)}&=\frac{n/2}{(z-w)^2}\Omega(w)+\frac{\Omega'(w)}{z-w}\,,
\\
\wick{\c I(z) \c \Omega(w)}&=\frac{n\Omega(w)}{z-w}\,.
\end{align}
The field $\Upsilon$ has weight $\tfrac{n+1}{2}$ and is primary with respect to $T$. It has $U(1)$ charge $n-1$:
\begin{align}
\wick{\c T(z) \c \Upsilon(w)}&=\frac{(n+1)/2}{(z-w)^2}\Upsilon(w)+\frac{\Upsilon'(w)}{z-w}\,,
\\
\wick{\c I(z) \c \Upsilon(w)}&=\frac{n-1}{z-w}\Upsilon(w)\,.
\end{align}
$\Omega$ and $\Upsilon$ are related by supersymmetry as follows:
\begin{align}
\wick{\c G^+(z) \c \Omega(w)}&=0 \,,
\\
\wick{\c G^-(z) \c \Omega(w)}&=\frac{2\Upsilon(w)}{z-w} \,,
\\
\wick{\c  G^+(z) \c \Upsilon(w)}&=\frac{n\Omega(w)}{(z-w)^2}+\frac{\Omega'(w)}{z-w} \,,
\\
\wick{\c  G^-(z) \c \Upsilon(w)}&=0 \,.
\end{align}

Similarly to the $G_2$ Shatashvili--Vafa algebra, the Odake $n=3$ algebra suffers from the associativity issue discussed in the vicinity of \eqref{eq:Wick}. Associativity of the OPEs only holds in the case $n=3$ modulo an ideal. Here this ideal is generated by the fields
\begin{equation} \label{eq:N1N2}
N^{1}=A'-\no{JB}\,,\qquad N^{2}=B'+\no{JA}\,.
\end{equation}

The $n=2$ Odake algebra is isomorphic, it turns out, to the $\mathcal{N}=4$ Virasoro algebra \cite{Odake:1988bh,Ali:1999ut,Ademollo:1976wv,Ademollo:1975an}---more precisely to one of three operator algebras bearing this name: the so-called \emph{little} $\mathcal{N}=4$ Virasoro algebra. It is automatically associative.

In appendices~\ref{app:Odake2}--\ref{app:Odake3} we consider individually $n=2$ and $n=3$ and give the OPEs involving only $\Omega$, $\Upsilon$ and their conjugates.

%
%
%

\chapter[Operator algebras for TCS $G_2$ manifolds]{Operator algebras for twisted connected sum $G_2$ manifolds} \label{chap:TCS}

\onehalfspacing
\minitoc
\vspace{25pt}

\doublespacing

The recent construction \cite{MR2024648,MR3109862,Corti:2012kd} of millions of examples of so-called \textit{twisted connected sum} (TCS) compact seven-dimensional manifolds with holonomy group $G_2$ (we will simply say ``$G_2$ manifolds'' from now on) sparked much interest in the physics community in particular for their role in supersymmetric compactifications \cite{Halverson:2014tya, Halverson:2015vta, Braun:2016igl, Guio:2017zfn, Braun:2017ryx, Braun:2017csz, Braun:2017uku, Braun:2018joh, Braun:2018fdp,Braun:2018vhk}.

Prior to this, the only reliable construction of $G_2$ manifolds was Joyce's original method \cite{MR1424428, Joyce2000} based on torus orbifolds. As described in chapter~\ref{chap:OA}, $G_2$-structures on $\mathbb{T}^7$ are approachable in conformal field theory thanks to the free field realisation of Shatashvili and Vafa \cite{Shatashvili:1994zw}. This realisation of $\text{SV}^{G_2}$ is the starting point of more detailed investigations of stringy properties of Joyce orbifolds, for example using discrete torsion \cite{Vafa:1986wx} to deal with resolutions of the quotient space \cite{Shatashvili:1994zw, Acharya:1996fx, Acharya:1997rh, Roiban:2002iv, Gaberdiel:2004vx}. Arguing from $\text{SV}^{G_2}$, conjectures were also made in \cite{Shatashvili:1994zw} about analogues of mirror symmetry and topological twists for exceptional geometries. This led to a surge of interest in this subject \cite{Shatashvili:1994zw, Papadopoulos:1995da, Becker:1996ay, Acharya:1996fx, Figueroa-OFarrill:1996tnk, Acharya:1997rh, Blumenhagen:2001jb, Gukov:2002jv, Roiban:2002iv, Eguchi:2001ip, Noyvert:2002mc, Sugiyama:2001qh, Sugiyama:2002ag, Eguchi:2003yy, deBoer:2005pt}.

Another possible approach to constructing $G_2$ manifolds is to start with the product of a Calabi--Yau three-fold with a circle and then take the quotient by an involution (and eventually desingularise) \cite{MR1424428}. $G_2$ manifolds admitting such a construction also closely correspond to a particular realisation of $\text{SV}^{G_2}$ using the Odake $n=3$ algebra, as first described in \cite{Figueroa-OFarrill:1996tnk}. Our introduction of $\text{Od}^{n=3}$ in section~\ref{sec:Odake} was in fact exploiting this result within free field algebras.

In this chapter we combine these elements and introduce the superconformal algebra appropriate for twisted connected sum $G_2$ holonomy manifolds. More precisely, we prove that the Shatashvili--Vafa $G_2$ algebra can be obtained from copies of Odake algebras and free fields in a way that mimics the TCS construction. A key step is to match the two halves of the construction with an algebra automorphism $F$ as explained more fully in the vicinity of equation~\eqref{eq:resultForIntro},
\begin{equation}
\text{SV}^{G_2} \longhookrightarrow \frac{\text{Od}^{n=2} \times (\text{Free})^3}{F} ~~ \Big/ ~ \langle N^1 , N^2\rangle \,.
\end{equation}
Here $N^1$ and $N^2$ refer to the null ideal of $\text{Od}^{n=3}$ modulo which our statements hold.

The clear relationship of our result with TCS manifolds calls readily for mirror symmetry applications in type II string theory, in the sense recently suggested by Braun and Del Zotto \cite{Braun:2017ryx,Braun:2017csz}. With this in mind, we describe in section~\ref{sec:AutomTCS} two consistent automorphisms of our algebraic construction. We propose to regard these as conformal field theoretic versions of $G_2$ mirror maps argued for from a geometrical point of view in \cite{Braun:2017csz}. In this sense, we expect our results to help venture away from the geometrical locus of type II string compactifications on TCS manifolds.

The rest of this chapter is organised as follows. We cover aspects of the TCS construction in section~\ref{sec:TCSgeometry}.
We exhibit our TCS algebra realisation in section~\ref{sec:Realisations}. Automorphisms of the realisation and mirror symmetry are discussed in section~\ref{sec:Automorphisms}. We close in section~\ref{sec:Conclusion} with a fuller discussion of our results.

\section{TCS geometry} \label{sec:TCSgeometry}

We begin with a short informal account of the twisted connected sum (TCS) construction of seven-dimensional manifolds with holonomy group $G_2$. This technique was pioneered by Donaldson and Kovalev \cite{MR2024648} and substantially improved and developped by Corti, Haskins, Nordstr\"om and Pacini \cite{MR3109862,Corti:2012kd}. We refer to the work of these authors for details. For basics about $G_2$-structure geometry, we recommend \cite{Joyce2000,Grigorian:2009ge,Corti:2012kd}.

First, we need a pair $X_\pm$ of asymptotically cylindrical Calabi--Yau three-folds. A complete Calabi--Yau $n$-fold $(X,g,\omega,\Omega)$ is called \textit{asymptotically (exponentially) cylindrical} (ACyl) if and only if it contains a compact $2n$-dimensional set whose complement is diffeomorphic to a Calabi--Yau half-cylinder $X_\infty=\mathbb{R}^+\times C^{2n-1}$ and if this diffeomorphism identifies sufficiently fast (along $\mathbb{R}^+$) the K\"ahler and holomorphic $n$-forms of $X$ with the corresponding forms $\omega_\infty$ and $\Omega_\infty$ of the half-cylinder. In the TCS case, the \emph{cross-section} $C^{2n-1}$ of $X_\infty$ is $C^{5}=S \times \mathbb{S}^1$, where $(S,g_S,\omega_S,\Omega_S)$ is a K3 surface. We parametrise $\mathbb{R}^+$ with $t$ and $\mathbb{S}^1$ with $\theta$. The structure tensors of $X$ asymptote along $\mathbb{R}^+$ as
\begin{align}
g &~ \overset{t\rightarrow\infty}{\widesim} ~ g_\infty = \dd t^2 + \dd\theta^2 + g_S\,, \label{eq:ginfty} \\
\omega &~ \overset{t\rightarrow\infty}{\widesim} ~ \omega_\infty = \dd t \wedge \dd\theta + \omega_S \,, \label{eq:omegainfty} \\
\Omega &~ \overset{t\rightarrow\infty}{\widesim} ~ \Omega_\infty = \left( \dd\theta - i \dd t \right) \wedge \Omega_S \,. \label{eq:Omegainfty}
\end{align}
A heuristic understanding of these asymptotic formulas is enough for our purposes. It suffices to picture a distinguished real direction in $X$ along which the manifold asymptotes to a translation-invariant Calabi--Yau $X_\infty$ with structure forms as given above. As explained, we need two such ACyl Calabi--Yaus $X_\pm$. We will add $\pm$ to all their associated quantities to distinguish them.

We note that the ambient Calabi--Yau structure induces a particular choice of K\"ahler form $\omega_S$ and holomorphic top form $\Omega_S$ on $S$. Writing $\omega_S = \omega^I$ and $\Omega_S = \omega^J+i\omega^K$, makes the hyper-K\"ahler structure of the K3 surface more manifest. Here the hyper-K\"ahler triplet of closed forms $(\omega^I, \omega^J, \omega^K)$ satisfies by definition the relations 
\begin{align} \label{eq:hyperkahlertriplet}
(\omega^I)^2 = (\omega^J)^2 = (\omega^K)^2 \,, \qquad
\omega^I\wedge\omega^J = \omega^J\wedge\omega^K = \omega^K\wedge\omega^I = 0 \,.
\end{align}

Now for any Calabi--Yau three-fold $(X,g,\omega,\Omega)$, it is always possible to equip $\mathcal{M}=X\times \mathbb{S}^1$, with a torsion-free $G_2$-structure (see e.g. \cite[prop.\ 11.1.9]{Joyce2007}). If $\xi$ is the coordinate along the $\mathbb{S}^1$, the associative and co-associative forms on the product are defined respectively as
\begin{equation}
\phi = \dd\xi \wedge \omega + \text{Re}(\Omega)\,, \qquad
\psi = \frac{1}{2}\omega\wedge \omega - \dd\xi \wedge \text{Im}(\Omega)\,. \label{eq:M=XxS1--phipsi}
\end{equation}
These are mutually Hodge dual with respect to the product metric
\begin{equation}
g_\mathcal{M} = g + d\xi^2 \,. \label{eq:M=XxS1--g}
\end{equation}
These formulas characterise a \textit{torsion-free $G_2$-structure}, meaning $\dd\phi=0$ and $\dd\psi=0$. An equivalent way to characterise the vanishing torsion is to say that the Riemannian holonomy group of $g_\mathcal{M}$ is contained in $G_2$; here this is a proper inclusion because $\text{Hol}(g_\mathcal{M})=SU(3)$ is only a subgroup of $G_2$. The procedure can be applied equally to $X_\pm$, leading to $(\mathcal{M}_{\pm},g_{\pm},\phi_{\pm},\psi_{\pm})$ and to their asymptotic models $X_{\infty\pm}$, leading to $(\mathcal{M}_{\infty\pm},g_{\infty\pm},\phi_{\infty\pm},\psi_{\infty\pm})$.

The next step is essentially to form the connected sum $\mathcal{M}_+ \,\#\, \mathcal{M}_-$ of the two open seven-dimensional manifolds by ``pasting'' the asymptotic models. However, this cannot be done in the most naive way because this would lead to a manifold with infinite fundamental group. It is an important fact in $G_2$ geometry that a compact manifold $\mathcal{M}$ with torsion-free $G_2$-structure has holonomy exactly $G_2$ if and only if $\pi_1(M)$ is finite \cite{Joyce2007}. This is where the ``twist'' plays a useful role. To describe this, we first assume the existence of a diffeomorphism of K3 surfaces\footnote{It is a non-trivial requirement about the ACyl manifolds that they should be compatible in this sense.}
\begin{equation}
\mathfrak{r}\, : S_+ \longrightarrow S_- \,,
\end{equation}
which is an isometry,
\begin{equation}
\mathfrak{r}^* \, g_{S-} = g_{S+}\,,
\end{equation}
and which satisfies
\begin{equation}
\mathfrak{r}^* \, \text{Im}(\Omega_{S-}) = -\text{Im}(\Omega_{S+})\,,\quad
\mathfrak{r}^* \, \text{Re}(\Omega_{S-}) = \omega_{S+}\,,\quad
\mathfrak{r}^* \, \omega_{S-} = \text{Re}(\Omega_{S+})\,.
\end{equation}

Such a map is called a \textit{hyper-K\"ahler matching}. We use it to fuse $\mathcal{M}_+$ and $\mathcal{M}_-$ together. More precisely, let $T\in \mathbb{R}^+$ and introduce a boundary at $t=T+1$ to define the manifolds $\mathcal{M}_\pm(T)$ with boundary $\partial \, \mathcal{M}_\pm(T) \simeq S_\pm \times \mathbb{S}^1_\pm \times \mathbb{S}^1_\pm$. $T$ is eventually taken large enough for existence theorems to apply. Let $I=[T,T+1]\subset \mathbb{R}^+$. Next, introduce a diffeomorphism
\begin{equation} \label{eq:GeomPastingMap}
F_T \quad : \quad \begin{matrix}
I\times S_+ \times \mathbb{S}^1_+ \times \mathbb{S}^1_+ & \longrightarrow & I\times S_- \times \mathbb{S}^1_- \times \mathbb{S}^1_- \\ 
(~t~, ~z~, ~\theta~, ~\xi~) & \longmapsto & (~2T+1-t~, ~\mathfrak{r}(z)~, ~\xi~, ~\theta~)\,.
\end{matrix}
\end{equation}
It is a simple exercise to show that the associative and co-associative forms are compatible under this map,
\begin{equation}
(F_T)^* \, \phi_{\infty-} = \phi_{\infty+}\,, \qquad (F_T)^* \, \psi_{\infty-} = \psi_{\infty+} \,.
\end{equation}
Since the $G_2$ metric can be derived from any of these, this implies in particular that $F_T$ is an isometry.

\begin{figure}[htbp]
\begin{center}
\begin{tikzpicture}[scale=0.85]
\node at (0,0) {\includegraphics[scale=0.238]{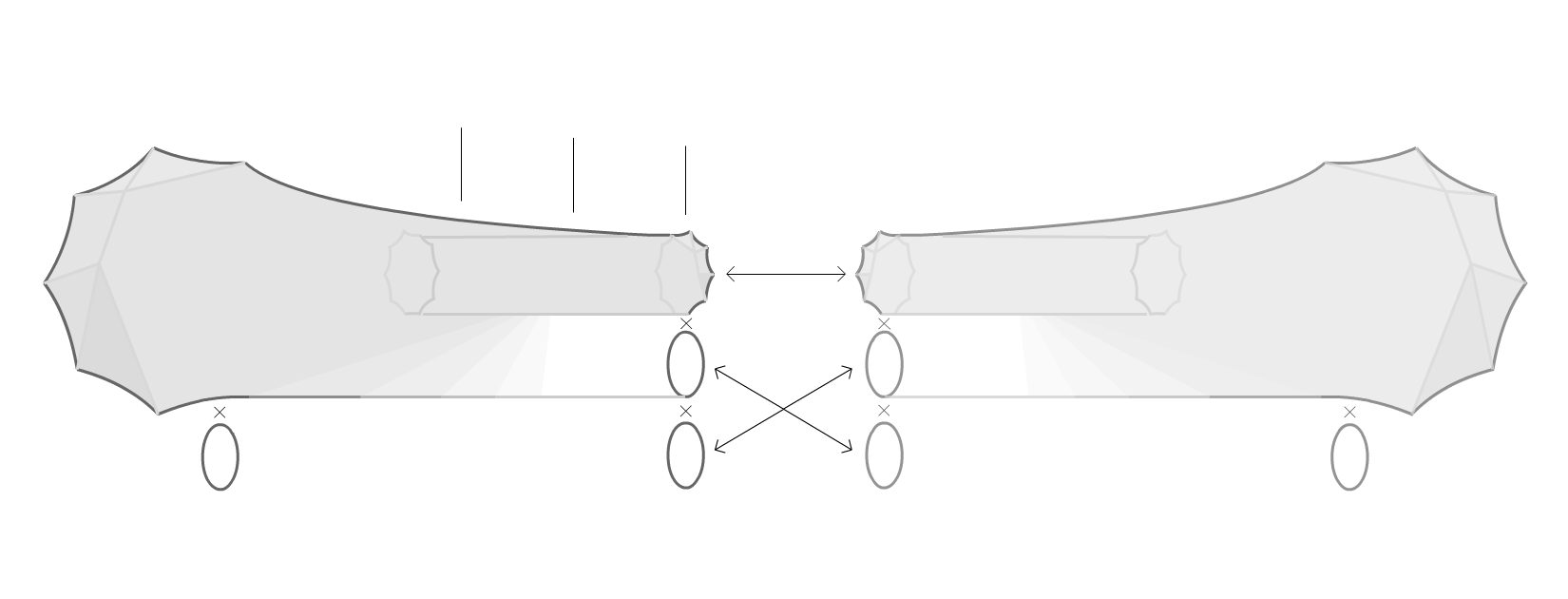}};
\node at (-6.25,0.225) {$X_+$};
\node at (6.3,0.22) {$X_-$};
\node at (-1.15,0.225) {$S_+$};
\node at (1.2,0.22) {$S_-$};
\node at (-1.48,-0.65) {$\mathbb{S}^1_\theta$};
\node at (-1.48,-1.65) {$\mathbb{S}^1_\xi$};
\node at (-6.31,-1.65) {$\mathbb{S}^1_\xi$};
\node at (-6.25,-2.5) {$\underbrace{\qquad\qquad\qquad}$};
\node at (-2.5,-2.5) {$\underbrace{\qquad\qquad\qquad\quad}$};
\node at (6.25,-2.5) {$\underbrace{\qquad\qquad\qquad}$};
\node at (2.5,-2.5) {$\underbrace{\qquad\qquad\qquad\quad}$};
\node at (-6.25,-3) {Region I\textsuperscript{$+$}};
\node at (-2.5,-3) {Region II\textsuperscript{$+$}};
\node at (2.5,-3) {Region II\textsuperscript{$-$}};
\node at (6.25,-3) {Region I\textsuperscript{$-$}};
\node at (0,0.5) {$\mathfrak{r}$};
\node at (-3.3,2) {\tiny $T-1$};
\node at (-2.2,1.85) {\tiny $T$};
\node at (-1,1.8) {\tiny $T+1$};
\end{tikzpicture}
\caption{The twisted connected sum construction of $G_2$ holonomy manifolds. $X_\pm$ are ACyl Calabi--Yau three-folds. $S_\pm$ are K3 surfaces. $\mathbb{S}^1$ are circles. Performing the connected sum requires a hyper-K\"ahler matching $\mathfrak{r}$. The resulting manifold has two types of regions: $X_\pm\times \mathbb{S}^1$ (regions $\text{I}^{\pm}$) and $S_\pm\times \mathbb{S}^1\times \mathbb{S}^1\times \mathbb{R}^+$ (regions $\text{II}^{\pm}$).}
\label{fig:TCSgeometry}
\end{center}
\end{figure}

The pasting is finished as follows. Endow $\mathcal{M}_\pm(T)$ with the $G_2$-structure interpolating from $(g_\pm,\phi_\pm,\psi_\pm)$ at $t=T-1$ to $(g_{\infty\pm},\phi_{\infty\pm},\psi_{\infty\pm})$ at $t=T$. Next, on $I=[T,T+1]$, use $F_T$ to establish the local isomorphism $\mathcal{M}_{\infty +} \simeq \mathcal{M}_{\infty -}$. This yields a compact $G_2$-structure manifold $\mathcal{M}=\mathcal{M}_+ \,\#_{\mathfrak{r}}\, \mathcal{M}_-$ with a neck-like region of length $\sim 2T$. There remains non-vanishing torsion ($\dd\phi\neq 0$, $\dd\psi \neq 0$) concentrated in $t\in[T-1,T]$ on both sides. Nevertheless there are theorems that guarantee, for sufficiently large $T$, the existence of a torsion-free and Ricci-flat perturbation of the $G_2$-structure $\phi$ of $\mathcal{M}$ within the same cohomology class (see\ \cite{Joyce2000} theorem 11.6.1, \cite{MR2024648} theorem 5.34 and \cite{Corti:2012kd} theorem 3.12). A sketch of a twisted connected sum is proposed on figure~\ref{fig:TCSgeometry}.

There are two qualitatively different regions in each of the two parts $\mathcal{M}_\pm$ of the TCS construction. One is approximated by the direct product of an ACyl Calabi--Yau three-fold with an $\mathbb{S}^1$. We label it as type $\text{I}^\pm$ in figure~\ref{fig:TCSgeometry}. The other one, type $\text{II}^\pm$, is the asymptotic region $\mathcal{M}_{\infty\pm}=\mathbb{R}^+ \times S_\pm \times \mathbb{S}^1 \times \mathbb{S}^1$.

Note that the larger the parameter $T$, the lesser the perturbation needed to reach vanishing torsion. A TCS with $T$ very large, i.e.\ with a highly stretched ``neck'', is said to be in a \textit{Kovalev limit}. This is often a useful limit to work with in practical applications. As is manifest in section~\ref{sec:Realisations}, our algebraic construction however has no notion of scale $T$, so a priori no Kovalev limit. This suggests that it remains adequate even in the limit of a short neck, i.e.\ in the phase of the conformal manifold where geometric control starts to disappear.

\section[Conformal algebras and twisted connected sums]{Conformal algebras and TCS} \label{sec:Realisations}

In this section, we explain how to realise the Shatashvili--Vafa $G_2$ algebra associated to twisted connected sum geometries. Achieving this involves two main steps: 
\begin{enumerate}
\item find realisations of $\text{SV}^{G_2}$ associated to distinctive local open subsets $U_\alpha$ in the TCS geometry;
\item match these local realisations on overlaps $U_\alpha\cap U_\beta$ in the TCS geometry by using invariances.
\end{enumerate}
This strategy may seem unusual in a number of ways because it mixes intuition from differential geometry and conformal field theory. We address this here, starting with point 1.

One often regards operator algebras as providing an exact quantum description of chiral symmetries of a conformal field theory used as internal sector in a $g_s$-perturbative string compactification. This is true whether or not the internal theory admits a geometric interpretation via a $\sigma$-model in the large radius limit.
In particular, this abstract framework should allow a description of
phases where no geometric interpretation exists, let alone the ``local open subsets'' of point 1. $\mathcal{N}=2$ Landau-Ginzburg phases are well-known examples.

We suggest to resolve this puzzle by assuming that we are working in an intermediate phase of the space of conformal field theories (\textit{conformal manifold}) between the perturbative non-linear $\sigma$-model, whose target manifold is approximately\footnote{A $G_2$ holonomy target does not lead to a scale-invariant model due to corrections in $\alpha'$. However it was argued not long ago \cite{Becker:2014rea} that such corrections only mildly perturb away from $G_2$ holonomy.} a TCS and the full-fledged quantum regime. In this intermediate region, there should be remnants of a geometric intuition about the would-be target space and thus it makes sense to refer to local open subsets. However we are seeking to venture into the stringy regime and the description is not in terms of a non-linear $\sigma$-model, but rather in terms of ``local'' operator algebras. This understanding is at least useful to bear in mind for now. We will show below that each region in the TCS construction ($\text{I}^+$, $\text{II}^+$, $\text{II}^-$ and $\text{I}^-$) admits a distinguished local realisation of $\text{SV}^{G_2}$ highly reminiscent of the associated geometric structure. 

Step 2 above also deserves more explanations. We take again a semi-classical point of view and borrow from the non-linear $\sigma$-model description. With $\sigma$-models, it is often customary to designate a \textit{compact} manifold $\mathcal{M}$ as target space. In the strictest of senses this cannot be done explicitly (i.e.\ with a Lagrangian) unless the target space has particular symmetries, such as for group manifolds. For a generic target space, a particular open patch $U_\alpha\subset \mathcal{M}$ and local coordinates must indeed be chosen before a Lagrangian $\mathcal{L}_\alpha$ can be formulated. We will see this in detail in chapter~\ref{chap:G-symmetries}.

This technical limitation is quite inconsequential in practise because all the local Lagrangians $\mathcal{L}_\alpha$ are sufficiently consistent with one another for a well-defined theory $\mathcal{L}=\bigcup_{U_\alpha\in \mathcal{M}}\mathcal{L}_\alpha$ to exist for the whole of $\mathcal{M}$. The key feature of $\mathcal{L}_\alpha$ allowing this global interpretation is the existence of symmetries (or more general formal invariances involving variations of the couplings of the $\sigma$-model). These invariances of $\mathcal{L}_\alpha$ admit target space interpretations, such as for example diffeomorphisms, isometries, gauge invariances, etc. To insist, it is these invariances of $\mathcal{L}_\alpha$ that inform us on how to interpret globally the theory.\footnote{A parallel can be drawn here with the discussion in \cite{QFT&Strings2} (lectures 17 and 18) motivating the introduction of a Seiberg--Witten curve to describe the low-energy dynamics of $\mathcal{N}=2$ gauge theories in four dimensions. In that case the $\sigma$-model is an $\mathcal{N}=2$ $U(1)$ gauge theory and the $SL(2,\mathbb{Z})$ invariance of the field-dependent coupling $\tau_{\text{IR}}(\Phi)$ suggests an elliptic fibration over the Coulomb branch.} The proposal of point 2 above is that invariances of the local realisations of the $G_2$ algebra should similarly allow us to construct a globally consistent realisation for the whole TCS geometry. We will make this fully precise in section~\ref{sec:Compatibility}.

Let us stress that the interpretation proposed here is very conjectural. It is suggested to explain the success of the operator algebraic result to be presented in the upcoming sections. Simpler instances of ``patching local CFTs'' than TCS manifolds are to be understood before this is made rigorous. For example, it is not clear if one can describe the compact CFT of a free boson on $\mathbb{S}^1$ in terms of two non-compact bosonic CFTs of free bosons on $\mathbb{R}$, regarded as local  open patches of $\mathbb{S}^1$.
It may be possible to strengthen our interpretation in the context of non-geometric backgrounds; see \cite{Plauschinn:2018wbo} and references therein. In this context, locally geometric solutions to string theory are assembled with standard geometric invariances as well as via string dualities. A simple example is when a $1$-cycle of radius $R$ in target space is patched to a cycle of radius $1/R$ via $T$-duality. We have not explored this direction further.

We describe next the realisations corresponding to the two types of regions.

\subsection{Type $\text{I}^\pm$: $X_\pm \times \mathbb{S}^1$} \label{sec:TypeI}

Clearly the operator algebra associated to the Calabi--Yau three-fold $X$ is a copy of the Odake algebra $\text{Od}^{n=3}$ (see section~\ref{sec:Odake}). Associated to the $\mathbb{S}^1$ factor, we can realise $\mathcal{N}=1$ Virasoro using free fields in the way described in section~\ref{sec:Free}. We parametrise the $\mathbb{S}^1$ by the free boson $\xi$ and define the associated $U(1)$ current as ${\mathscr{I}}_\xi=i\xi'$. Its fermionic superpartner is denoted by $\psi_{\xi}$. The $\mathcal{N}=1$ generators are given explicitly by
\begin{align}
T_{\xi}&=\tfrac{1}{2}\no{{\mathscr{I}}_\xi {\mathscr{I}}_\xi}+\tfrac{1}{2}\no{\psi_{\xi}' \psi_{\xi}}\,, \label{eq:Tflat}\\
G_{\xi}&=\no{{\mathscr{I}}_\xi \psi_{\xi}}\,. \label{eq:Gflat}
\end{align}

These two algebras---$\text{Free}$ and $\text{Od}^{n=3}$---can be combined to provide a realisation of the $G_2$ algebra. It will be regarded as the local form taken by $\text{SV}^{G_2}$ in the geometric regions $\text{I}^\pm$ in figure~\ref{fig:TCSgeometry}. As first shown in \cite{Figueroa-OFarrill:1996tnk}, the following generators\footnote{Whenever confusion between generators of different algebras might occur, we use indices to distinguish them. Here $T_\mathcal{M}$ and $T$ are respectively energy-momentum tensors for the $G_2$ and Odake $n=3$ algebras.} satisfy the $G_2$ OPE relations, up to the ideal generated by $N^{1}$ and $N^{2}$:
\begin{align}
T&_\mathcal{M}=T+T_{\xi}\,, \label{eq:T7} \\
G&_\mathcal{M}=G+G_{\xi}\,, \label{eq:G7} \\
\Phi&=\no{J \psi_{\xi}}+A\,, \label{eq:Phi7} \\
X&=\no{B \psi_{\xi}}+\tfrac{1}{2}\no{JJ}-\tfrac{1}{2}\no{\psi_{\xi}' \psi_{\xi}}\,, \label{eq:X7} \\
K&=C+\no{J {\mathscr{I}}_\xi}+\no{G^3 \psi_{\xi}}\,, \label{eq:K7}\\
M&=\no{D \psi_{\xi}}-\no{B {\mathscr{I}}_\xi}+\tfrac{1}{2}\no{{\mathscr{I}}_\xi \psi'_{\xi}}-\tfrac{1}{2}\no{{\mathscr{I}}'_\xi \psi_{\xi}}+\no{JG^3}-\tfrac{1}{2}G'\,. \label{eq:M7}
\end{align}
We summarise this key statement as
\begin{equation} \label{eq:InclusionI}
\text{SV}^{G_2} \overset{\text{I}}{\longhookrightarrow} \left(\text{Od}^{n=3}\times \text{Free}\right) ~~ \Big/ ~ \langle N^1 , N^2\rangle \,.
\end{equation}
Moreover, the null field $N$ in the $G_2$ algebra (see\ \eqref{eq:NullFieldinG2algebra}) belongs to the ideal of $\text{Od}^{n=3}$:
\begin{equation}
\langle N \rangle \subset \langle N^1 , N^2\rangle\,.
\end{equation}

In the work of \cite{Figueroa-OFarrill:1996tnk}, this realisation was motivated by the construction of $G_2$ holonomy manifolds as desingularisations of $\mathbb{Z}_2$ quotients of Calabi--Yau three-folds times a circle \cite{MR1424428, Joyce2000}. It is a property of the realisation that it is left invariant by a $\mathbb{Z}_2$ map corresponding to the quotient and hence it does descent to the quotient as stated in \cite{Figueroa-OFarrill:1996tnk}. We have verified however that the OPEs themselves do not rely on the $\mathbb{Z}_2$  identification and therefore hold already for $X_\pm \times \mathbb{S}^1$ (even though the holonomy of this space is only \emph{contained} in $G_2$).

We point out a minor discrepancy between our generator $M$ in \eqref{eq:M7} and the one given in \cite{Figueroa-OFarrill:1996tnk}. It has been confirmed by the author of this reference that this is due to a sign error on their part.

Let us also appreciate the similarity between the field theoretic definitions \eqref{eq:T7} and \eqref{eq:G7} of $T_\mathcal{M}$ and $G_\mathcal{M}$ with the direct product metric \eqref{eq:M=XxS1--g} on $\mathcal{M_\pm}$. The same comment applies to $\phi$ and $\psi$ in \eqref{eq:M=XxS1--phipsi} which are essentially identical to $\Phi$ and $X$, up to the identifications
\begin{equation}
\psi_\xi \leftrightarrow \dd \xi \,, \qquad \no{~} \leftrightarrow \wedge \,, \qquad (J,A,B) \leftrightarrow (\omega , \text{Re}(\Omega) , \text{Im}(\Omega)) \,. \label{eq:GeometryVsalgebra}
\end{equation}
The only difference is the quantum correction in the case of $X$. All of this is consistent with the discussion in chapter~\ref{chap:OA}.

The embedding of the $G_2$ algebra described by \eqref{eq:InclusionI} is not unique. This is due to the automorphism $A+iB\longmapsto e^{iu}(A+iB)$, $u\in \mathbb{R}$, (with a similar transformation for $\Upsilon$) of the Odake algebra, which can be used to rotate the $\{A,B\}$ basis into itself. We will discuss automorphisms more fully in section~\ref{sec:Automorphisms}.

\subsection{Type $\text{II}^\pm$: $\mathbb{R}^+ \times S_\pm \times \mathbb{S}^1 \times \mathbb{S}^1$} \label{sec:TypeII}

In this region, the geometric intuition calls first for a realisation of a copy of $\text{Od}^{n=3}$ associated to the translation-invariant Calabi--Yaus $X_{\infty\pm}$ in terms of the algebras corresponding to  $\mathbb{R}^+$, $\mathbb{S}^1$ and the K3 surfaces $S_\pm$. Then, by the results of section~\ref{sec:TypeI}, we will be able to define a realisation of the $G_2$ algebra associated to $\mathcal{M}_{\infty\pm}=X_{\infty\pm}\times \mathbb{S}^1$. We start by examining how to realise the $\text{Od}^{n=3}$ for $X_{\infty\pm}$.

Each flat factor in the geometry leads again to free fields and Virasoro generators using \eqref{eq:Tflat}--\eqref{eq:Gflat}. We will keep denoting by $\xi$ the coordinate on the external $\mathbb{S}^1$ and we reuse the coordinates $t$, $\theta$ from section~\ref{sec:TCSgeometry} for $\mathbb{R}^+$ and the internal $\mathbb{S}^1$ respectively.

To the K3 surface, a Calabi--Yau two-fold, corresponds the $n=2$ Odake algebra described in section~\ref{sec:Odake} and appendix~\ref{app:Odake2}. Considering now the asymptotic formulas~\eqref{eq:omegainfty} for $g_\infty$, $\omega_\infty$ and $\Omega_\infty$ in region II and taking stock of the correspondences between geometry and algebras stated in \eqref{eq:GeometryVsalgebra}, we are naturally led to some ans\"atze for generators of $\text{Od}^{n=3}$. We take
\begin{align}
T&=T_t+T_S+T_{\theta} \,,\label{eq:T3}\\
G&=G_t+G_S+G_{\theta} \,,\label{eq:G3}\\
J&=\no{\psi_t \psi_{\theta}}+J_S \,,\label{eq:J3}\\
A+iB&=\no{(\psi_{\theta}-i\psi_t) (A_S+iB_S)} \,,\label{eq:A3+iB3}
\end{align}
where the subscript $S$ differentiates the generators of $\text{Od}^{n=2}$ from those of $\text{Od}^{n=3}$. The remaining generators can easily be derived by taking OPEs with $G$ in \eqref{eq:G3}. We find
\begin{align}
G^3&=G^3_S+\no{\psi_{\theta}{\mathscr{I}}_t}-\no{{\mathscr{I}}_\theta \psi_t} \,, \label{eq:G33}\\
C&=\no{{\mathscr{I}}_t B_S}+\no{{\mathscr{I}}_\theta A_S}-\no{\psi_tD_S}-\no{\psi_\theta C_S} \,, \label{eq:C3}\\
D&=-\no{{\mathscr{I}}_t A_S}+\no{{\mathscr{I}}_\theta B_S}+\no{\psi_tC_S}-\no{\psi_\theta D_S} \,. \label{eq:D3}
\end{align}

It can then be checked that these generators indeed give rise to the Odake $n=3$ algebra. Some of the OPEs hold exactly, while others only hold modulo the ideal generated by the fields in \eqref{eq:N1N2}, themselves written in terms of  $\text{Od}^{n=2}$ and free generators:
\begin{align}
N^1 &= \no{\psi_t\no{J_SA_S}}+\no{\psi_tB_S'}-\no{\psi_\theta\no{J_SB_S}}+\no{\psi_\theta A_S'}\,, \label{eq:N1II} \\
N^2 &= \no{\psi_t\no{J_SB_S}}-\no{\psi_tA_S'}+\no{\psi_\theta\no{J_SA_S}}+\no{\psi_\theta B_S'}\,. \label{eq:N2II}
\end{align}
We write this result as
\begin{equation} \label{eq:InclusionIICY}
\text{Od}^{n=3} \overset{\text{II}_X}{\longhookrightarrow} \left(\text{Od}^{n=2} \times (\text{Free})^2\right) ~~ \Big/ ~ \langle N^1 , N^2\rangle \,.
\end{equation}

As an example of the implications of the ideal, let us examine the OPE of $A$ with itself. The expected result in the Odake $n=3$ algebra is
\begin{equation}
\wick{ \c A(z) \c A(w)} = -\frac{4}{(z-w)^3}-\frac{2\no{JJ}(w)}{z-w}\,.
\end{equation}
Computing directly with \eqref{eq:A3+iB3}, the result at order $1$ differs from the expectation by
\begin{equation} \label{eq:AAmismatch}
\no{A_SA_S}+\no{B_SB_S}-2\no{J_SJ_S}\,.
\end{equation}

That this should vanish nicely reflects the hyper-k\"ahler structure of K3 surfaces. The geometric interpretation of $J_S$, $A_S$ and $B_S$ is as a triplet $(\omega^I,\omega^J,\omega^K)$ of $(1,1)$-forms related by the condition~\eqref{eq:hyperkahlertriplet}:
\begin{equation}
(\omega^I)^2 = (\omega^J)^2 = (\omega^K)^2 \,.
\end{equation}
We can actually derive this hyperk\"ahler condition directly from the structure of the conformal algebra owing to the identification $\no{~}\leftrightarrow \wedge$. Since we should consider the algebra only modulo the ideal $\langle N^1, N^2\rangle$ generated by the fields in \eqref{eq:N1N2}, any field appearing in OPEs involving $N^1$ or $N^2$ can be taken to vanish. After some work, we find
\begin{align}
\frac{1}{2} \, \wick{\no{A_S \c \psi_\theta}(z) \; \c N^1(w)}&=\frac{\no{J_SJ_S}(w)-\no{B_SB_S}(w)}{(z-w)^2}+\ldots \,,\\
\frac{1}{2} \, \wick{\no{B_S \c \psi_\theta}(z) \; \c N^2(w)}&=\frac{\no{J_SJ_S}(w)-\no{A_SA_S}(w)}{(z-w)^2}+\ldots \,,
\end{align}
proving in particular that \eqref{eq:AAmismatch} belongs to the ideal.

With the Odake algebra associated to $X_{\infty\pm}$ now established, we can proceed as in section~\ref{sec:TypeI} and define $G_2$ algebra generators by substituting the $n=3$ Odake generators \eqref{eq:T3}--\eqref{eq:D3} in \eqref{eq:T7}--\eqref{eq:M7}. For instance, we get
\begin{align} \label{eq:Phi7eg}
\Phi=\no{\no{\psi_t\psi_{\theta}}\psi_{\xi}}+\no{J_S\psi_{\xi}}+\no{\psi_{\theta}A_S}+\no{\psi_t  B_S} \,.
\end{align}
Again the $G_2$ OPEs are realised up to the ideal generated by $N^1$ and $N^2$. This yields the realisation
\begin{equation} \label{eq:InclusionII}
\text{SV}^{G_2} \overset{\text{II}}{\longhookrightarrow} \left(\text{Od}^{n=2} \times (\text{Free})^3\right) ~~ \Big/ ~ \langle N^1 , N^2\rangle \,.
\end{equation}

These realisations are not unique due to the automorphisms of $\text{Od}^{n=2}$ and $\text{Od}^{n=3}$ mentioned at the end of section~\ref{sec:TypeI}. There is in fact a two-parameter family of possible realisations of the $G_2$ algebra in region II and we have selected one for definitiveness.

\vfill

\subsection{Compatibility at junctions} \label{sec:Compatibility}

We distinguish three regions in the TCS geometry where the local patches supporting a known realisation of $\text{SV}^{G_2}$ overlap: $\text{I}^+\cap \text{II}^+$, $\text{II}^+\cap \text{II}^-$ and $\text{I}^-\cap \text{II}^-$. We now wish to establish transition maps on such overlaps. Let us start with $\text{I}^+\cap\text{II}^+$. $\text{I}^-\cap\text{II}^-$ is treated identically.

It should be rather clear from our exposition in section~\ref{sec:Realisations} that the realisations on $\text{I}^\pm\cap \text{II}^\pm$ are compatible. There is a clear map between the abstract generators $\{T,G,G^3,J,A,B,C,D\}$ in region $\text{I}^\pm$ and the explicit generators which they asymptote to in region $\text{II}^\pm$ as $t\rightarrow\infty$, see\ \eqref{eq:T3}--\eqref{eq:D3}. Under this correspondence, which is linear and bijective, the OPE relations of $\text{Od}^{n=3}$ and thus of the $G_2$ algebra, are invariant. Such a map preserving the algebraic structure defined by the OPEs will henceforth be called an \textit{automorphism} \label{p:Autom} of the operator algebra.

It remains to ascertain algebraic compatibility over the patch in the TCS geometry where the two parts merge into one another. On each side ($+/-$) of the joint, the $G_2$ algebra is realised in the same way (type $\text{II}$). We use $\pm$ labels to distinguish the generators. Geometrically the isomorphism $F_T$ in \eqref{eq:GeomPastingMap} allows to assemble the two sides. Algebraically this should correspond again to an automorphism of the $G_2$ algebra. We thus consider the map

\newpage
\vfill
\begin{equation}
F:\begin{cases}
({\mathscr{I}}_{t+},\psi_{t+})           &\longmapsto \quad (-{\mathscr{I}}_{t-},-\psi_{t-})         \\
\\
(T_{S+},G_{S+})              &\longmapsto \quad (T_{S-},G_{S-})              \\
\qquad\quad J_{S+}           &\longmapsto \quad A_{S-}                       \\
\qquad\quad A_{S+}           &\longmapsto \quad J_{S-}                       \\
\qquad\quad B_{S+}           &\longmapsto \quad -B_{S-}                      \\
\qquad\quad G^3_{S+}         &\longmapsto \quad C_{S-}                       \\
\qquad\quad C_{S+}           &\longmapsto \quad G^3_{S-}                     \\
\qquad\quad D_{S+}           &\longmapsto \quad -D_{S-}                      \\
\\
({\mathscr{I}}_{\xi+},\psi_{\xi+})       &\longmapsto \quad ({\mathscr{I}}_{\theta-},\psi_{\theta-}) \\
({\mathscr{I}}_{\theta+},\psi_{\theta+}) &\longmapsto \quad ({\mathscr{I}}_{\xi-},\psi_{\xi-})\,. 
\end{cases} \label{eq:PastingAuto}
\end{equation}

It is easy to see that $T_{\mathcal{M}+}\longmapsto T_{\mathcal{M}-}$ and $G_{\mathcal{M}+}\longmapsto G_{\mathcal{M}-}$ under these transformations (see\ \eqref{eq:T7}, \eqref{eq:G7} and \eqref{eq:T3}--\eqref{eq:D3}). We can also straightforwardly verify that the generator $\Phi$ in~\eqref{eq:Phi7eg} is preserved:
\begin{equation}
\Phi_+ \longmapsto \left(-\no{\no{\psi_t\psi_{\xi}}\psi_{\theta}}+\no{A_S\psi_{\theta}}+\no{\psi_{\xi}J_S}+\no{\psi_t B_S}\right)_- = \Phi_-\,,
\end{equation}
where the last equality follows from associativity of normal ordering and skew-commutativity for free fermions. Since $T_\mathcal{M}$, $G_\mathcal{M}$ and $\Phi$ can be used to define the remaining generators ($X$, $K$, $M$), it may seem clear from here that $F$ defines an automorphism by sending all generators to themselves. However some more work is needed to prove this statement\footnote{We remove the $\pm$ to avoid cluttering the proof.} since the fields $N^1$, $N^2$ (see\ \eqref{eq:N1II}--\eqref{eq:N2II}) get acted upon non-trivially by the map $F$:
\begin{align}
N^1 &\longmapsto \tilde{N}^1 \equiv -\no{\psi_t\no{A_SJ_S}}+\no{\psi_tB_S'}+\no{\psi_\xi\no{A_SB_S}}+\no{\psi_\xi J_S'}\,, \\
N^2 &\longmapsto \tilde{N}^2 \equiv \no{\psi_t\no{A_SB_S}}+\no{\psi_tJ_S'}+\no{\psi_\xi\no{A_SJ_S}}-\no{\psi_\xi B_S'}\,.
\end{align}
Recall that modding out by these fields is essential for associativity, so the status of associativity is a priori unclear.

We can show however that the original ideal $\langle N^1, N^2\rangle$ generated by $N^1$ and $N^2$ is in fact isomorphic to the ideal $\langle \tilde{N}^1, \tilde{N}^2\rangle$. It is also straightforward to see explicitly that all the remaining generators \{$X$, $K$, $M$\} are preserved by the map \eqref{eq:PastingAuto}, which guarantees that $F$ is indeed an automorphism of the $G_2$ algebra.

To prove the equivalence of the ideals, we notice first from \eqref{eq:N1II} and \eqref{eq:N2II} that $N^1$ and $N^2$ are contained in the ideal generated by
\begin{align}
N_S^1 &= A_S'-\no{J_SB_S}\,,
\\
N_S^2 &= B_S'+\no{J_SA_S}\,.
\end{align}
Thus $\langle N^1, N^2\rangle \subseteq \langle N_S^1, N_S^2\rangle$. On the other hand, $N_S^1$, $N_S^2$ may be found in $\langle N^1, N^2\rangle$, which proves the other inclusion. Indeed we have
\begin{align}
\wick{ \c \psi_\theta(z) \c N^1(w)}&=\frac{N_S^1(w)}{z-w}\,, \\
\wick{ \c \psi_\theta(z) \c N^2(w)}&=\frac{N_S^2(w)}{z-w}\,.
\end{align}

Hence $\langle N^1, N^2\rangle = \langle N_S^1, N_S^2\rangle$. We can then prove $\langle \tilde{N}^1, \tilde{N}^2\rangle = \langle N_S^1, N_S^2\rangle$ as follows. First we note that $\no{[A_S,J_S]}=2B_S'$ (following from $\text{Od}^{n=2}$). From this we see that $\tilde{N}^1$ and $\tilde{N}^2$ are normal ordered products,
\begin{align}
\tilde{N}^1 &= \no{\psi_t N^2_S}+\no{\psi_\theta N^3_S}\,, \\
\tilde{N}^2 &= \no{\psi_t N^3_S}-\no{\psi_\xi N^2_S}\,,
\end{align}
of free fermions with $N^2_S$ and the field
\begin{equation}
N^3_S=J_S'+\no{A_SB_S}\,,
\end{equation}
which itself arises as
\begin{equation}
\frac{1}{2} \, \wick{\no{A_S \c \psi_\theta}(z) \; \c N^2(w)} =\frac{N^3_S(w)}{(z-w)^2}+\ldots \,.
\end{equation}
This shows $\langle \tilde{N}^1, \tilde{N}^2\rangle \subseteq \langle N_S^1, N_S^2\rangle$. We get the remaining inclusion as follows:
\begin{align}
\frac{1}{2} \, \wick{\no{J_S \c \psi_\xi}(z) \; \c {\tilde{N}^2}(w)} &=\frac{N_S^1(w)}{(z-w)^2}+\ldots \,, \\
\wick{ \c \psi_\xi(z) \c {\tilde{N}}^2(w)} &=\frac{N_S^2(w)}{z-w}\,.
\end{align}

Summarising, in spite of the fact that $N^1$ and $N^2$ are not invariant under $F$, the ideal they generate is. We write this as
\begin{equation}
\langle N^1 , N^2 \rangle \subset \frac{\text{Od}^{n=2}\times (\text{Free})^3}{F}  \,.
\end{equation}
Moreover all the generators of the $G_2$ algebra are invariant, which guarantees that the map acts as the identity automorphism at the level of the $G_2$ algebra. Alternatively, the realisation defined by \eqref{eq:InclusionII} sits in the quotient by $F$:
\begin{equation} \label{eq:resultForIntro}
\text{SV}^{G_2} \longhookrightarrow \frac{\text{Od}^{n=2} \times (\text{Free})^3}{F} ~~ \Big/ ~ \langle N^1 , N^2\rangle \,.
\end{equation}

\vfill

\section{Automorphisms and $G_2$ mirror symmetry} \label{sec:Automorphisms}

Multiple conformal algebras are delicately incorporated in the TCS realisation of the $G_2$ algebra presented in the last section. Each of these admit automorphisms and some of them are known to be related to T-duality and Calabi--Yau mirror symmetry in type II string theory \cite{Lerche:1989uy}. For example, some of the early hints of the existence of Calabi--Yau mirrors were revealed by applying the automorphism
\begin{equation} \label{eq:Calabi-YauMiSYAut}
J\longmapsto -J\,, \qquad G^3\longmapsto -G^3 \,,
\end{equation}
on, say, the right-chiral sector of $\mathcal{N}=(2,2)$ SCFTs (keeping the left-movers invariant).

It is natural to ask if similar automorphisms are associated to mirror symmetry in the $G_2$ case, at least in the examples suggested of this phenomenon. Mirror symmetry for Joyce orbifolds received most of the attention \cite{Acharya:1996fx,Acharya:1997rh,Papadopoulos:1995da,Gaberdiel:2004vx}. In particular in \cite{Gaberdiel:2004vx}, by generalising the concept of discrete torsions \cite{Vafa:1994rv}, the authors were able to interpret all the then-known $G_2$ mirror dualities as the result of an automorphism $\bm{\mathcal{M}}$ applied on the (right-chiral) $G_2$ algebra. In the examples they covered, this automorphism was generated by T-dualities along the flat directions.

Equipped with our construction from section~\ref{sec:Realisations}, we are now in a position to get a similar conformal field theoretic understanding of mirror symmetry for twisted connected sums \cite{Braun:2017ryx,Braun:2017csz}. To this end, we consider in this section some automorphisms of the algebras discussed above: free fields, $\text{Od}^{n=2}$, $\text{Od}^{n=3}$ and $\text{SV}^{G_2}$. Then we address how these automorphisms may be composed together while respecting the compatibility maps established in section~\ref{sec:Compatibility}. This yields two ``mirror'' maps valid for the whole TCS realisation. We find striking similarities with the geometric maps $\mathcal{T}_4$ and $\mathcal{T}_3$ proposed in \cite{Braun:2017ryx,Braun:2017csz} and we connect with the interpretation in \cite{Gaberdiel:2004vx}.

\subsection{Basic automorphisms and their relations} \label{sec:AutomBasic}

We start by listing some simple automorphisms of the algebras used previously which hold regardless of any particular realisation. Almost all of them consist in sign shifts so it is convenient to display them in tables. As an example of how to read the tables, the \textit{parity} automorphism in the first row of the first table translates in
\begin{equation}
\mathbb{\mathbf{P}}_\xi~:~\begin{cases}
{\mathscr{I}} & \longmapsto +{\mathscr{I}} \\
\psi & \longmapsto -\psi \,.
\end{cases}
\end{equation}
All of these automorphisms can be checked easily by considering the explicit OPE relations in appendix~\ref{app:algebras}.

\subsubsection*{Free:}

$\qquad$\renewcommand\arraystretch{1.2}
\begin{tabular}{ r | c | c }
 & ${\mathscr{I}}_\xi$ & $\psi_\xi$ \\\hline
Parity $\textbf{P}_\xi$ & $+$ & $-$ \\\hline
T-duality $\textbf{T}_\xi$ & $-$ & $-$
\end{tabular}
\label{tab:Free}

\vspace{10pt}

As the name suggests, when $\xi$ parametrises an $\mathbb{S}^1$, the second map above applied on right-movers is a T-duality.

\vspace{5pt}

\subsubsection*{Odake $n=2$:}

$\qquad$\renewcommand\arraystretch{1.2}
\begin{tabular}{ r | c | c | c | c | c | c | c | c }
 & $T$ & $G$ & $G^3$ & $J$ & $A$ & $B$ & $C$ & $D$ \\\hline
Parity $\textbf{P}_S$ & $+$ & $-$ & $-$ & $+$ & $+$ & $+$ & $-$ & $-$ \\\hline
Mirror symmetry $\textbf{M}_S$ & $+$ & $+$ & $-$ & $-$ & $+$ & $-$ & $+$ & $-$ \\\hline
Phase $\textbf{Ph}^{\phi=\pi}_S$ & $+$ & $+$ & $+$ & $+$ & $-$ & $-$ & $-$ & $-$
\end{tabular}

\vspace{10pt}

Note that the so-called \textit{phase} automorphism in the last table is a special case of the rotation
\begin{equation}
\textbf{Ph}^\phi : \Omega\longmapsto e^{i\phi}\Omega\,,\qquad \Upsilon\longmapsto e^{i\phi}\Upsilon\,,
\end{equation}
which was briefly pointed out in section~\ref{sec:TypeI}.

\vspace{5pt}

\subsubsection*{Odake $n=3$:}

$\qquad$\renewcommand\arraystretch{1.2}
\begin{tabular}{ r | c | c | c | c | c | c | c | c || c | c }
 & $T$ & $G$ & $G^3$ & $J$ & $A$ & $B$ & $C$ & $D$ & $N^1$ & $N^2$ \\\hline
Parity $\textbf{P}$ & $+$ & $-$ & $-$ & $+$ & $-$ & $-$ & $+$ & $+$ & $-$ & $-$ \\\hline
Mirror symmetry $\textbf{M}$ & $+$ & $+$ & $-$ & $-$ & $+$ & $-$ & $+$ & $-$ & $+$ & $-$ \\\hline
Phase $\textbf{Ph}^{\phi=\pi}$ & $+$ & $+$ & $+$ & $+$ & $-$ & $-$ & $-$ & $-$ & $-$ & $-$ 
\end{tabular}

\vspace{10pt}

\subsubsection*{$G_2$ algebra:}

$\qquad$\renewcommand\arraystretch{1.2}
\begin{tabular}{ r | c | c | c | c | c | c || c }
 & $T$ & $G$ & $\Phi$ & $X$ & $K$ & $M$ & $N$ \\\hline
Parity $\bm{\mathcal{P}}$ & $+$ & $-$ & $-$ & $+$ & $+$ & $-$ & $-$ \\\hline
GK mirror symmetry $\bm{\mathcal{M}}$ & $+$ & $+$ & $-$ & $+$ & $-$ & $+$ & $+$
\end{tabular}

\vspace{10pt}

The automorphism $\bm{\mathcal{M}}$ of $\text{SV}^{G_2}$ was first observed in \cite{Becker:1996ay} (see also \cite{Roiban:2002iv}). It was interpreted as mirror symmetry of Joyce orbifolds by Gaberdiel and Kaste (GK) in \cite{Gaberdiel:2004vx}, as mentioned above.

\vspace{10pt}

We now point out a few relations between these automorphisms when the algebras are assembled together as per the TCS construction given in section~\ref{sec:Realisations}. Starting in region I, we notice that the T-duality map applied to the free algebra labelled by $\xi$, when combined with $\mathbf{Ph}^\pi$ on the $\text{Od}^{n=3}$ factor, induces an involution of $\text{Od}^{n=3}\times \text{Free}_\xi$. It can be checked explicitly with \eqref{eq:T7}--\eqref{eq:M7} that this involution restricts to a well-defined automorphism of the $G_2$ algebra sitting inside (see\ \eqref{eq:InclusionI}). The map defined in this way is the  Gaberdiel--Kaste automorphism. We write this as
\begin{equation} \label{FOFautomT}
\mathbf{T}_\xi \circ \mathbf{Ph}^\pi \overset{\text{I}}{\dashrightarrow} \bm{\mathcal{M}} \,.
\end{equation}
Note that all the automorphisms in the tables above are involutive, so we can compose them in any order we want.

Another way to produce the Gaberdiel--Kaste automorphism in region $\text{I}$ is with a mirror automorphism of the Odake factor:
\begin{equation} \label{FOFautomM}
\mathbf{M} \circ \mathbf{Ph}^\pi \overset{\text{I}}{\dashrightarrow} \bm{\mathcal{M}} \,.
\end{equation}
Combining these two, we get the identity map
\begin{equation} \label{FOFautomM+T}
\mathbf{M} \circ \mathbf{T}_\xi \overset{\text{I}}{\dashrightarrow} \textbf{1} \,.
\end{equation}

For the realisation of $\text{Od}^{n=3}$ in region $\text{II}$ defined by the embedding $\text{II}_X$ in \eqref{eq:InclusionIICY}, we identify the following relations:
\begin{align}
\mathbf{M}_S \circ \mathbf{T}_\theta \circ \mathbf{Ph}^\pi_S &\overset{\text{II}_X}{\dashrightarrow} \mathbf{M} \,, \label{eq:automoM->M} \\
\mathbf{Ph}^\phi_S &\overset{\text{II}_X}{\dashrightarrow} \mathbf{Ph}^\phi \,. \label{eq:automoPh->Ph}
\end{align}

\subsection{Two TCS automorphisms} \label{sec:AutomTCS}

We finally discuss algebra automorphisms consistent all over the twisted connected sum geometry.

Starting from one side of the construction, say the $+$ side, we can perform for example the map \eqref{FOFautomM+T} locally in region $\text{I}^+$. At the level of the $G_2$ algebra, this acts as the identity despite being generated by non-trivial maps. The geometric interpretation is clearly a T-duality on the external $\mathbb{S}^1_\xi$ and mirror symmetry on the ACyl Calabi--Yau $X_+$.

The map $\mathbf{M}\circ\mathbf{T}_\xi$ from \eqref{FOFautomM+T} in region I$^+$, via the transition relations \eqref{eq:T3}--\eqref{eq:D3} described in section­~\ref{sec:Compatibility}, gives rise uniquely to an automorphism of the realisation in region $\text{II}^+$. As an example of how this is done, consider $J$ in region $\text{I}^+$, which becomes $\no{\psi_t \psi_\theta}+J_S$ in region $\text{II}^+$. For $J$ to change sign, we need $J_S$ to change sign, as well as either $\psi_t$ or $\psi_\theta$. Continuing in this way, we find how all generators of $\text{Od}^{n=2}\times (\text{Free})^3$ should transform. The result is
\begin{equation}
\mathbf{M}_S \circ \mathbf{Ph}^\pi_S \circ \mathbf{T}_\theta \circ \mathbf{T}_\xi \,.
\end{equation}
Geometrically, this corresponds to taking the mirror of the K3 surface $S_+$ and T-duals of the circles $\mathbb{S}^1_\theta$ and $\mathbb{S}^1_\xi$. Interestingly, a phase automorphism is also necessary.

Proceeding similarly as we progress from region $\text{II}^+$ to region $\text{II}^-$, now with the pasting automorphism $F$ in \eqref{eq:PastingAuto}, we find again a unique consistent map in region $\text{II}^-$,
\begin{equation}
\mathbf{Ph}^\pi_S \circ \mathbf{M}_S \circ \mathbf{T}_\xi \circ \mathbf{T}_\theta \,.
\end{equation}
It is worth pointing out that the phase and mirror symmetry automorphisms in region $\text{II}^+$ get swapped as we get to region $\text{II}^-$. Finally, we reuse the asymptotic formulas \eqref{eq:T3}--\eqref{eq:D3} to reach a unique local form of the automorphism in region $\text{I}^-$.

Over the whole TCS construction, we have thus built a rigid arrangement of local maps starting from the choice \eqref{FOFautomM+T} in region $\text{I}^+$. It turns out that the automorphism generated on $\text{SV}^{G_2}$ is the same in all regions. By the relations \eqref{FOFautomM+T} and \eqref{eq:automoM->M} it is easy to see that it is the identity. We summarise these results in table~\ref{tab:TCS_Id}.

\begin{table}[h]
\begin{center}
\renewcommand\arraystretch{1.5}
\begin{tabular}{ c | c | c | c | c }

Region & $\text{I}^+$ & $\text{II}^+$ & $\text{II}^-$ & $\text{I}^-$ \\\hline

Autom. of factors & \pbox{20cm}{$\mathbf{M}$ \\ $\mathbf{T}_\xi$} & \pbox{20cm}{~\vspace{8pt} \\ $\mathbf{M}_S$ \\ $\mathbf{Ph}^\pi_S$ \\ $\mathbf{T}_\theta$ \\ $\mathbf{T}_\xi$ \\} & \pbox{20cm}{~\vspace{8pt} \\ $\mathbf{Ph}^\pi_S$ \\ $\mathbf{M}_S$ \\ $\mathbf{T}_\xi$ \\ $\mathbf{T}_\theta$ \\} & \pbox{20cm}{~\vspace{-6pt} \\ $\mathbf{M}$ \\ $\mathbf{T}_\xi$} \\\hline

Autom. of $\text{SV}^{G_2}$ & $\textbf{1}$ & $\textbf{1}$ & $\textbf{1}$ & $\textbf{1}$
\end{tabular}
\caption{TCS mirror automorphism $\mathcal{T}_4$}
\label{tab:TCS_Id}
\end{center}
\end{table}

By the geometrical interpretation given above, it seems clear that this scenario is a stringy version of the mirror map $\mathcal{T}_4$ defined in \cite{Braun:2017csz}. The authors of this paper choose for this map a special Lagrangian fibration of the ACyl Calabi--Yau manifolds and T-dualise along the $\mathbb{T}^3$ fibres and the external $\mathbb{S}^1_\xi$. As a consequence, the Calabi--Yau $2$- and three-folds get replaced by their mirrors by the SYZ argument \cite{Strominger:1996it}. We presently learn that this map $\mathcal{T}_4$ corresponds to the identity automorphism of the Shatashvili--Vafa algebra.\footnote{This fact has been verified explicitely for certain TCS obtained as Joyce orbifolds in \cite{Braun:2017csz}. See also \cite{Braun:2019lnn}.}

Note that our point of view however does not rely on any SYZ fibration. Only the Calabi--Yau mirror automorphism enters our consideration. In fact, we need no geometric assumptions at all suggesting that $\mathcal{T}_4$ persists after stringy corrections. Another advantage of our approach is that we can systematically classify all possible combinations of automorphisms that are mutually consistent in the TCS realisation and lead to global $G_2$ automorphisms. We have not pursued this in full generality, but we can immediately identify a second map besides $\mathcal{T}_4$.

Instead of \eqref{FOFautomM+T}, let us now start with \eqref{FOFautomM} in region $\text{I}^+$. Consistency with the transition maps again allows to progress inside the geometry and through the pasting isomorphism to produce the results of table~\ref{tab:TCS_M}. This is the only possibility consistent with \eqref{FOFautomM} in region $\text{I}^+$.

\begin{table}[h]
\begin{center}
\renewcommand\arraystretch{1.5}
\begin{tabular}{ c | c | c | c | c }

Region & I$^+$ & II$^+$ & II$^-$ & I$^-$ \\\hline

Autom. of factors & \pbox{20cm}{$\mathbf{M}$ \\ $\mathbf{Ph}^\pi$} & \pbox{20cm}{~\vspace{8pt} \\ $\mathbf{M}_S$ \\ $\mathbf{T}_\theta$ \\} & \pbox{20cm}{~\vspace{8pt} \\ $\mathbf{Ph}^\pi_S$ \\ $\mathbf{T}_\xi$ \\} & \pbox{20cm}{~\vspace{-2pt} \\ $\mathbf{Ph}^\pi$ \\ $\mathbf{T}_\xi$} \\\hline

Autom. of $\text{SV}^{G_2}$ & $\bm{\mathcal{M}}$ & $\bm{\mathcal{M}}$ & $\bm{\mathcal{M}}$ & $\bm{\mathcal{M}}$
\end{tabular}
\caption{TCS mirror automorphism $\mathcal{T}_3$}
\label{tab:TCS_M}
\end{center}
\end{table}

The geometric interpretation is now much different. Mirror symmetry only acts on one side of the construction, here $X_+$ and its asymptotic cross-section $S_+\times \mathbb{S}^1_\theta$, while T-duality acts on the other side (along with a phase shift). The induced $G_2$ automorphism is again identical over the whole geometry, but this time it is the Gaberdiel--Kaste mirror map $\bm{\mathcal{M}}$.

This case corresponds to the mirror map $\mathcal{T}_3$ from \cite{Braun:2017csz}. Geometrically, only the $+$ side ACyl Calabi--Yau was assumed to be SYZ fibered. On the other side, the K3 surface $S_-$ was assumed to be elliptically fibered.

\section{Conclusion} \label{sec:Conclusion}

We have explained in this chapter how to take advantage of relations like \eqref{eq:GeometryVsalgebra} and inspiration from existing realisations of $\text{SV}^{G_2}$ to establish a realisation for twisted connected sums. Our realisation could perhaps seem unsurprising based on the mathematically established existence of Ricci-flat metrics on TCSs, which suggests a fixed point of the RG flow. However, we should recall that a true CFT corresponds to a Ricci-flat target manifold only to leading orders in worldsheet perturbation theory. Moreover, the aforementioned ambiguities in the correspondence between CFT and geometry makes worthwhile an explicit construction.

Our approach gives a certain exhaustivity in the identification of mirror maps for TCS geometries. A preliminary analysis was sufficient to recover a stringy version of both mirror maps $\mathcal{T}_4$ and $\mathcal{T}_3$ suggested by Braun and Del Zotto \cite{Braun:2017csz}. It is quite possible that a more systematic inspection of the automorphisms of the algebras we used---$G_2$ algebra, Odake $n=3$ and Odake $n=2$---would reveal new mirror maps by following the logic of section~\ref{sec:AutomTCS}.

Perhaps the most interesting future direction to pursue with our TCS realisation is the highest weight representation theory of these symmetry algebras. Interesting lessons about mirror symmetry could be learned from a firmer grasp on the Hilbert space of the CFT. A prime example is to determine the effect of the mirror maps $\mathcal{T}_3$ and $\mathcal{T}_4$ on the Betti numbers of the TCS manifold. This would be in analogy with the exchange of $h^{1,1}$ and $h^{2,1}$ in Calabi--Yau mirror symmetry, which follows from the automorphism \eqref{eq:Calabi-YauMiSYAut}.

The key step for this task is the correspondence between Ramond--Ramond ground states and cohomology groups explained in \cite{Shatashvili:1994zw}. This correspondence remains ambiguous, partly due to the generalised mirror conjecture stated in the introduction of this thesis. In the case of the $G_2$ algebra, only the sum
\begin{equation}
b^2+b^3
\end{equation}
of Betti numbers can be obtained from the algebra; not the indvidual values. We can suggest however that the explicit realisation of the $G_2$ algebra we have given will allow a more detailed understanding of the ground states and, in turn, of the Betti numbers.

Carrying out this program explicitly requires the representation theory of the local conformal algebras used in regions $\text{I}$ and $\text{II}$. In spite of the known representation theory of the Odake and free algebras, this is technically complicated because adding generators substantially alters the labelling of highest weight states and the creation operators and thus the whole tower of descendants. Subsequently, one would have to devise transition maps for the ground states following the philosophy of section~\ref{sec:Compatibility} and understand how to make global statements about the topology. Comparison with the geometric approach based on a Mayer--Vietoris sequence \cite{Corti:2012kd} would certainly be useful for this.

We should admit the possibility that our presentation in this chapter of the twisted connected sum realisation of the $G_2$ CFT is perhaps not the best way to interpret our result. We trust however that it promises a better control on twisted connected sum manifolds from a worldsheet angle. A clearer view would certainly arise from a specific $G_2$ CFT built on the principles of this paper. We may for example try to use Gepner models for all Calabi--Yau factors in the TCS construction and assemble them together following the ideas in section~\ref{sec:Compatibility}.\footnote{This suggestion was made by S. Sch\"afer-Nameki.} This could yield a Gepner-type model for $G_2$ manifolds in the spirit of the discussion in \cite{Roiban:2001cp,Roiban:2002iv} (see also \cite{Eguchi:2001ip, Eguchi:2003yy,  Blumenhagen:2001jb, Noyvert:2002mc, Sugiyama:2001qh, Sugiyama:2002ag}).

Finally, we wish to point out the recent work \cite{Braun:2018joh} where a construction of $Spin(7)$-holonomy manifolds is given based on the twisted connected sum principle. We expect there to exist a corresponding realisation of the $Spin(7)$ superconformal algebra, also due to Shatashvili and Vafa \cite{Shatashvili:1994zw}. Mirror maps for these geometries would be an interesting by-product.

\begin{appendices}
\chapter{Explicit OPE relations} \label{app:algebras}

\section{Odake $n=2$} \label{app:Odake2}

Here we discuss the OPEs involving $\Omega$, $\Upsilon$ and their conjugates in the case $n=2$ of $\text{Od}^n$. There are a priori $10$ remaining OPEs. In real basis, they are $AA$, $AB$, $AC$, $AD$, $BB$, $BC$, $BD$, $CC$, $CD$ and $DD$. In fact, associativity of the OPE can be used to deduce them all in terms of the first three, for instance. We thus start with these and take
\begin{align}
\wick{\c A(z) \c A(w)}&=-\frac{2}{(z-w)^2}\,,
\\
\wick{\c B(z) \c B(w)}&=-\frac{2}{(z-w)^2}\,, \label{eq:AB}
\\
\wick{\c A(z) \c B(w)}&=-\frac{2J(w)}{z-w}\,.
\end{align}

As it turns out, the Odake $n=2$ algebra is the \emph{small $\mathcal{N}=4$ Virasoro in disguise} \cite{Odake:1988bh,Ali:1999ut,Ademollo:1976wv}. This algebra has four superymmetry currents $G^{0}$, $G^{1}$, $G^{2}$, $G^{3}$ and, as compared to the basic multiplet $(\tfrac{1}{2}G^{0},T)$, each new supercurrent introduces a new $U(1)$ current: $\mathcal{J}^{1}$, $\mathcal{J}^{2}$, $\mathcal{J}^{3}$, where $\mathcal{J}^{3}$ is proportional to the generator that we called $J$ previously. These currents form an \emph{$SU(2)$ Kac--Moody algebra} at level $k=c/6=1$ \cite{Ali:1999ut}:
\begin{equation}
\wick{\c {\mathcal{J}}^i(z) \c {\mathcal{J}}^j(w)}=\frac{k/2}{(z-w)^2}\delta_{ij}+\frac{i\epsilon_{ijk}\mathcal{J}^k(w)}{z-w}\,.
\end{equation}

The OPEs given in \eqref{eq:AB} are in fact consistent with the following identifications:
\begin{equation}
\mathcal{J}^{1}=\frac{A}{2i}\,,\qquad \mathcal{J}^{2}=\frac{B}{2i}\,,\qquad \mathcal{J}^{3}=\frac{J}{2i}\,.
\end{equation}
As for $\Upsilon$, it is clear by dimensional analysis, that it is related to the two new supersymmetry currents $G^1$, $G^2$.
From this, associativity is enough to fix the remaining OPEs:
\begin{align}
\wick{\c A(z) \c C(w)}&=\frac{G^0(w)}{z-w}\,,
\\
\wick{\c A(z) \c D(w)}&=-\frac{G^3(w)}{z-w}\,,
\\
\wick{\c B(z) \c C(w)}&=\frac{G^3(w)}{z-w}\,,
\\
\wick{\c B(z) \c D(w)}&=\frac{G^0(w)}{z-w}\,,
\\
\wick{\c C(z) \c C(w)}&=\frac{4}{(z-w)^3}+\frac{2T(w)}{z-w}\,,
\\
\wick{\c C(z) \c D(w)}&=\frac{2J}{(z-w)^3}+\frac{J'(w)}{z-w}\,,
\\
\wick{\c D(z) \c D(w)}&=\frac{4}{(z-w)^3}+\frac{2T(w)}{z-w}\,.
\end{align}

\section{Odake $n=3$} \label{app:Odake3}
\begin{align}
\wick{\c A(z) \c A(w)}&=-\frac{4}{(z-w)^3}+\frac{2\no{JJ}(w)}{z-w}
\\
\wick{\c B(z) \c B(w)}&=-\frac{4}{(z-w)^3}+\frac{2\no{JJ}(w)}{z-w}
\\
\wick{\c A(z) \c B(w)}&=-\frac{4J(w)}{(z-w)^2}-\frac{2J'(w)}{z-w}
\end{align}
\begin{align}
\wick{\c A(z) \c C(w)}&=-\frac{2G^0(w)}{(z-w)^2}-\frac{2\no{JG^3}(w)}{z-w}
\\
\wick{\c A(z) \c D(w)}&=\frac{2G^3(w)}{(z-w)^2}-\frac{2\no{JG^0}(w)}{z-w}
\\
\wick{\c B(z) \c C(w)}&=-\frac{2G^3(w)}{(z-w)^2}+\frac{2\no{JG^0}(w)}{z-w}
\\
\wick{\c B(z) \c D(w)}&=-\frac{2G^0(w)}{(z-w)^2}-\frac{2\no{JG^3}(w)}{z-w} 
\end{align}
\begin{align}
\wick{\c C(z) \c C(w)}&=-\frac{12}{(z-w)^4}+\frac{(2\no{JJ}-4T)(w)}{(z-w)^2}+\frac{(2\no{J'J}-2T')(w)}{z-w}
\\
\wick{\c C(z) \c D(w)}&=-\frac{8J(w)}{(z-w)^3}-\frac{4J'(w)}{(z-w)^2}+\frac{(2\no{GG^3}-4\no{TJ})(w)}{z-w}
\\
\wick{\c D(z) \c D(w)}&=-\frac{12}{(z-w)^4}+\frac{(2\no{JJ}-4T)(w)}{(z-w)^2}+\frac{(2\no{J'J}-2T')(w)}{z-w}
\end{align}

Associativity of the OPE is only realised modulo the ideal generated by the null weight-$5/2$ fields \cite{Odake:1988bh}
\begin{equation}
N^1=A' -\no{JB}\,,\qquad N^2=B' +\no{JA}\,.
\end{equation}

\section{$G_2$ Shatashvili--Vafa} \label{app:SV}

The Shatashvili--Vafa $G_2$ superconformal algebra is an extension of $\mathcal{N}=1$ Virasoro $(\tfrac{1}{2}G,T)$ with central charge $c=21/2$.
\begin{align}
\wick{\c T(z) \c T(w)}&=\frac{c/2}{(z-w)^4}+\frac{2T(w)}{(z-w)^2}+\frac{T'(w)}{z-w}\,,
\\
\wick{\c T(z) \c G(w)} &=\frac{3/2}{(z-w)^2}G(w)+\frac{G'(w)}{z-w}\,,
\\
\wick{\c G(z) \c G(w)} &=\frac{2c/3}{(z-w)^3}+\frac{2T(w)}{z-w}\,.
\end{align}
The extra multiplets are $(\Phi,K)$ and $(X,M)$ where the lowest components have respectively weight $3/2$ and $2$. We start with the OPEs with the $\mathcal{N}=1$ generators:

\begin{align}
\wick{\c T(z) \c \Phi(w)}&=\frac{3/2}{(z-w)^2}\Phi(w)+\frac{\Phi'(w)}{z-w} \,,
\\
\wick{\c T(z) \c X(w)}&=-\frac{7/4}{(z-w)^4}+\frac{2X(w)}{(z-w)^2}+\frac{X'(w)}{z-w} \,,
\\
\wick{\c T(z) \c K(w)}&=\frac{2K(w)}{(z-w)^2}+\frac{K'(w)}{z-w} \,,
\\
\wick{\c T(z) \c M(w)}&=-\frac{1/2}{(z-w)^3}G(w)+\frac{5/2}{(z-w)^2}M(w)+\frac{M'(w)}{z-w} \,,
\\
\wick{\c G(z) \c \Phi(w)}&=\frac{K(w)}{z-w} \,,
\\
\wick{\c G(z) \c X(w)}&=-\frac{1/2}{(z-w)^2}G(w)+\frac{M(w)}{z-w} \,,
\\
\wick{\c G(z) \c K(w)}&=\frac{3\Phi(w)}{(z-w)^2}+\frac{\Phi'(w)}{z-w} \,,
\\
\wick{\c G(z) \c M(w)}&=-\frac{7/2}{(z-w)^4}+\frac{\left(T+4X\right)(w)}{(z-w)^2}+\frac{X'(w)}{z-w} \,.
\end{align}

We now provide the OPEs of the new generators with each other:

\begin{align}
\wick{\c \Phi(z) \c \Phi(w)}&=-\frac{7}{(z-w)^3}+\frac{6X(w)}{z-w} \,,
\\
\wick{\c \Phi(z) \c X(w)}&=-\frac{15/2}{(z-w)^2}\Phi(w)-\frac{5/2}{(z-w)}\Phi'(w) \,,
\\
\wick{\c \Phi(z) \c K(w)}&=-\frac{3G(w)}{(z-w)^2}-\frac{3}{z-w}\left(M+\frac{1}{2}G'\right)(w) \,,
\\
\wick{\c \Phi(z) \c M(w)}&=\frac{9/2}{(z-w)^2}K(w)+\frac{1}{z-w}\left(3\no{\Phi G}-\frac{1}{2} K'\right)(w) \,,
\\
\wick{\c X(z) \c X(w)}&=\frac{35/4}{(z-w)^4}-\frac{10X(w)}{(z-w)^2}-\frac{5X'(w)}{z-w} \,,
\\
\wick{\c X(z) \c K(w)}&=-\frac{3K(w)}{(z-w)^2}-\frac{3\no{\Phi G}(w)}{z-w} \,,
\end{align}
\begin{align}
\wick{\c X(z) \c M(w)}&=-\frac{9/2}{(z-w)^3}G(w)-\frac{1}{(z-w)^2}\left(5M+\frac{9}{4}G'\right)(w) \\
&\qquad\qquad\qquad\qquad\qquad
 +\frac{1}{z-w}\left(4\no{XG}+\frac{1}{2}M'+\frac{1}{4} G''\right)(w) \,, \nonumber
\\
\wick{\c K(z) \c K(w)}&=-\frac{21}{(z-w)^4}+\frac{6(X-T)(w)}{(z-w)^2}+\frac{3(X'-T')(w)}{z-w} \,,
\\
\wick{\c K(z) \c M(w)}&=-\frac{15\Phi(w)}{(z-w)^3}-\frac{11/2}{(z-w)^2}\Phi'(w)+\frac{\left(3\no{GK}-6\no{TP}\right)(w)}{z-w} \,,
\\
\wick{\c M(z) \c M(w)}&=-\frac{35}{(z-w)^5}+\frac{(20X-9T)(w)}{(z-w)^3}+\frac{1}{(z-w)^2}\left(10X'-\frac{9}{2}T'\right)(w) \nonumber\\
&\qquad+\frac{1}{z-w}\left(\frac{3}{2}(X''-T'')-4\no{GM}+8\no{TX}\right)(w)  \,.
\end{align}

Associativity of the OPE is only realised modulo the ideal generated by the null weight-$7/2$ field \cite{Figueroa-OFarrill:1996tnk}
\begin{equation}
N=4\no{GX}-2\no{\Phi K}-4M'-G''\,.
\end{equation}
\end{appendices}

\part{Non-linear $\sigma$-models} \label{part:II}

\chapter{$\mathcal{G}$-structure symmetries} \label{chap:G-symmetries}

\adjustmtc
\onehalfspacing
\minitoc
\vspace{25pt}

\doublespacing

Part I of the thesis was mainly from a bootstrap angle, focusing on various extended chiral symmetry algebras that two-dimensional conformal field theories may sometimes be expected to have. Heavily constrained by associativity (axiom 3, p.\pageref{eq:Jacobilike}), these symmetries offer unparalleled glimpses into exact quantum aspects of the theory. Meanwhile the Lagrangian-based formalism, when it exists, remains arguably the most concrete handle on the theory. It is of interest, when possible, to compare these frameworks.

In two dimensions, a good reason to study this connection
stems from the very intimate relationship of supersymmetric Lagrangians, or \emph{non-linear $\sigma$-models}, with the geometry of their space $\mathcal{M}$ of field configurations.
This is reflected in the operator algebras that mainly occupied us so far ($\text{SV}^{Spin(7)}$, $\text{SV}^{G_2}$, $\text{Od}^n$). They clearly have to do with certain $\mathcal{G}$-structures---understood here as the existence of nowhere vanishing differential forms on $\mathcal{M}$. However we remained somewhat vague about this relationship previously.
It is natural to ask for a characterisation of the precise target manifolds $\mathcal{M}$ related to a given operator algebra, at least to leading orders in $\sigma$-model perturbation theory. At the classical level, a long-established result due to Howe and Papadopoulos \cite{Howe:1991ic,Howe:1991vs,Howe:1991im} sheds light on this question in the context of massless $(1,0)$ non-linear models with generic target space metric as well as B-field \cite{Hull:1985jv}. Their result is a correspondence between conserved currents associated to certain symmetries on the one hand and differential forms $\Phi$ on $\mathcal{M}$, on the other hand, preserved by a connection $\nabla^+$,
\begin{equation}\label{eq:NablaPhi---Intro}
\nabla^+\Phi = 0\,,
\end{equation}
with symbols $\Gamma^+=\Gamma+\tfrac{1}{2}\dd B$ twisted by the flux.

In this chapter, largely based on \cite{delaOssa:2018azc}, we describe a simple but enlightening generalisation of this result, which we refer to as {\it extended $\mathcal{G}$-structure symmetry}. We assume minimal $(1,0)$ supersymmetry and we include a Fermi sector in the $\sigma$-model, allowing us to incorporate a vector bundle $\mathcal{V}\rightarrow\mathcal{M}$ with gauge field $A$ and curvature $F$, while keeping a general metric and B-field background. We also allow a mass term \cite{AlvarezGaume:1983ab,Hull:1993ct} coupled through a section $S$ of $\mathcal{V}^*$. The symmetry is described in sections~\ref{ssec:RevHP}--\ref{ssec:GstructureSym}. It holds provided we impose \eqref{eq:NablaPhi---Intro} and further geometric constraints, to be discussed extensively later:
\begin{align} \label{eq:iFPhi---Intro}
&i_F(\Phi)=0\,,
&i_{\dd_A S}(\Phi)=0\, .
\end{align}
The \textit{insertion operator} here is defined as follows. Let $P$ be a $p$-form with values in the tangent bundle of $\cal M$. The insertion operator $i_P$ is a derivation of degree $p-1$ on the space of forms, perhaps valued in a vector bundle $\mathcal{E}$ over $\mathcal{M}$. It is given by
\begin{equation}
\begin{split}
i_P~:~ \Omega^k({\cal M}, {\cal E}) &~~\longrightarrow~~ \Omega^{k+p-1}({\cal M}, {\cal E})
\\
\alpha ~~\quad&~~~\mapsto ~~~ i_P(\alpha) = P^i\wedge\alpha_i\, , 
\end{split}\label{eq:insert}
\end{equation}
where
\begin{equation}
\alpha_i = \frac{1}{(k-1)!}\, \alpha_{ij_1\cdots j_{k-1}}\, 
\dd x^{j_1\cdots j_{k-1}}\,.
\end{equation}

We comment on the systems $(\mathcal{M}, \mathcal{V}\,;\,G,B,A,S)$ solving the condition \eqref{eq:iFPhi---Intro} in section~\ref{ssec:Geometry}. These geometries are closely related to supersymmetric backgrounds of heterotic supergravity. Often in this chapter we will refer to the natural application of our results to heterotic compactifications. Meanwhile our statements are very general---we need only minimal supersymmetry---and valid regardless of the role played by the $\sigma$-model.

In section~\ref{sec:Anomalies}, we examine whether these $\mathcal{G}$-structure symmetries are anomalous. We prove (sect.~\ref{sec:HPSymAnomaly}) that the one-loop quantum effective action corresponding to $(1,0)$ $\sigma$-models \cite{Hull:1986xn} is invariant provided we assign order-$\alpha'$ quantum corrections to the conditions mentioned above. In particular, there must be a connection $\Theta$ on $T\mathcal{M}$ satisfying a curvature condition analogous to \eqref{eq:iFPhi---Intro},
\begin{equation}
i_{R^{\Theta}}(\Phi)=0\,,
\end{equation}
and the torsion in \eqref{eq:NablaPhi---Intro} must be replaced by the gauge-invariant combination
\begin{equation}
\dd B + \frac{\alpha'}{4}\left(\text{CS}_3(A)-\text{CS}_3(\Theta)\right)\,.
\end{equation}
This result is beautifully consistent with the worldsheet Green--Schwarz mechanism in heterotic string theory \cite{Green:1984sg,Hull:1986xn}, reviewed in section~\ref{sec:GreenSchwarz}. We discuss how sensitive these results are to counterterm ambiguities.

We also comment on gauge-invariance at order $\alpha'$ in relation with $(1,0)$ superconformal symmetry. Again, with the effective action, we show how to $\alpha'$-correct the supercurrent when flux is turned on. We connect with familiar results on conformal anomalies. The next section introduces our conventions about non-linear $\sigma$-models, which are also used in chapter~\ref{chap:MarginalDeformations}.

\section{Two-dimensional $(1,0)$ non-linear $\sigma$-model} \label{sec:2d NLSM}

\subsection{Conventions} \label{sec:Conventions}

Our $\sigma$-model conventions are as follows \cite{Hull:1985jv,Hull:1986xn,Lambert:1995hs}. We work on a compact worldsheet without boundary in Lorentzian signature and use lightcone coordinates $\sigma^+,\sigma^-$. To avoid cluttering formulas, we omit some of the usual Lorentz indices when no confusion is possible. The Grassmann direction is parametrized by $\theta$ and generic  superspace coordinates are denoted by $\varsigma^\mu = (\sigma^+,\sigma^-,\theta)$. We write the superspace measure as $d^{2|1}\varsigma = \dd \sigma^+ \dd \sigma^- \dd\theta$. The \textit{superderivative} and \textit{supercharge} are given by
\begin{equation} \label{eq:DQSuperspace}
D = \partial_\theta+i\theta\partial_+\, ,\qquad Q = \partial_\theta-i\theta\partial_+\, ,
\end{equation}
where by convention
\begin{equation}
\partial_+\sigma^+=\partial_-\sigma^-=\partial_\theta \theta=1\, .
\end{equation}
They satisfy $-Q^2 = D^2 = i\partial_+$ and both have weights $(h_+,h_-)= (1/2,0)$.

We need two types of superfields,
\begin{equation}
X^i = x^i +  \theta\psi^i\,,\qquad
\Lambda^\alpha = \lambda^\alpha + \theta f^\alpha\, .
\end{equation}
The \textit{Bose superfields} $X$ locally\footnote{Target space will always be described with a preferred set of local coordinates, hence we ignore dynamical effects involving more than one coordinate patches. Assembling local patches into a global geometry is usually done via invariances, under diffeomorphisms and gauge transformations for instance. This is not different for $\sigma$-models. In section \ref{sec:GreenSchwarz}, we examine gauge invariance more closely at the $\sigma$-model quantum-level.} define a map $X : \Sigma\,\longrightarrow\, \mathcal{M}$ from superspace $\Sigma$ to a $d$-dimensional target space $\mathcal{M}$ and have weights $(0,0)$. Their leading components are ordinary bosonic fields, while $\psi$ are left-moving Majorana--Weyl fermions. The \textit{Fermi superfields} $\Lambda$ have weights $(0,1/2)$ and form a section of the bundle $\sqrt{K}_-\otimes {X}^*\mathcal{V}$, where $\mathcal{V}$ is a vector bundle with connection $A$ on the target space $\mathcal{M}$ and $\sqrt{K}$ is the spin bundle over the worldsheet. The Majorana--Weyl fermions $\lambda$ are right-moving and $f$ are auxiliary fields.

The most general renormalizable action preserving $(1,0)$ supersymmetry \cite{Hull:1985jv, Sen:1985qt} that can be written for these fields follows from dimensional analysis. Allowing also for a mass term, we shall consider $\text{S}=\text{S}_{\mathcal{M}}+\text{S}_{\mathcal{V}}+\text{S}_{\mathcal{S}}$, where
\begin{align}
\text{S}_{\mathcal{M}}[X] &= \int_\Sigma\frac{\dd^{2|1}\varsigma}{4\pi\alpha'} ~(- i)\, M_{ij}(X)\,  DX^i\partial_-X^j\, , \label{eq:S_M} \\
\text{S}_{\mathcal{V}}[X,\Lambda] &= \int_\Sigma\frac{\dd^{2|1}\varsigma}{4\pi\alpha'} ~  \tr ( \Lambda  D_A\Lambda) \, , \label{eq:S_V} \\
\text{S}_{\mathcal{S}}[X,\Lambda] &= \int_\Sigma\frac{\dd^{2|1}\varsigma}{4\pi\alpha'} ~  m\, \tr( S(X)\, \Lambda) \, . \label{eq:S_S}
\end{align}
Here $M(X)$ is a $d\times d$ matrix whose symmetric and anti-symmetric parts are the target space metric and Kalb--Ramond field: $M_{ij}=G_{ij}+B_{ij}$. We also use the gauge covariant superspace derivative
\begin{equation}
D_A\Lambda^\alpha = D\Lambda^\alpha + \hat A^\alpha{}_\beta\Lambda^\beta\, .
\end{equation}
Here and later, we add hats to operators constructed by appending factors of superderivatives of the Bose superfields to expressions with form indices. For example $
\hat A^\alpha{}_\beta = A_i{}^\alpha{}_\beta(X) \, D X^i\, .
$
The trace over bundle-valued forms is taken with respect to the bundle metric $h_{\alpha\beta}(X)$, so in the expression for the action this means 
\begin{equation}
\tr(\Lambda D_A\Lambda)  
=  h_{\alpha\beta}(X)\, \Lambda^\alpha D_A\Lambda^\beta\, .
\end{equation}
We choose, without loss of generality, the bundle metric $h_{\alpha\beta}$ to be constant, i.e.\ $h_{\alpha \beta}=\delta_{\alpha \beta}$. Finally, $m$ is a \emph{constant} parameter of mass dimension one and $S(X)$ is a section of ${\cal V}^*$. The associated term is a potential for the bosonic fields introduced in \cite{AlvarezGaume:1983ab,Hull:1993ct}. It may be used to cure infrared divergences \cite{AlvarezGaume:1981hn,AlvarezGaume:1983ab} and is related to solitonic effects \cite{Papadopoulos:1994kj} and Landau--Ginzburg theories \cite{Witten:1994tz}.

\vfill

\subsection{General variations}\label{sec:Vars}

We begin by considering general \emph{variations} $\delta \text{S}$ of the action \eqref{eq:S_M}--\eqref{eq:S_S} to prepare the ground for analysing local symmetries. In this chapter, we shall write expressions like
\begin{equation} \label{eq:deltaS}
\delta \text{S} = \int \frac{\dd^{2|1}\varsigma}{4\pi\alpha'} ~ \left(\frac{\delta \text{S}}{\delta X^i}\, \delta X^i + \frac{\delta \text{S}}{\delta \Lambda^\alpha}\, \delta \Lambda^\alpha\right)\,.
\end{equation}
It should be stressed that \emph{functional derivatives} with respect to the superfields must be interpreted as being evaluated at a point $\varsigma=(\sigma^+,\sigma^-,\theta)$ in superspace. For example, the functional derivative in the first term of the integrand is more properly denoted
\begin{equation} \label{eq:EoMFuncDer}
\frac{\delta }{\delta X^i(\varsigma)}\text{S}[X,\Lambda]\,,
\end{equation}
and it is in particular a field as opposed to the integral of a field. All of our fields will usually have their arguments suppressed.

To illustrate what we mean here, consider the toy example $\text{S}[X] = \int \frac{\dd^{2|1} \varsigma}{4\pi\alpha'} X(\varsigma)$. Then, taking its functional derivative gives 
\begin{equation} \label{eq:Toy}
\frac{\delta}{\delta X(\varsigma_0)} \text{S}[X]
= \int \frac{\dd^{2|1} \varsigma}{4\pi\alpha'} \delta^{2|1}(\varsigma - \varsigma_0) = 1\,. 
\end{equation}
Note that this is different from the functional derivative of the Lagrangian $\text{L}(X)\equiv X$, which is the formal Dirac delta $\delta^{2|1}(\varsigma-\varsigma_0)$.

Functional derivatives of an action are easily read off from the result \eqref{eq:deltaS} of an infinitesimal perturbation. In the toy example, the value $1$ in \eqref{eq:Toy} is easily read off from
\begin{equation}
\delta \text{S} = \int \frac{\dd^{2|1} \varsigma}{4\pi\alpha'} ~~ 1 ~~ \delta X^i \,.
\end{equation}
In order to put the variation in the form \eqref{eq:deltaS}, we must usually integrate by parts. Boundary terms must also be discarded on the account that we work on a closed worldsheet.\footnote{Strictly speaking, the argument here is flawed because of our choice to stick to a local patch of coordinates $(\sigma^+, \sigma^-)$ on the worldsheet. This patch is non-compact and infinitely big as measured by the local Lorentzian worldsheet metric. Its size in fact leads to infrared divergences, discussed in section~\ref{sec:Caveat}. It would be better to account for boundary terms more precisely. At genus $0$, one can for example work on a non-compact worldsheet, but impose as a boundary condition that $|\sigma|\rightarrow \infty$ maps to the single point $x^i_\infty$ in the target space. Alternatively one can attempt to formulate the physics globally on the worldsheet. These issues are in no way specific to our problem---they arise in textbook string theory---and we shall not comment further. The open string case, where boundaries are welcome, is also beyond our scope but very interesting in the context.} Setting \eqref{eq:EoMFuncDer} to zero gives a classical equation of motion.

Returning to the non-linear $\sigma$-model, we find 
\begin{align}
\frac{\delta \text{S}_{\mathcal{M}}}{\delta X^i} &= 2i\, G_{ij}\left( D\partial_- X^j 
+ \Gamma^+{}^{\, j}{}_{kl}\, \partial_-X^k D X^l\right) \, ,\label{eq:EoMX}
\\[3pt]
\frac{\delta \text{S}_{\mathcal{V}}}{\delta X^i} &= \tr\big(\Lambda F_{ij}D X^j\Lambda\big)
+ 2 h_{\alpha\beta}
\big(
 D_A\Lambda^\alpha 
\big) A_i{}^\beta{}_\delta \Lambda^\delta\, ,
\\[3pt]
\frac{\delta \text{S}_{\mathcal{S}}}{\delta X^i} &=  m \, \tr \big((\partial_i S) \Lambda\big)\, ,
\end{align}
where $F$ is the curvature 2-form of $A$,
\begin{equation}
F = \dd A + A\wedge A\, ,
\end{equation}
and we have defined a connection $\nabla^+$  on $T\mathcal{M}$ with symbols $\Gamma^+$ given by
\begin{equation}
\Gamma^+{}^i{}_{jk} = \Gamma^{\, i}{}_{jk} + \frac{1}{2}\, (\dd B)^i{}_{jk}
= \Gamma^-{}^i{}_{kj}\, , \label{eq:GammaPlus}
\end{equation}
where $\Gamma$ represents the Levi--Civita connection symbols.\footnote{We take covariant derivatives to act as $\nabla_i\Phi_j=\partial_i\Phi_j-\Gamma^k{}_{ij}\Phi_k$ on 1-forms.}

The variations with respect to the Fermi superfields are
\begin{equation}
\frac{\delta \text{S}_{\mathcal{M}}}{\delta \Lambda^\alpha} = 0\, ,
\qquad
\frac{\delta \text{S}_{\mathcal{V}}}{\delta \Lambda^\alpha} = 2\, h_{\alpha\beta} \,
 D_A\Lambda^\beta 
\,,
\qquad
\frac{\delta \text{S}_{\mathcal{S}}}{\delta \Lambda^\alpha} =  m\, S_\alpha\,.
\end{equation}

It will be easier to describe symmetries if we write these expressions in terms of {\it covariant} perturbations $\delta_A\Lambda$ of $\Lambda$ \cite{Hull:1993ct}, that is
\begin{equation}
\delta_A \Lambda^\alpha =
\delta\Lambda^\alpha 
+ A_i{}^\alpha{}_\beta \Lambda^\beta\, \delta X^i\,.\label{eq:CovdeltaL}
\end{equation}

In terms of this, a general variation of the action can be written as
\begin{equation} \label{eq:generalCovariantVariationAction}
\delta \text{S} = \int \frac{\dd^{2|1}\varsigma}{4\pi\alpha'} \left(\frac{\Delta \text{S}}{\Delta X^i}\, \delta X^i + \frac{\Delta \text{S}}{\Delta \Lambda^\alpha}\, \delta_A \Lambda^\alpha\right)
\, ,
\end{equation}
where we have reorganised the expressions above to define
\begin{align}
\frac{\Delta \text{S}}{\Delta X^i} &= \frac{\delta \text{S}_{\mathcal{M}}}{\delta X^i} + \tr\big(\Lambda F_{ij} D X^j \Lambda\big)
+  m \, \tr \big((\dd_A S)_i\, \Lambda\big)\, , \label{eq:delX}
\\[5pt]
\frac{\Delta \text{S}}{\Delta \Lambda^\alpha} &= 
2\, h_{\alpha\beta}\, D_A\Lambda^\beta 
+ m \, S_\alpha = \frac{\delta \text{S}}{\delta \Lambda^\alpha}
\, .\qquad\qquad\qquad\quad
\label{eq:delLamb}
\end{align}
Here 
\begin{equation}
\dd_A S = \dd S - SA = (\partial_i S - S A_i)\, \dd x^i~
\end{equation}
is the appropriate covariant exterior derivative for the  section $S$ of ${\cal V}^*$.

\vfill

\section{Chiral and superconformal symmetries} \label{sec:SemilocalSyms}

\subsection{Chiral symmetries}

We give here a brief account of continuous \textit{chiral symmetries} of the action \eqref{eq:S_M}--\eqref{eq:S_S}, of which the $\cal G$-structure symmetry described in section~\ref{sec:Extended G-structure symmetry} below is an example. Such symmetries sit somewhere between global and local symmetries in that they are parametrized, in their infinitesimal version, by a small function $\epsilon(\varsigma)$ depending on some, but not all, of the superspace coordinates. Equivalently $\epsilon(\varsigma)$ is a constrained parameter. It could be Grassmann even or odd. A symmetry transformation is generally given as
\begin{align}
\delta X^i &= \overline{\delta X^i}(\epsilon,X,\Lambda) \, , \label{eq:BarSymmetry1}\\
\delta \Lambda^\alpha &= \overline{\delta \Lambda^\alpha}(\epsilon,X,\Lambda)\, , \label{eq:BarSymmetry2}
\end{align}
where the right hand sides are specific expressions involving the infinitesimal parameter, the fundamental fields and, in general, their (super)derivatives. The statement of symmetry is that the induced variation of the action can be recasted in the form
\begin{equation} \label{eq:Noether}
\overline{\delta \text{S}} 
= \int \frac{d^{2|1}\varsigma}{4\pi\alpha'} \left(\frac{\delta \text{S}}{\delta X^i}\, \overline{\delta X^i} 
+ \frac{\delta \text{S}}{\delta \Lambda^\alpha}\, \overline{\delta \Lambda^\alpha}\right) = \int \frac{d^{2|1}\varsigma}{4\pi\alpha'} ~ \partial_\mu \epsilon(\varsigma) J^\mu\, ,
\end{equation}
where we must still define the right hand side. The first equality here is general for arbitrary variations, while the second is specific to the ``barred'' symmetry variations. The index $\mu$ covers all directions $(\sigma^+,\sigma^-,\theta)$ of superspace, but in the case of chiral symmetries, at least one of the superfields $J^+$, $J^-$, $J^\theta$ vanishes identically. Because of this, $\overline{\delta S}=0$ to leading order if and only if we impose the constraint $\partial_{\bar{\mu}} \epsilon(\varsigma)=0$ for all directions $\bar{\mu}$ with non-vanishing $J^{\bar{\mu}}$.

Integrating by parts, \eqref{eq:Noether} is Noether's theorem in $(1,0)$ superspace. When $\epsilon(\varsigma)$ is freed from its constraint, i.e.\ made fully local on superspace, then $\overline{\delta S}\neq 0$, but the second equality in \eqref{eq:Noether} still holds. The familiar local current conservation rule $\partial_{\bar{\mu}}J^{\bar{\mu}}\approx 0$  then follows from the fact that $\epsilon(\varsigma)$ is made to depend on \textit{all} integration variables (including $\theta$) and from the equations of motion. We use curly equal signs for equations holding on-shell. Note that the current $J^{\bar{\mu}}$ is not uniquely determined. In particular, adding to it any function $f$ satisfying $\partial_{\bar{\mu}} f=0$ does not change the conservation equation.

\subsection{Superconformal symmetry} \label{sec:sconfsymm}

Arguably the most important examples of chiral symmetries are conformal and supersymmetry transformations, combined here in a single $\mathcal{N}=1$ \textit{superconformal symmetry}. We focus on the supersymmetric side ($+$) of the massless $\sigma$-model \eqref{eq:S_M}--\eqref{eq:S_V}, although there is also a conformal symmetry on the non-supersymmetric $(-)$ side. Consider the transformation 
\begin{align}
\delta^\epsilon X^i &= i \epsilon \partial_+ X^i + \frac{1}{2} D \epsilon D X^i\, ,\label{eq:susy1}
\\
\delta^\epsilon_A \Lambda^\alpha &= i\epsilon \partial_{+A} \Lambda^\alpha 
+ \frac{1}{2} D \epsilon D_A \Lambda^\alpha\, ,\label{eq:susy2}
\end{align}
where $\epsilon(\varsigma)$ is an infinitesimal function of the worldsheet coordinates and
\begin{equation}
D_A \Lambda = D \Lambda +  \hat{A} \Lambda\, , \qquad
\partial_{+A} \Lambda = \partial_+ \Lambda + (A_i\,\partial_+ X^i)\Lambda\, .
\end{equation}
The massless action is invariant under these superconformal transformations 
whenever  $\epsilon=\epsilon(\sigma^+, \theta)$.

The chiral supercurrent associated to this symmetry is given by 
\begin{equation} \label{eq:T+}
{\cal T} =  G_{ij}\, \partial_+ X^i D X^j 
-i\widehat{\dd B}\, .
\end{equation}
Note that this is the same current as the one obtained when $\Lambda = 0$. The leading component $G$ generates chiral supersymmetry, while the other current $2T$ is the left-moving part of the energy-momentum tensor.

We can make contact for example with chiral supersymmetry by letting $\epsilon = 0+\theta 2\varepsilon (\sigma^+)$. The transformations \eqref{eq:susy1}--\eqref{eq:susy2} become
\begin{align}
\delta^\epsilon X^i &= \varepsilon (\sigma^+) Q X^i \, ,
\\
\delta^\epsilon \Lambda^\alpha &= \varepsilon (\sigma^+) Q \Lambda^\alpha\, .
\end{align}

\section{Extended $\mathcal{G}$-structure symmetry} \label{sec:Extended G-structure symmetry}

Suppose the target space manifold $\mathcal{M}$,  with ${\rm dim}\, {\cal M} = d$, admits a globally-defined nowhere vanishing $p$-form $\Phi$. The existence of such a form amounts to a reduction of the {structure group} $GL(d)$ of the frame bundle of $\cal M$ to a subgroup $\cal G$.  
Howe and Papadopoulos \cite{Howe:1991ic,Howe:1991vs,Howe:1991im} showed that if $\Phi$ satisfies a certain constraint, then the $\sigma$-model with action $\text{S}_{\cal M}$ (that is, in the case where $\Lambda = 0$) has an extra symmetry.
In this section we generalise this symmetry to include the bundle and the mass terms by $\text{S}_\mathcal{V}$ and $\text{S}_\mathcal{S}$. 

\subsection{Review of the Howe--Papadopoulos symmetry}\label{ssec:RevHP}

Let $\epsilon(\varsigma)$ be a general function over superspace with left-moving weight $h_+=(1-p)/2$. It has even/odd Grassmann parity depending on whether $p$ is odd/even.
Consider the transformation
\begin{align} \label{eq:Howe-Pap}
&\delta^{\Phi} X^i = \,
\frac{\epsilon(\varsigma)}{(p-1)!}\, 
\,\Phi^i{}_{i_2\ldots i_p}(X) D X^{i_2\ldots i_p}
= \epsilon(\varsigma)\, \hat\Phi^i\,,
\, 
\end{align}
where $DX^{i_2\ldots i_p}$ is a shorthand for $DX^{i_2}\ldots DX^{i_p}$.

The variation of the $\sigma$-model action $\text{S}_{\cal M}$ induced by \eqref{eq:Howe-Pap} follows from the analysis of section \ref{sec:Vars}. Only the first terms in \eqref{eq:generalCovariantVariationAction} and in \eqref{eq:delX} participate and after integrating by parts we find
\begin{equation}
\delta \text{S}_{\cal M} = \int \frac{\dd^2z}{4\pi\alpha'}\, \epsilon(\varsigma) \dd\theta
\, (-2i) \nabla^+_i\hat\Phi\,\partial_- X^i   
+ \  \int_\Sigma \frac{\dd^2z}{4\pi\alpha'}~ \partial_-\epsilon(\varsigma)\,  \dd\theta\,  \hat \Phi\, , 
\label{eq:HPdeltaSM}
\end{equation}
where we have defined
\begin{equation}
\hat \Phi =  \frac{1}{p}\, D X^i\,  \hat\Phi_i = \frac{1}{p!}\,  \Phi_{i_1\cdots i_p}\,  D X^{i_1\cdots i_p}\,,
\end{equation}
and
\begin{equation}
\nabla^+_i\hat\Phi = 
\frac{1}{p!}\, \nabla^+_i\Phi_{j_1\ldots j_p}\, DX^{j_1\ldots j_p} \,.
\end{equation}
It is easy to see that the first term of \eqref{eq:HPdeltaSM} gives the main result of \cite{Howe:1991ic}. If $\Phi$ is parallel under the connection $\nabla^+$ with torsion\footnote{The torsion $T$ of a covariant derivative $\nabla$ which has connection symbols $\Gamma^i{}_{jk}$ is defined as $T^i{}_{jk} = 2 \Gamma^i{}_{[jk]}$.} $T=\dd B$, i.e.
\begin{equation}
\nabla^{+}_i\Phi_{j_1j_2\ldots{j_p}}=0\qquad \text{(} T = \dd B \text{)}\,,
\end{equation}
then the first term vanishes identically. Moreover if $\epsilon=\epsilon(\sigma^+,\theta)$ is purely left-moving, the last term in equation \eqref{eq:HPdeltaSM} also vanishes and we conclude that \eqref{eq:Howe-Pap} is an infinitesimal chiral symmetry of ${\text S}_\mathcal{M}$. Note that this symmetry is non-linear for $p\geq 3$. The last term in equation \eqref{eq:HPdeltaSM} corresponds to the current \cite{Howe:1991ic}
\begin{equation} \label{eq:HPcurrent}
J^- = 2i(-1)^{p}\hat {\Phi}\, , 
\end{equation}
which is simply the operator naturally associated to the differential form $\Phi$. Its conservation equation is the statement that, up to the equation of motion for $X$, it is left-moving:
\begin{equation}
\partial_-\hat{\Phi} \approx 0\,.
\end{equation}
The overall normalization in \eqref{eq:HPcurrent} is consistent with the definition of section~\ref{sec:SemilocalSyms}, but is largely a matter of conventions.

Chiral currents such as \eqref{eq:HPcurrent} are fundamental to the quantum description of conformal field theories in terms of operator algebras. The $p$-form current here and the classical stress-tensor in \eqref{eq:T+} are classical limits of corresponding generators in appropriate operator algebras. An importance difference between them to bear in mind is that a choice of normal ordering must be made when quantum operators are build from classical composite fields. In interacting theories, such as our non-linear $\sigma$-models, there is, to our knowledge, no canonical resolution to this ordering ambiguity.

\subsection{The extended $\cal G$-structure symmetry} \label{ssec:GstructureSym}
We now proceed to generalise the Howe--Papadopoulos symmetry to the full model, including the gauge and mass sectors. As a
first step, we set $\delta_A\Lambda = 0$ and focus on the variation of $\text{S}$ induced only by the Howe--Papadopoulos transformation of $X$ given by \eqref{eq:Howe-Pap}. The ansatz $\delta^\Phi_A\Lambda = 0$ will be relaxed below.\footnote{Recall that the covariant variation of $\Lambda$ is given in 
equation \eqref{eq:CovdeltaL}. Nevertheless $\delta^\Phi_A\Lambda^\alpha=0$ means that $\Lambda$ does transform according to  $\delta^\Phi\Lambda^\alpha = -A_i{}^\alpha{}_\beta\Lambda^\beta \delta^\Phi X^i$.}
All the terms in \eqref{eq:delX} now participate and we find, for the full $\sigma$-model variation,
\begin{equation}
\begin{split}
\delta \text{S} &= \int \frac{\dd^2z}{4\pi\alpha'}\, \epsilon(\varsigma) \dd\theta
\left( 
(-2i) \nabla^+_i\hat\Phi\,\partial_- X^i  
+   \tr\Big(
\Lambda F_{ij}\, D X^j\hat \Phi^i \Lambda
+ (-1)^{p-1} m(\dd_A S)_i\,\hat \Phi^i \Lambda\Big)
\right)
\\[3pt]
&\qquad\qquad
+ \  \int_\Sigma \frac{\dd^2z}{4\pi\alpha'}~ \partial_-\epsilon(\varsigma)\,  \dd\theta\,  (-2i)\hat \Phi\,.
\end{split} \label{eq:HPdelSX}\raisetag{20pt}
\end{equation}

The vanishing of the first term in \eqref{eq:HPdelSX} gives back of course the results reviewed in section \ref{ssec:RevHP}, but there are now two extra terms. As it is manifest in \eqref{eq:HPdelSX}, we have a symmetry if the following geometric conditions are satisfied
\begin{align}
&\nabla^{+}_i\Phi_{j_1j_2\ldots{j_p}}=0\qquad \text{(} T = \dd B \text{)}\,,\label{eq:HPcondition} \\ 
&F_{i[j_1}\Phi^i{}_{j_2\ldots j_p]}=0\, , \label{eq:Fisetcondition}\\
& (\dd_A S)_i\, \Phi^i{}_{j_2\ldots j_{p}} = 0\, . \label{eq:Mcondition}
\end{align}

As we will see, this already constitutes an interesting extension of the Howe--Papadopoulos symmetry, but we can generalise it one step further. We keep \eqref{eq:Howe-Pap}, but we now also assign a covariant variation to the Fermi superfields:
\begin{align}
&\delta^{\Phi} X^i = \frac{\epsilon(\varsigma)}{(p-1)!}\,\Phi^i{}_{i_2\ldots i_p}(X) D X^{i_2\ldots i_p}
= \epsilon(\varsigma)\, \hat\Phi^i\,, \label{eq:NonZerodelX} \\
&\delta_A^\Phi\Lambda^\alpha  =  \epsilon(\varsigma)\hat\Upsilon^\alpha{}_\beta\, (2 D_A\Lambda^\beta + m\, S^\beta)  =  \epsilon(\varsigma)\hat\Upsilon^\alpha{}_\beta \frac{\Delta{\text S}}{\Delta \Lambda_\beta}\,.\label{eq:NonZerodelL}
\end{align}
Here, a priori, the superfield
\begin{align}
\hat\Upsilon^\alpha{}_\beta\,
=\frac{1}{(p-2)!}\, 
\,\Upsilon^\alpha{}_\beta{}_{i_1\ldots i_{p-2}}(X) D X^{i_1\ldots i_{p-2}} \label{eq:Ups}
\end{align}
corresponds to an arbitrary $\text{End}(\mathcal{V})$-valued differential $(p-2)$-form. It is easy to see from \eqref{eq:generalCovariantVariationAction} that the variation $\delta \text{S}$ corresponding to \eqref{eq:NonZerodelX}--\eqref{eq:NonZerodelL} is composed of \eqref{eq:HPdelSX} as well as the extra term
\begin{equation}
\int \frac{\dd^{2|1}\varsigma}{4\pi\alpha'} ~ \frac{\Delta \text{S}}{\Delta \Lambda^\alpha}\, \delta_A^\Phi \Lambda^\alpha = \int \frac{\dd^{2|1}\varsigma}{4\pi\alpha'} ~ \epsilon(\varsigma)\hat{\Upsilon}_{\alpha\beta}\frac{\Delta \text{S}}{\Delta \Lambda_\alpha}\frac{\Delta \text{S}}{\Delta \Lambda_\beta}
\,.\label{eq:HPdelSL}
\end{equation}
The vanishing of this term is achieved if the endomorphism-valued form satisfies $\Upsilon_{(\alpha\beta)}=0$, in other words, whenever $\Upsilon\in\Omega^{p-2}(\mathcal{M},\bigwedge^2\mathcal{V})$. Summarising,

\vspace{5pt}

\noindent\textit{
\eqref{eq:NonZerodelX}--\eqref{eq:NonZerodelL} is a symmetry of the full $\sigma$-model \eqref{eq:S_M}--\eqref{eq:S_S} if the geometric conditions \eqref{eq:HPcondition}--\eqref{eq:Mcondition} and $\Upsilon_{(\alpha\beta)}=0$ are satisfied.
}

\vspace{5pt}

This transformation was in fact considered in \cite{Howe:1988cj,Hull:1993ct} in the case ${\cal G} = U(d/2)$ and $p= 2$.  In these references, the Howe--Papadopoulos symmetry is constructed such that it is a new supersymmetry transformation, hence enhancing the superconformal symmetry from $(1,0)$ to $(2,0)$. In this case, the form $\Upsilon$ is a section of ${\rm End}({\cal V})$ and it corresponds to a complex structure on $\cal V$. We want to encompass this particular case in our general symmetry, hence the inclusion of $\hat{\Upsilon}_{\alpha\beta}$ in our considerations.

The constraints needed for extended $\mathcal{G}$-structure symmetries, \eqref{eq:Fisetcondition} and \eqref{eq:Mcondition},  can be written nicely in terms of \textit{insertion operators} defined on p.\pageref{eq:insert}.
The constraint \eqref{eq:Fisetcondition} can be written as
\begin{equation}
i_F(\Phi) = 0 \,,\label{eq:InstGeneral}
\end{equation}
where in this equation $F$ is interpreted as a 1-form with values in ${T{\cal M}\otimes\rm End}({\cal V})$,
\begin{equation}
F^i = G^{ij}\, F_{jk}\, \dd x^k\,.
\end{equation} 
We shall say that a connection $A$ which satisfies condition \eqref{eq:InstGeneral} is a {\it $\sigma$-model quasi-instanton}. As we illustrate in section \ref{ssec:Geometry}, in some {definite} examples this condition agrees on the nose with the notion of a gauge bundle instanton arising in heterotic supergravity.

Similarly, using insertion operators, the constraint \eqref{eq:Mcondition} can  be written as
\begin{equation}
i_{\dd_A S}(\Phi) = 0 \,.\label{eq:iMcondition}
\end{equation}

In summary, the conditions for our extended $\mathcal{G}$-structure symmetry to hold are written as
\begin{equation}
\nabla^+\, \Phi = 0\, , \quad i_F(\Phi)= 0\,,
\quad
i_{\dd_A S}(\Phi) = 0\, , \quad \Upsilon_{(\alpha\beta)} = 0\,.\label{eq:AllConditions}
\end{equation}
In section \ref{sec:Anomalies} we consider the potential anomalies of this symmetry and show that the one-loop effective action is invariant as long as we assign appropriate $\alpha'$-corrections to these conditions.

We return below to a description of these geometric constraints. Before doing so, it is worth noting the following remarkable fact:

\vspace{5pt}

\noindent\textit{The conserved current for the extended $\mathcal{G}$-structure symmetry is the same as for the corresponding Howe--Papadopoulos symmetry: the bundle sector and the mass terms do not affect the current.}

\vspace{5pt}

\noindent This follows from \eqref{eq:HPdelSX} and \eqref{eq:HPdelSL}. 
In the classical limit, this fact explains why the heterotic bundle sector, while present, is not manifest in the conformal field theoretic framework. This framework is mainly concerned with the existence of currents; not their explicit definition, which is where the bundle manifests itself.

To further illustrate this, we mention \cite{Melnikov:2017yvz}, where the authors identify the internal superconformal algebras preserving various amounts of supersymmetry in Minkowski spacetimes of low dimensions $10-d$ after compactifying critical heterotic string theory. Focusing on minimal spacetime supersymmetry, they find $\text{SV}^{Spin(7)}$ in the case $d=8$ and $\text{SV}^{G_2}$ for $d=7$. In the original work \cite{Shatashvili:1994zw}, these algebras were argued to describe type II string compactifications on $Spin(7)$ and $G_2$ \textit{holonomy} manifolds without NS--NS flux. These are very restricted types of geometries from a heterotic viewpoint, where $Spin(7)$- and $G_2$-\textit{structure} manifolds are instead necessary. Obviously no vector bundles arise either in type II. Nevertheless the Shatashvili--Vafa operator algebras apply in all these cases. One may thus similarly believe that Odake algebras are relevant to heterotic compactifications on $SU(n)$-structure manifolds more generic than the restricted class of Calabi--Yau $n$-folds, as is usually asserted. We have however not explored this possibility in detail.

Another heterotic application of $\text{SV}^{Spin(7)}$ and $\text{SV}^{G_2}$ will be described in chapter~\ref{chap:MarginalDeformations}. The algebras will be used to define a worldsheet BRST operator whose cohomology contains infinitesimal marginal deformations of the conformal field theory. We will see that the heterotic vector bundle and B-field can be included almost without cost in the framework, again as if the operator algebras ``knew'' about this general geometric data.

\subsection{Geometrical constraints on $(\cal M, \cal V)$}\label{ssec:Geometry}

We now turn to a geometrical explanation of the conditions \eqref{eq:AllConditions} in the context where the target space $\mathcal{M}$ is a real $d$-dimensional closed Riemannian manifold on which heterotic supergravity is compactified.

As discussed earlier,  the existence of a well-defined nowhere vanishing $p$-form $\Phi$ on $\mathcal{M}$ amounts to a reduction of the structure group of $\mathcal{M}$ to ${\cal G}\subset SO(d)$. Interesting examples are $U(d/2)$, $SU(d/2)$, $Sp(d/4)$, $Sp(1) \times Sp(d/4) / \mathbb{Z}_2$, $G_2$ and $Spin(7)$. In some cases (as for example $Spin(7)$, $G_2$ and $SU(d/2)$) the target manifold admits at least one nowhere vanishing spinor, which is the basic topological condition on the target manifold necessary to obtain spacetime supersymmetric effective field theories. 


Consider for example an eight-dimensional target space with structure group 
$\mathcal{G} = Spin(7)$, or a seven-dimensional manifold with ${\cal G} = G_2$.
In both of these cases, the form $\Phi$ is of degree $p=4$ and such target spaces admit one well-defined nowhere vanishing spinor.  Compactifying heterotic supergravity on a manifold with a $G_2$-structure gives rise to three-dimensional Yang--Mills  $N=1$ supergravity \cite{Gunaydin:1995ku,Gauntlett:2001ur,Friedrich:2001nh,Friedrich:2001yp,Gauntlett:2003cy,Ivanov:2003nd}, while compactification on a manifold with a $Spin(7)$-structure gives a two-dimensional $(1,0)$ supergravity theory \cite{Ivanov:2003nd,Ivanov:2001ma}. 

Other interesting examples are $\mathcal{G}=U(d/2)$ and $\mathcal{G}=SU(d/2)$.  The case where $\mathcal{G}=U(d/2)$ corresponds to even-dimensional almost Hermitian target spaces where $\Phi=\omega$ is the Hermitian 2-form. Heterotic string compactifications on almost Hermitian manifolds are not supersymmetric (the group $U(d/2)\subset SO(d)$ does not leave any invariant spinors) unless the structure group is reduced further to $G=SU(d/2)$.  In this case there is another nowhere vanishing form $\Phi= \Omega$ of degree $n=d/2$ and the corresponding target spaces are almost Hermitian with vanishing first Chern class. Compactifying heterotic supergravity on a manifold with such an $SU(d/2)$-structure is not yet sufficient to obtain a supersymmetric spacetime supergravity.  For example, it was shown in \cite{Hull:1986kz,Strominger:1986uh} that when $d=6$,  one needs to demand further that the almost complex structure is integrable to obtain spacetime Yang--Mills  $N=1$ supergravity.  Furthermore, as mentioned in section \ref{ssec:GstructureSym}, the Howe--Papadopoulos symmetry in \cite{Howe:1988cj,Hull:1993ct} corresponds precisely to an enhancement of the superconformal symmetry to $(2,0)$.  This is of course beautifully consistent with the work of \cite{Banks:1987cy} in which it is shown that the worldsheet quantum field theory corresponding to a four-dimensional supersymmetric spacetime theory obtained from superstring compactifications must have $\mathcal{N}=2$ superconformal invariance.

Manifolds with a $\mathcal{G}$-structure admit connections $\nabla$ which are metric and compatible with the  $\mathcal{G}$-structure, that is $\nabla \Phi = 0$. These connections have an {\it intrinsic} torsion $T(\Phi)$ which is uniquely determined by the $\mathcal{G}$-structure $\Phi$. Equation \eqref{eq:HPcondition} says that the form $\Phi$ needs to be covariantly constant with respect to a connection with totally antisymmetric torsion $T(\Phi) = \dd B.$
Note that this relation ties the target space geometry with the physical NS--NS flux.

Not all manifolds with a given $\mathcal{G}$-structure admit such a connection with a totally antisymmetric torsion except in the case of ${\cal G} = Spin(7)$ \cite{Ivanov:2001ma}.
For instance, when $\mathcal{G} = G_2$, taking $\Phi$ to be the co-associative 4-form, the necessary and sufficient condition for the existence of a $G_2$-compatible connection with totally antisymmetric torsion is that the 5-form $\dd \Phi$ is in the ${\bf 7}$ dimensional representation\footnote{A 5-form on a manifold with a $G_2$-structure decomposes into the $G_2$ irreducible representations $\bm{7}\oplus\bm{14}$.} of $G_2$. In fact, in this case there is a unique $G_2$-compatible connection with totally antisymmetric torsion \cite{Bryant:2005mz}. In even dimensions, with $\mathcal{G}=U(d/2)$, there exists a unique metric connection compatible with the $U(d/2)$ structure with totally antisymmetric torsion which is called the {\it Bismut} connection \cite{MR1006380, Friedrich:2001nh}.\footnote{Note that the complex structure does not need to be integrable for this statement to be true.}

We now turn to the $\sigma$-model ``quasi instanton'' connection $A$ on the bundle
\begin{equation}
i_F(\Phi) = 0 \,.\label{eq:InstGeneral2}
\end{equation}
For the examples pertaining to heterotic compactifications with ${\cal G} = Spin(7), G_2$, or $SU(n)$, we want to see to what extent this corresponds to the \textit{instanton} condition obtained by demanding a supersymmetric background in heterotic supergravity (gaugino Killing spinor equation). We first need minimal information about form decompositions. Differential forms on a manifold endowed with a $\mathcal{G}$-structure decompose into irreducible representations (\emph{irreps}) of $\mathcal{G}$. For the following groups of interest, the irreps are
\begin{align}
\mathcal{G}&=Spin(7) : \bm{1}, \bm{7}, \bm{8}, \bm{21}, \ldots\,, \\
\mathcal{G}&=G_2 : \qquad\, \bm{1}, \bm{7}, \bm{14}, \bm{27},\ldots\,, \\
\mathcal{G}&=SU(3) : ~~ \bm{1}, \bm{3}, \bar{\bm{3}}, \bm{8}, \ldots\,.
\end{align}
For a $p=2$ form (perhaps valued in some bundle) on a manifold with these $\mathcal{G}$-structures, we have the decompositions
\begin{align}
\mathcal{G}&=Spin(7) : \bm{28} \longrightarrow \bm{7} \oplus \bm{21}\,, \\
\mathcal{G}&=G_2 : \qquad~ \bm{21} \longrightarrow \bm{7} \oplus \bm{14}\,, \\
\mathcal{G}&=SU(3) : ~~ \bm{15} \longrightarrow \bm{1} \oplus \bm{3} \oplus \bar{\bm{3}} \oplus \bm{8}\,.
\end{align}

Suppose the target manifold admits a $Spin(7)$- or  a $G_2$-structure. It is a well known fact about the geometry of these manifolds that \eqref{eq:InstGeneral2} is equivalent to $F\in\Omega^2_{\bm{21}}({\cal M}, {\rm End}({\cal V}))$
in the case of $Spin(7)$ and to $F\in\Omega^2_{\bm{14}}({\cal M}, {\rm End}({\cal V}))$ for $G_2$ \cite{MR2164593}. In both cases, one can in fact write this condition using an appropriate projection operator on $F$ into the appropriate irreducible representation of $\cal G$.
\begin{equation}
(2\delta^{kl}_{ij} +  \Phi^{kl}{}_{ij}) F_{kl} = 0\, ,
\end{equation}
or equivalently,
\begin{equation}
F+F\lrcorner \Phi = 0\, . \label{eq:Inst78}
\end{equation}

The \emph{contraction operator} $\lrcorner$ here is defined in general for a $p$-form $\Pi$ and a $q$-form $\Psi$, $q\geq p$, by
\begin{equation}
\Pi \lrcorner \Psi = \frac{1}{p!(q-p)!} \Pi_{i_1 \ldots i_p}\Psi^{i_1 \ldots i_p}{}_{j_{m+1} \ldots j_q}  \dd x^{j_{m+1} \ldots j_q} \,.
\end{equation}

In the case of $U(n)$-structures (where the dimension of $\cal M$ is $d = 2n$),
the condition \eqref{eq:InstGeneral2} constrains the bundle $\cal V$ to be holomorphic, that is, we have
\begin{equation}
i_F(\omega) = 0\, \quad\iff\quad F^{(0,2)} = 0\,.\label{eq:InstU(n)}
\end{equation}
To see this, note that 
\begin{equation}
i_F(\omega) = F^i\wedge\omega_i = - F_{ki}J^k{}_j \dd x^{ij}\,, 
\qquad J^i{}_j = G^{ik}\omega_{jk}\, ,
\end{equation}
where $J$ is the almost complex structure. 
Therefore, 
\begin{equation}
i_F(\omega)= 0~\iff~F_{ij} =  J^k{}_i J^l{}_j\, F_{kl}\, ,
\end{equation}
and the result \eqref{eq:InstU(n)} follows. If moreover the structure group reduces to $SU(n)$, then there is a further constraint on $\cal V$ due to the existence of a second $n$-form $\Omega$, which is
\begin{equation}
 i_F(\Omega) = 0\,\quad\iff\quad \omega\lrcorner F = 0\, . \label{eq:InstSU(n)}
\end{equation}
that is,  $F$ must be a \textit{primitive} 2-form.
To see this equivalence, note first that, when $F^{(0,2)} = 0$, the three form  $i_F(\Omega)$ must be type $(n,0)$. Then 
\begin{equation}
i_F(\Omega)= 0~\iff~\overline\Omega\lrcorner\, i_F(\Omega) = 0
\, . 
\end{equation}
Noting that the $(n,0)$  form $\Omega$
satisfies
\begin{equation}
\overline\Omega^{ik_1\cdots k_{n-1}}\,\Omega_{jk_1\cdots k_{n-1}}
\propto \delta^i{}_j - i\, J^i{}_j\, , 
\end{equation}
we obtain 
\begin{equation}
i_F(\Omega)= 0~\iff~\overline\Omega\lrcorner\, i_F(\Omega) = 0
~\iff~ \omega\lrcorner F = 0\, . 
\end{equation}
Together, the conditions \eqref{eq:InstU(n)} and \eqref{eq:InstSU(n)}, are equivalent to $F\in \Omega_{\bm{8}}^2({\cal M}, {\rm End}({\cal V}))$ (for $n=3$).
One can also show that this is equivalent to 
\begin{equation}
F+F\lrcorner \, \rho = 0\,,\qquad \rho= \frac{1}{2}\, \omega\wedge\omega\, .
\end{equation}
Note the similarity with equation \eqref{eq:Inst78}.

In summary, for the examples $\mathcal{G}= Spin(7), G_2$ and $SU(3)$, we have that
\begin{equation}
F\in \Omega^2_{\bf adj} ( {\cal M}, {\rm End}({\cal V}))\, ,
\end{equation}
where $\bf adj$ is the adjoint representation of $\mathcal{G}$, or equivalently
\begin{equation}
F+F\lrcorner \Phi = 0\, ,
\end{equation}
where $\Phi$ is the Cayley 4-form for $\mathcal{G}= Spin(7)$ structure, the co-associative 4-form for $\mathcal{G}= G_2$ and, for $\mathcal{G}= SU(n)$, we have $\Phi = \rho$.  We say that the bundle connection $A$ satisfying this condition is an {\it instanton}. The curvature $F$ of such an instanton connection satisfies the Yang--Mills equation\footnote{The paper by Harland and N\"olle \cite{Harland:2011zs} contains a very good discussion about instantons as solutions of the Yang--Mills equation in various dimensions.} 
\begin{equation}
\dd_A^\dagger F = - F\lrcorner \dd^\dagger \Phi\,.\label{eq:GenInst}
\end{equation}

The final constraint \eqref{eq:iMcondition} can be written as
\begin{equation}
\dd_A S = 0\, ,\label{eq:flatS}
\end{equation}
for the examples at hand, because $\dd_A S$ is a 1-form. This means that the section $S$ must be a flat section. We will not be concerned with this condition any further.

We add here one last comment about the minimally supersymmetric  heterotic compactifications we have been discussing in this section. In order for these supersymmetric solutions to satisfy the supergravity equations of motion to first order in $\alpha'$, it is also necessary that there is a connection $\Theta$ on the tangent bundle $T{\cal M}$ which is an instanton \cite{Ivanov:2009rh}. In the $(1,0)$ $\sigma$-model, the connection $\Theta$ appears in order to cancel the gravitational anomalies. We will see in the following section that to first order in $\alpha'$,  the 
one-loop effective action is invariant under the $\cal G$-structure symmetries, provided that $\Theta$ is a $\sigma$-model quasi-instanton together with the usual corrections to the torsion involving the Chern--Simons forms for $A$ and $\Theta$.

\section{Anomalies} \label{sec:Anomalies}

In this part, we propose an analysis of anomalies of $\mathcal{G}$-structure and superconformal symmetries from the angle of effective actions.

In the absence of anomalies, a standard argument shows formally that the $\sigma$-model quantum effective action \cite{Howe:1986vm,Hull:1986xn,Hull:1986hn}, denoted $\Upgamma$, obeys the Slavnov--Taylor identity
\begin{equation} \label{eq:SlavnovTaylor}
\int \frac{d^{2|1}\varsigma}{4\pi\alpha'}\left(\frac{\delta \Upgamma[X,\Lambda]}{\delta X^i}\langle{ \overline{\delta X^i} }\rangle + \frac{\delta \Upgamma[X,\Lambda]}{\delta \Lambda^\alpha}\langle{ \overline{\delta \Lambda^\alpha} }\rangle\right)=0\,,
\end{equation}
where the expectation values are taken in the presence of background sources for the dynamical fields and where the bars refer to specific symmetry variations as explained near \eqref{eq:BarSymmetry1}--\eqref{eq:BarSymmetry2}. In the presence of chiral fermions, the functional measure in the path integral generically transforms anomalously, leading to a non-vanishing right hand side in \eqref{eq:SlavnovTaylor}. For linear symmetries, this can be probed by a first order variation of the effective action since $\langle{ \overline{\delta X^i} }\rangle = \overline{\delta X^i}$ and similarly for the Fermi superfield.

General $\mathcal{G}$-structure symmetries are however non-linear which complicates their analysis. Very few research efforts appear to have gone into this problem; in fact, we are only aware of \cite{Howe:2006si}. In this work, anomalies of $\mathcal{G}$-structure symmetries of $(1,1)$ models without torsion ($\dd B = 0$) were examined within a BV--BRST framework and multiple related difficulties were highlighted.

We will take a simplified approach which nevertheless yields results consistent with supergravity.
We assume an expansion in powers of $\alpha'$ of the form
\begin{equation}
\langle{ \overline{\delta X^i} }\rangle = \overline{\delta X^i}+\alpha'\overline{\delta X^i_{(1)}}+O(\alpha'{}^2) \,,
\end{equation}
and address anomalies perturbatively, i.e. we allow for the possibility of $\alpha'$-corrections to $\mathcal{G}$-structure symmetries.
We also use an effective action computed order by order using the background field method \cite{DeWitt:1967uc,Honerkamp:1971sh,AlvarezGaume:1981hn,Braaten:1985is,Henty:1987wc,Gates:1986ez}. This implies the usual limitations: target space curvature and fluxes must be small and slowly varying in string units.\footnote{Allowing non-trivial fluxes makes this assumption questionable and it is important to verify self-consistency.}

Our method is analogous to the treatment of $\sigma$-model anomalies \cite{Moore:1984dc,AlvarezGaume:1985yb,Bagger:1985pw} for target space gauge and Lorentz transformations \cite{Hull:1985jv,Sen:1985tq,Sen:1986nm}, especially as covered in \cite{Hull:1986xn}. We start by reviewing this discussion in order to prepare the ground for our treatment of $\mathcal{G}$-structure anomalies in section~\ref{sec:HPSymAnomaly}. The corresponding analysis of conformal anomalies is then presented for comparison in section~\ref{sec:Superconformal anomalies}.

\subsection{Effective action $\Upgamma$ and Green--Schwarz mechanism} \label{sec:GreenSchwarz}

In 1986, there were some questions as to whether the worldsheet implementation of the target space Green--Schwarz cancellation mechanism \cite{Green:1984sg,Hull:1985jv} was consistent with $(1,0)$ supersymmetry \cite{Sen:1985tq, Sen:1986nm}. Hull and Townsend addressed the issue in \cite{Hull:1986xn} by calculating a worldsheet one-loop effective action directly in superspace and used it to cancel the anomaly supersymmetrically. They found that the non-local
and gauge non-invariant contributions can be packaged as
\begin{equation} \label{eq:San}
\text{S}_{\text{an}}^{(A)}[X]=\int \frac{d^{2|1}\varsigma}{4\pi\alpha'}~\frac{\alpha'}{4}\text{tr}\left( \partial_- \hat{A} \,\frac{1}{\partial_+}D\hat{A}\right) ,
\end{equation}
where $\hat{A}=A_i(X)DX^i$ and where the trace is over the gauge indices of $A$. There is also a similar term due to gravitational anomalies,
\begin{equation} \label{eq:SanTheta}
\text{S}_{\text{an}}^{(\Theta)}[X]=\int \frac{d^{2|1}\varsigma}{4\pi\alpha'} ~(-1)\, \frac{\alpha'}{4}\text{tr}\left(\partial_- \hat{\Theta}\,  \frac{1}{\partial_+}D\hat{\Theta}\right),
\end{equation}
where $\Theta$ is the spin connection on $T\mathcal{M}$ associated to a covariant derivative $\nabla^-$ with connection symbols $\Gamma^-$  defined by\footnote{We remark that $\nabla^-$ is not compatible with the $\cal G$-structure ($\nabla^-\Phi \ne 0$).  It is however metric because the torsion is totally antisymmetric.} 
\begin{equation}
\Gamma^-{}^i{}_{jk} = \Gamma^i{}_{jk} - \frac{1}{2}\, (\dd B)^i{}_{jk}\, ,\label{eq:GammaMinus}
\end{equation}
and where we must now trace over 
spin connection indices.
The analysis of $\text{S}_{\text{an}}^{(\Theta)}$ is entirely parallel to that of $\text{S}_{\text{an}}^{(A)}$ so we will mostly omit it for simplicity of the discussion.
The formal inverse in \eqref{eq:San} is defined using the Green's function of $\partial_+$. We refer to \cite{Polchinski:1998rq} for more details. For our purposes it suffices to know that an analogue of integration by parts holds for such operators and that it can be commuted through ordinary differential operators.

We stress that $\text{S}_{\text{an}}^{(A)}$ correctly describes gauge non-invariance only up to quadratic order in the gauge connection. This is important for example when describing the Green--Schwarz mechanism, which we now review. Under a general variation of $\hat{A}$, $\text{S}_{\text{an}}^{(A)}$ transforms as
\begin{equation} \label{eq:deltaSan}
\delta \text{S}_{\text{an}}^{(A)}=  \int \frac{d^{2|1}\varsigma}{4\pi\alpha'} ~ (-1)\, \frac{\alpha'}{2} \text{tr} \left[\left(\partial_-\frac{1}{\partial_+}D\hat{A}\right)\delta \hat{A}\right].
\end{equation}
In order to check gauge anomalies, we substitute in~\eqref{eq:deltaSan} the target space gauge variation
\begin{equation}
\updelta \hat{A} = \big(\partial_i\Xi+[A_i,\Xi]\big)DX^i = D\Xi+[\hat{A},\Xi] \,,
\end{equation}
where we use $\updelta$ to distinguish gauge from generic variations and where $\Xi(X)$ is the gauge parameter. We obtain
\begin{equation}
\updelta \text{S}_{\text{an}}^{(A)}=- i \int \frac{d^{2|1}\varsigma}{4\pi\alpha'} ~ \frac{\alpha'}{2} \text{tr}\left( A_i(\partial_j\Xi) \,DX^i\partial_- X^j\right),
\end{equation}
which is finite and local. Clearly, this has the same form as the classical action \eqref{eq:S_M} and can thus be cancelled by assigning a compensating gauge variation to $M_{ij}$,
\begin{equation}
\updelta M_{ij}(X) = - \frac{\alpha'}{2}\text{tr}\left(A_i\,\partial_j\Xi \right)\,.
\end{equation}

This process cancels the anomaly but introduces a gauge-\textit{variant} metric since $\delta M = \delta G + \delta B$. One generally prefers to work with invariant objects. This can be achieved as follows. The effective action is only well-defined up to finite local counterterms. Consider then adding the metric counterterm
\begin{equation} \label{eq:Sadd}
\text{S}_{\text{add}}^{(A)}[X] =i\int \frac{d^{2|1}\varsigma}{4\pi\alpha'}~\frac{\alpha'}{4}\text{tr}(\hat{A}A_j\partial_-X^j)\,.
\end{equation}

Under a gauge transformation, this varies by
\begin{equation}
\updelta \text{S}_{\text{add}}^{(A)}=i\int \frac{d^{2|1}\varsigma}{4\pi\alpha'}~\frac{\alpha'}{2} \text{tr}\left(A_{(i}\partial_{j)}\Xi )\right) DX^i\partial_- X^j\,.
\end{equation}
This is enough to cancel the symmetric part of the anomaly,
\begin{equation}
\updelta (\text{S}_{\text{an}}^{(A)}+\text{S}_{\text{add}}^{(A)})= - i\int \frac{d^{2|1}\varsigma}{4\pi\alpha'}~\, \frac{\alpha'}{2}\text{tr}\left(\Xi\partial_{[i}A_{j]}\right)DX^i\partial_- X^j\,,
\end{equation} 
so that we may simply assign an anomalous gauge transformation to the B-field,
\begin{equation}
\updelta B_{ij}= - \frac{\alpha'}{2}\,\tr\left(\Xi\,\partial_{[i}A_{j]}\right)\,,
\end{equation}
and leave the metric gauge-invariant. Repeating this argument for \eqref{eq:SanTheta}, we obtain the usual Green--Schwarz mechanism and heterotic Bianchi identity. The associated gauge-invariant field strength involves Chern--Simons 3-forms for the gauge and tangent bundle connections:
\begin{equation} \label{eq:H}
H=\text{d}B+ \frac{\alpha'}{4}\left(\text{CS}_3(A)-\text{CS}_3(\Theta)\right).
\end{equation}
We briefly note that \cite{Hull:1985dx} points at an ambiguity in this  cancellation scheme, whereby any connection $\Theta$ can be used in the Bianchi identity. This ambiguity is lifted in the literature by various other conditions, such as conformal invariance or $(2,0)$ supersymmetry \cite{Howe:1987nw}. As described recently in \cite{Melnikov:2014ywa}, the correct choice is ultimately the Hull connection \eqref{eq:GammaMinus}.

\subsection{$\mathcal{G}$-structure symmetries and $\alpha'$-corrections} \label{sec:HPSymAnomaly}

We now investigate the effects of a $\mathcal{G}$-structure transformation on $\text{S}_{\text{an}}^{(A)}+\text{S}_{\text{add}}^{(A)}$. The gauge field varies here purely through the chain rule,
\begin{equation}
\delta^\Phi \hat{A} = \partial_iA_j(\delta^\Phi X^i)DX^j + A_iD(\delta^\Phi X^i)\,.
\end{equation}
Substituting this in \eqref{eq:deltaSan}, we find the non-local result
\begin{equation} \label{eq:deltaSanBasic}
\delta^\Phi \text{S}_{\text{an}}^{(A)}= \int \frac{d^{2|1}\varsigma}{4\pi\alpha'} ~ \frac{\alpha'}{2} \text{tr} \left[-2\left(\frac{\partial_-D}{\partial_+}\hat{A}\right)\partial_{[i} A_{j]}DX^j+i(\partial_- \hat{A}) A_i\right]\delta^{\Phi}X^i\,.
\end{equation} 
Next we vary the local counterterm. We obtain
\begin{align}
\delta^{\Phi} \text{S}_{\text{add}}^{(A)}= i \int \frac{d^{2|1}\varsigma}{4\pi\alpha'} & ~ \frac{\alpha'}{2} \tr \left[ -A_i A_j\partial_-DX^j \right.
\label{eq:deltaSaddBasic}\\
& ~~ \left. + \left(A_j\partial_{[i}A_{k]}+A_k\partial_{[i}A_{j]}-A_i\partial_{(j}A_{k)}\right)\partial_- X^kDX^j \right]\delta^{\Phi}X^i \, . \nonumber
\end{align}

In the sum of \eqref{eq:deltaSanBasic} and \eqref{eq:deltaSaddBasic}, the following combination appears:
\begin{equation} \label{eq:AdelAforHP}
2\left(A_j\partial_{[i}A_{k]}+A_k\partial_{[i}A_{j]}+A_i\partial_{[k}A_{j]}\right) = -(A\text{d}A)_{ijk}+4A_k\partial_{[i}A_{j]}\, .
\end{equation}
The first term on the right hand side is minus the Chern--Simons 3-form in the approximation where cubic powers of the gauge field are discarded. In the full variation $\delta^{\Phi}(\text{S}+\text{S}_{\text{an}}^{(A)}+\text{S}_{\text{add}}^{(A)})$, it naturally couples to $\Gamma^+_{ijk}$ in \eqref{eq:generalCovariantVariationAction}--\eqref{eq:delX} and \eqref{eq:EoMX} and redefines its torsion to be the gauge-invariant combination:
\begin{equation} \label{eq:redefinitionCS}
\begin{split}
\delta^{\Phi}(\text{S}&+\text{S}_{\text{an}}^{(A)}+\text{S}_{\text{add}}^{(A)}) 
\\[5pt]
&= \int \frac{d^{2|1}\varsigma}{4\pi\alpha'} ~ 2i \left(\Gamma+\frac{1}{2}\dd B+
\frac{\alpha'}{8}\text{CS}_3(A)\right)_{ijk}\partial_- X^jDX^k\delta^\Phi X^i + \ldots \,.
\end{split}
\end{equation}
The complete field $H$ as in \eqref{eq:H} is generated when repeating this derivation starting with the term $\text{S}_{\text{an}}^{(\Theta)}$ also included in the effective action.

Finally we notice that the remaining term in \eqref{eq:AdelAforHP} shares with the non-local term in \eqref{eq:deltaSanBasic} a crucial factor. Ignoring the Chern--Simons term,
\begin{align} \label{eq:vanishingAnomaly}
\delta^\Phi (\text{S}_{\text{an}}^{(A)}+\text{S}_{\text{add}}^{(A)})\quad& \nonumber \\
\quad= \int \frac{d^{2|1}\varsigma}{4\pi\alpha'} \frac{\alpha'}{2} \tr &\left( -\frac{\partial_-D}{\partial_+}\hat{A} + i\,A_k\partial_- X^k\right)2\partial_{[i}A_{j]}\delta^\Phi X^iDX^j+\ldots \,.
\end{align}
Keeping in mind that only quadratic terms in the gauge field are accounted for, we identify $2\partial_{[i}A_{j]}=F_{ij}$ and realise the following remarkable fact:
\vskip3pt
\textit{The term \eqref{eq:vanishingAnomaly}, which is non-local, is exactly cancelled if we assume the \textit{same} geometric condition \eqref{eq:Fisetcondition} that was necessary for our generalisation \eqref{eq:NonZerodelX}--\eqref{eq:NonZerodelL} of the classical Howe--Papadopoulos symmetry.}
\vskip3pt
\noindent We see this by recalling $\delta^{\Phi}X^i=\epsilon\hat{\Phi}^i$ and because \eqref{eq:vanishingAnomaly} has the factor
\begin{equation}
i_{F}(\Phi)=0\,.
\end{equation}
We knew about this constraint on $\mathcal{V}$ from the classical symmetry. Repeating the analysis with the Lorentz anomalous term $\text{S}_{\text{an}}^{(\Theta)}$ given by \eqref{eq:SanTheta}, we now obtain the same constraint on the tangent bundle $T\mathcal{M}$ as a quantum condition,
\begin{equation} \label{eq:ThetaInstanton}
i_{R^{\Theta}}(\Phi)=0\,,
\end{equation}
that is, {\it $\Theta$ must be a $\sigma$-model quasi-instanton}. Geometrically, the appearance of this condition is reasonable and in fact gives credence to our $\sigma$-model approach.  For the examples discussed in section \ref{ssec:Geometry}, where ${\cal G} = Spin(7), G_2, SU(n)$ (including both forms $(\omega,\Omega)$ in the $SU(n)$ case), the connection $\Theta$ becomes in fact a gauge-bundle instanton.  This 
extra condition is necessary for  a supersymmetric solution of a heterotic string compactification on $({\cal M}, {\cal V})$ to satisfy the supergravity equations of motion to first order in $\alpha'$ (see for example \cite{Ivanov:2009rh}). 

We conclude from this analysis that the $\mathcal{G}$-structure symmetry \eqref{eq:NonZerodelX}--\eqref{eq:NonZerodelL} is strictly-speaking anomalous. However it can be corrected at first order in $\alpha'$ provided we impose the new target space constraint \eqref{eq:ThetaInstanton} and provided we change the torsion from $\text{d}B$ to $H$ in the classical condition,
\begin{equation}
\nabla^{+}_i\Phi_{i_1\ldots i_p}=0\,\qquad \text{(} T=H \text{)}\,.
\end{equation}
This is consistent with the redefinition induced by \eqref{eq:redefinitionCS}.

\subsubsection{Current}

Revisiting the analysis of section~\ref{sec:HPSymAnomaly}, we remark that we did not use that the infinitesimal parameter $\epsilon$ of the transformation is purely left-moving ($\partial_-\epsilon = 0$). This has implications for the $\alpha'$-corrected current associated to the symmetry (see appendix~\ref{sec:SemilocalSyms}). As explained above, the variation of $\text{S}_{\text{an}}^{(A)}+\text{S}_{\text{add}}^{(A)}$ contains a term proportional to $i_F(\Phi)$ and a term which redefines the classical torsion found in $\delta^\Phi \text{S}$, eq.~\eqref{eq:generalCovariantVariationAction}--\eqref{eq:delX}. Assuming $i_F(\Phi)=0$ this means that the full $\mathcal{G}$-structure variation of $\text{S}+\text{S}^{(A)}_{\text{an}}+\text{S}^{(A)}_{\text{add}}$ (with general $\epsilon$) has exactly the same form \eqref{eq:HPdelSX} as the variation of $\text{S}$. The only difference is the redefined torsion $\dd B\rightarrow H$. We conclude that, after using all the appropriate geometric conditions,
\begin{equation}
\delta^\Phi (\text{S}+\text{S}_{\text{an}}+\text{S}_{\text{add}}) = \int \frac{\dd^2z}{4\pi\alpha'}~ \partial_-\epsilon(\varsigma)\,  \dd\theta\,  (-2i)\hat \Phi\,.
\end{equation}
Therefore, remarkably, {\it the tree-level current proportional to $\hat{\Phi}$ persists at one-loop}.
Furthermore all results thus far are true regardless of whether $\delta^\Phi_A\Lambda$ vanishes or otherwise.
The current is now conserved up to the non-local equation of motion derived from $\text{S}+\text{S}^{(A)}_{\text{an}}+\text{S}^{(A)}_{\text{add}}$, which is easy to write down from our formulas. This equation of motion should be interpreted as the one-loop approximation to the operator equation
\begin{equation}
\frac{\delta \Upgamma}{\delta X^i}=0\,,
\end{equation}
where $\Upgamma$ is the exact quantum effective action. Note that the corresponding equation obtained by varying with respect to $\Lambda$ is not necessary in the conservation statement. Its role is solely to impose constraints; with no contribution to the current.

\subsubsection{Counterterms}

Effective actions are only well-defined up to finite local counterterms. These arise in particular when different schemes are used to regulate ultraviolet divergences.  In our discussion so far, we have made implicit choices when writing the effective action. We now briefly reconsider our discussion of $\mathcal{G}$-structure symmetries at order $\alpha'$ in light of these ambiguities.

The original action \eqref{eq:S_M}--\eqref{eq:S_S} is the most general covariant renormalizable $(1,0)$ supersymmetric functional. Hence, counterterms must have the same form in order not to spoil these properties \cite{Hull:1986kz,Sen:1986nm}. All the couplings in the $\sigma$-model (metric, $B$-field, gauge field and $S$) have corresponding counterterms
$\Delta G_{ij}$, $ \Delta B_{ij}$, $\Delta A_i{}^\alpha{}_\beta$ and $\Delta S_\alpha$. We define $\widetilde{G}_{ij}=G_{ij}+\Delta G_{ij}$ and similarly for the others and add tildes to identify quantities constructed from such redefined tensors. In this section, we also write explicitly the dependence of action functionals on target space tensors: for example, the allowed counterterms are collectively written as $\text{S}_{\text{c.t.}}=\text{S}(\Delta G,\Delta B,\Delta A,\Delta S)$. The one-loop effective action, with counterterms, is then taken as
\begin{equation} \label{eq:CTEffAction}
\text{S}(\widetilde{G},\widetilde{B},\widetilde{A},\widetilde{S})+\text{S}^{(A)}_{\text{an}}(A)+\text{S}^{(A)}_{\text{add}}(A)\, ,
\end{equation}
where we still ignore $\text{S}_{\text{an}}^{(\Theta)}$ to simplify the discussion.

It is important to distinguish carefully the gauge fields in \eqref{eq:CTEffAction} from the gauge field in the symmetry variation. We must use the same symmetry as before, namely \eqref{eq:NonZerodelX}--\eqref{eq:NonZerodelL},
\begin{align}
&\delta^\Phi X^i = \epsilon\hat{\Phi}^i\,,
&\delta^\Phi\Lambda^\alpha + A_i{}^\alpha{}_\beta\Lambda^\beta\delta^\Phi X^i = \epsilon\hat{\Upsilon}^\alpha{}_\beta\,\frac{\Delta \text{S}(A,S)}{\Delta \Lambda_\beta}\,, 
\end{align}
but computations are simpler if we write this variation of $\Lambda$ as
\begin{equation}
\delta^\Phi_{\widetilde{A}}\Lambda^\alpha = (\Delta A)_i{}^\alpha{}_\beta\Lambda^\beta\delta^\Phi X^i+\epsilon\hat{\Upsilon}^\alpha{}_\beta\,\frac{\Delta \text{S}(A,S)}{\Delta \Lambda_\beta}\,. \label{eq:FunnyDeltaLambda}
\end{equation}

Focusing for the moment on the symmetry variation of $\text{S}(\widetilde{G},\widetilde{B},\widetilde{A},\widetilde{S})$ with respect to $X$, we find equation~\eqref{eq:HPdelSX} again, this time written in terms of tilde tensors
\begin{align}
\delta \text{S} &= \int \frac{\dd^2z}{4\pi\alpha'}\, \epsilon(\varsigma) \dd\theta
\bigg[
-\frac{2i}{p} \partial_- X^i DX^{j_1} \widetilde{\nabla}_i^+(\widetilde{G}_{jj_1}\hat{\Phi}^j)
\nonumber \\[3pt]
&\qquad\qquad \qquad \qquad \qquad +   \tr\Big(
\Lambda \widetilde{F}_{ij}\, D X^j\hat \Phi^i \Lambda
+ (-1)^{p-1} m\, (\dd_{\widetilde{A}} \widetilde{S})_i\,\hat \Phi^i \Lambda\Big)
\bigg]
\nonumber \\[3pt]
&\quad
+ \  \int_\Sigma \frac{\dd^2z}{4\pi\alpha'}~ \partial_-\epsilon(\varsigma)\,  \dd\theta\,  (-2i)\frac{1}{p!}\widetilde{G}_{jj_1}\Phi^j{}_{j_2\ldots j_p}DX^{j_1\ldots j_p}\,. \label{eq:CTeasy}
\end{align}

Meanwhile $\text{S}^{(A)}_{\text{an}}+\text{S}^{(A)}_{\text{add}}$ is independent of $\Lambda$ and its variation with respect to $X$ is exactly as in section~\ref{sec:HPSymAnomaly}. It produces the non-local term \eqref{eq:vanishingAnomaly} and a term which redefines the torsion in \eqref{eq:CTeasy} to be
\begin{equation}
T = \text{d}\widetilde{B}+ \frac{\alpha'}{4}\text{CS}_3(A)\,,
\end{equation}
as in \eqref{eq:redefinitionCS}.

Finally, we account for the variation of $\text{S}(\widetilde{G},\widetilde{B},\widetilde{A},\widetilde{S})$ due to \eqref{eq:FunnyDeltaLambda}. We find
\begin{align}
\delta\text{S} &= \int \frac{\dd^{2|1}\varsigma}{4\pi\alpha'} \bigg[\frac{\Delta \text{S}(A,S)}{\Delta\Lambda_\alpha}\bigg((\Delta A)_{i\alpha\beta}\Lambda^\beta\delta^\Phi X^i - \big(2\,\widehat{\Delta A}^\beta{}_\gamma\Lambda^\gamma +  m\, \Delta S^\beta\big)\epsilon\hat{\Upsilon}_{\alpha\beta} \bigg) \nonumber \\[2pt]
&\qquad\qquad\qquad\qquad\qquad\qquad\qquad + m\, \Delta S^\alpha(\Delta A)_{i\alpha\beta}\Lambda^\beta\delta^\Phi X^i \bigg]\,.\label{eq:CTcomplicated}
\end{align}

To obtain this we used $\Upsilon_{(\alpha\beta)}=0$ and
\begin{equation}
\frac{\Delta \text{S}(\widetilde{A},\widetilde{S})}{\Delta\Lambda_\alpha} = 2\, D_{\widetilde{A}}\Lambda^\alpha 
+  m\, \widetilde{S}^\alpha = \frac{\Delta \text{S}(A,S)}{\Delta\Lambda_\alpha} + 2\,\widehat{\Delta A}^\alpha{}_\beta\Lambda^\beta + m\, \Delta S^\alpha \,.
\end{equation}

The next step is to group the terms in the full variation (the sum of \eqref{eq:CTeasy}, \eqref{eq:vanishingAnomaly} and \eqref{eq:CTcomplicated}) sharing the same powers of the fundamental superfields and their derivatives. From their prefactors, it is straightforward to read off constraints on counterterms and target space tensors ensuring preservation of the symmetry. We leave the general case to the reader and focus here on the most commonly encountered counterterms $\Delta G$ and $\Delta B$.

If we set $\Delta A = 0$ and $\Delta S = 0$, then \eqref{eq:CTcomplicated} does not interfere with \eqref{eq:CTeasy} and we can read off, much like before, the condition
\begin{equation}
i_F(\Phi)=0\,,
\end{equation}
from the non-local term and, from \eqref{eq:CTeasy},
\begin{align}
&\widetilde{\nabla}_i^+(\widetilde{G}_{j[j_1}\Phi^j{}_{j_2\ldots j_p]})=0 \qquad 
\left( T = \text{d}\widetilde{B}+ \frac{\alpha'}{4}\text{CS}_3(A) \right) \,, \label{eq:CTcondition}\\ 
&i_{\dd_A S}(\Phi)=0\,,
\end{align}
and the current
\begin{equation}
2i(-1)^p\frac{1}{p!}\widetilde{G}_{jj_1}\Phi^j{}_{j_2\ldots j_p}DX^{j_1\ldots j_p} = 2i(-1)^p\left( \hat{\Phi} + \frac{1}{p}\Delta G_{ij}DX^{i}\hat{\Phi}^j \right) \,.
\end{equation}

\subsection{Superconformal anomalies} \label{sec:Superconformal anomalies}

Section~\ref{sec:SemilocalSyms} gives a short account of $(1,0)$ superconformal symmetry in our non-linear $\sigma$-model (for vanishing mass). It is tantalising to try on this symmetry the anomaly analysis presented in the last section, using the effective action. A good motivation to treat all symmetries on the same footing is in anticipation to study the algebra they form. This is particularly interesting at the quantum level. In the case of superconformal symmetries, we have a prejudice on the outcome based on the substantial literature on conformal anomalies in two dimensional $\sigma$-models (see e.g.\ \cite{Hull:1987yi, Hull:1985rc, Sen:1985qt, Lambert:1995hs, Hull:1986hn, Hull:1986kz, Hull:1985zy, Hull:1987pc, Hull:1987yi, Callan:1989nz, AlvarezGaume:1983ab}). Nevertheless the method we use, based on the effective action \eqref{eq:SlavnovTaylor}, is non-standard in this context. As a complement to our discussion of $\mathcal{G}$-structure anomalies, it is worthwhile to connect our angle of analysis with classical string theory lore.

Our main result is simply that the superconformal variation of the one-loop effective action \eqref{eq:SanTheta} vanishes,
\begin{equation} \label{eq:deltaSan=0(conf)}
\delta^\epsilon \text{S}_{\text{an}}^{(A)} = 0\,.
\end{equation}
This fact will be proven shortly. It follows after some algebraic manipulations only, without using any equations of motion and without imposing any constraints on the $\sigma$-model couplings.

Naively the conclusion is that superconformal symmetries are not anomalous at one-loop, which is consistent with the expectation that a nearby superconformal fixed point exists in the universality class of $\text{S}$. However, this should only be true for certain configurations of the $\sigma$-model couplings: those which satisfy effective target space equations of motion \cite{Callan:1985ia,Hull:1986kz}. There appears to be a contradiction.

To reconcile \eqref{eq:deltaSan=0(conf)} with the literature, it is useful to reconsider the calculation of the effective action itself. Along the way, ultraviolet divergences are generated and are renormalized away in redefined couplings \cite{Friedan:1980jf,Howe:1986vm}. This generates beta functionals for the metric, B-field and gauge field, which must be trivial (not necessarily zero) to guarantee scale invariance. It is at this step that the familiar constraints on the couplings arise. Only for those configurations satisfying the target space equations of motion is the model scale invariant.

After renormalization, there remains in the effective action ultraviolet-finite terms only, which are all expressed in terms of renormalized quantities. The term $\text{S}_{\text{an}}^{(A)}$ that we have been using and the whole discussion of this section, were in terms of renormalized objects. At this level, the fact that we find $\delta \text{S}_{\text{an}}^{(A)}=0$ and thus no further restrictions by imposing conformal symmetry, can essentially\footnote{Strictly speaking it is best to revisit \cite{Mavromatos:1988jj} Zamolodchikov's theorem when working with non-linear $\sigma$-models. Assumptions sometimes fail, such as discreteness of the spectrum for noncompact target manifolds and unitarity for Lorentzian signature \cite{Polchinski:1987dy}.} be understood from the argument that scale invariant theories in two dimensions are automatically conformal \cite{Zamolodchikov:1986gt}.

We now prove \eqref{eq:deltaSan=0(conf)}. It is useful to break the proof into two steps. First we show that the superconformal variation is local. Then we show that it vanishes. The derivation starts as in section~\ref{sec:HPSymAnomaly} and we can reuse \eqref{eq:deltaSanBasic}
now for a superconformal transformation \eqref{eq:susy1}.
Focusing on the non-local part of the variation, we notice that
\begin{equation}
2\partial_{[i}A_{j]}DX^j\delta^\epsilon X^i = D(\epsilon DA_iDX^i)\,.
\end{equation}
Integrating by parts with $D$, we find the local variation
\begin{equation} \label{eq:deltaSanConfIntermediate}
\delta^\epsilon \text{S}_{\text{an}}^{(A)}=i \int \frac{d^{2|1}\varsigma}{4\pi\alpha'} ~ \frac{\alpha'}{2}\,\text{tr}\left[\partial_-\hat{A}(\epsilon DA_iDX^i+A_i\delta^\epsilon X^i)\right].
\end{equation}

We now show that this vanishes. It is useful to define the operator
\begin{equation}
D_\epsilon = \epsilon D+\frac{1}{2}D\epsilon \,,
\end{equation}
so that $\delta^\epsilon X^i = D_\epsilon DX^i$ and identify in~\eqref{eq:deltaSanConfIntermediate}
\begin{equation}
\epsilon DA_iDX^i+A_i\delta^\epsilon X^i = D_\epsilon \hat{A}\,.
\end{equation}
Then, integrating by parts with $\partial_-$,
\begin{equation} \label{eq:deltaSanConf-}
\delta^\epsilon \text{S}_{\text{an}}^{(A)}=-i\int \frac{d^{2|1}\varsigma}{4\pi\alpha'} ~ \frac{\alpha'}{2}\,\text{tr}(\hat{A}\partial_-D_\epsilon\hat{A})\,.
\end{equation}
Alternatively we can integrate by parts with $D_\epsilon$. Indeed it is easy to prove that, given two superfields $U$ and $V$
\begin{equation}
(D_\epsilon U) V + (-1)^F\, UD_\epsilon V = D(\epsilon\, UV)\, ,
\end{equation}
where $F= +1$ if $U$ is a commuting superfield and $F=-1$ if it is anticommuting.
From \eqref{eq:deltaSanConfIntermediate} this yields
\begin{equation} \label{eq:deltaSanConf+}
\delta^\epsilon \text{S}_{\text{an}}^{(A)}=i \int \frac{d^{2|1}\varsigma}{4\pi\alpha'} ~ \frac{\alpha'}{2}\,\text{tr}\big((D_\epsilon\partial_-\hat{A})\hat{A}\big)
=i\int \frac{d^{2|1}\varsigma}{4\pi\alpha'} ~ \frac{\alpha'}{2}\,\text{tr}(\hat{A}D_\epsilon\partial_-\hat{A})\,,
\end{equation}
where we have used cyclicity of the trace in the last step. We complete the proof of \eqref{eq:deltaSan=0(conf)} by comparing \eqref{eq:deltaSanConf+} and \eqref{eq:deltaSanConf-} and by using $[\partial_-,D_\epsilon]=0$, which follows from $\partial_-\epsilon = 0$.

For a general symmetry parameter, we have instead
\begin{equation} \label{eq:deltaSanConfFinal}
\delta^\epsilon \text{S}_{\text{an}}^{(A)}=i\int \frac{d^{2|1}\varsigma}{4\pi\alpha'} ~ \frac{\alpha'}{4}\,\text{tr}(\hat{A}[D_\epsilon,\partial_-]\hat{A})
=-i\int \frac{d^{2|1}\varsigma}{4\pi\alpha'} ~ \frac{\alpha'}{4}\partial_-\epsilon \, \text{tr}(\hat{A}D\hat{A})\,.
\end{equation}

\vfill

\subsubsection{Gauge-invariant supercurrent at order $\alpha'$}

As an application of the proof above, we now derive the $\alpha'$-correction to the left-moving stress-tensor and supersymmetry currents of generic massless $(1,0)$ $\sigma$-models \eqref{eq:S_M}--\eqref{eq:S_V}. To the best of the author's knowledge, this calculation is new. Classically the Noether procedure yields the superfield \eqref{eq:T+}
\begin{equation} \label{eq:T+text}
{\cal T} =  G_{ij}\, \partial_+ X^i D X^j 
- i\, \widehat{\dd B}\, .
\end{equation}
This supercurrent is right-moving on shell, $\partial_-\mathcal{T}\approx 0$ and is composed of the supersymmetry current $G$ and stress-tensor $T$. The second term in \eqref{eq:T+text} is often discarded in the literature. At order $\alpha'$, $\dd B$ is not gauge-invariant, as reviewed in section~\ref{sec:GreenSchwarz}. It is natural to ask if our considerations from this section can fix this issue.

It turns out they do. To see this, we extract from \eqref{eq:deltaSanConfFinal} the contribution of $\text{S}_{\text{an}}^{(A)}$ to $\mathcal{T}$,
\begin{equation}
- i\,\frac{\alpha'}{4}\,\tr(\hat{A}D\hat{A})\,.
\end{equation}
Substituting $D\hat{A}=\hat{F}+ iA_i\partial_+X^i$
this is composed of two terms. The first one immediately yields the Chern--Simons correction necessary to make $\mathcal{T}$ gauge-invariant:
\begin{equation}
\tr(\hat{A}\hat{F})=\widehat{\text{CS}_3(A)}\,,
\end{equation}
up to corrections cubic in $\hat{A}$. The second term can be absorbed by the variation of $\text{S}_{\text{add}}^{(A)}$. A particularly easy way to see this is to remember that $\text{S}_{\text{add}}^{(A)}$ is a metric counterterm, so we can read off its contribution to the current directly from \eqref{eq:T+text}. With $\Delta G_{ij}= - \frac{\alpha'}{4}\,\tr(A_i A_j)$, this is
\begin{equation}
\Delta G_{ij}\partial_+ X^i DX^j = - \frac{\alpha'}{4}\,\tr(A_i A_j)\partial_+ X^i DX^j\,.
\end{equation}

More generally, the impact of changing counterterms is easy to analyse for superconformal transformations. Assuming counterterms of the form of the classical action, with $G$ replaced by $\Delta G$ and similarly for the other couplings, superconformal invariance cannot be spoiled. Indeed no assumption on the couplings are made to prove classical superconformal invariance. As for the current, the modifications are as discussed in the case of $\text{S}_{\text{add}}^{(A)}$. The most general form of the $\alpha'$-corrected supercurrent, including counterterms, is
\begin{equation}
{\cal T} =  (G_{ij}+\Delta G_{ij})\, \partial_+ X^i D X^j 
-i (\widehat{H} + \widehat{\dd (\Delta B)})\, .
\end{equation}

\subsection{A caveat: gauge-invariant contributions to $\Upgamma$} \label{sec:Caveat}

It should be stressed that our analysis of $\alpha'$-corrections in this section has turned out to be much simpler than it should perhaps have been. There is an important caveat to our analysis, which we now point out even if it seems to be largely unimportant given the sensible results obtained so far in section~\ref{sec:Anomalies}.

As they were primarily interested in the Green--Schwarz mechanism, the authors of \cite{Hull:1986xn} focused only on Yang--Mills and Lorentz non-covariance in the $\sigma$-model one-loop effective action, leading to what we called $\text{S}_{\text{an}}$. Analyses of gauge anomalies at higher loops have been performed \cite{Hamada:1987ph,Grignani:1987gk,Foakes:1988wy,Ellwanger:1988cc,Lambert:1995hs}. However, we have not been able to locate in the existing literature a more complete calculation of the effective action which would include all covariant terms.\footnote{One particular covariant but infrared divergent term was reported in \cite{Hull:1986xn}. We have not included it in our present analysis given that further terms on the same footing are expected to exist and should be analysed together.} Such terms are crucial to our analysis because they may lead to anomalies of $\mathcal{G}$-structure (and superconformal) symmetries even if they do not produce gauge and gravity anomalies.

The fact that our results at order $\alpha'$ so far nicely align with supergravity expectations suggests that this problem in fact does not arise. More precisely, we conjecture that gauge and Lorentz invariant contributions to the effective action are automatically invariant under $\mathcal{G}$-structure symmetries --- up to the usual target space conditions \eqref{eq:HPcondition}--\eqref{eq:Mcondition}.
We hope to report on this conjecture more fully in a future communication.

\section{Conclusion}

Our main result in this chapter is the generalisation \eqref{eq:NonZerodelX}--\eqref{eq:NonZerodelL} of the symmetry of \cite{Howe:1991im} holding for general $(1,0)$ non-linear $\sigma$-models with non-Abelian background gauge fields turned on and also possibly a mass term. This symmetry is defined with a target space $p$-form $\Phi$ as well as a tensor $\Upsilon\in \Omega^{p-2}(\mathcal{M},\bigwedge^2\mathcal{V})$, which may or may not be chosen to vanish identically. The constraints \eqref{eq:AllConditions} on these tensors and the couplings of the $\sigma$-model are strongly reminescent of the supersymmetry conditions appearing in the context of heterotic compactifications. In fact, for the cases of $Spin(7)$ and $G_2$ compactifications that we discussed more closely, these conditions are equivalent. Contrastingly in the $SU(3)$ case, our symmetry does not require an integrable complex structure, but this can be enforced by demanding that it generates with $(1,0)$ superconformal symmetry the $(2,0)$ algebra.

We have demonstrated moreover how a modified version of our $\mathcal{G}$-structure symmetry persists quantum-mechanically. There remains caveats to this statement: crucially, a complete calculation of the $\sigma$-model one-loop effective action at first order in $\alpha'$ is necessary for definitive conclusions. Nevertheless our analysis based only on the non-local term $S_{\text{an}}$ has already produced quantum conditions impressively close to the supergravity expectations, such as the quasi-instanton condition $i_{R^{\Theta}}(\Phi)=0$ on $T\mathcal{M}$.

The conserved current for all the $\mathcal{G}$-structure symmetries that we considered is the operator $\hat{\Phi}$ naturally associated to the differential $p$-form. This remains true when including $\alpha'$-corrections but can be affected by metric counterterms.

Superconformal transformations were also discussed from the angle of the quantum effective action and compared with string theory. Our results at order $\alpha'$ are all consistent with Green--Schwarz gauge-invariance and the heterotic Bianchi identity.

It is likely that there will be some connections of our results with the considerations of \cite{Ekstrand:2009zd,Ekstrand:2010wu}, where the Chiral de Rham complex \cite{Malikov:1998dw} was likened to a formal quantization of the $(1,1)$ nonlinear $\sigma$-model. In these papers, $\lambda$-brackets (see also p.\pageref{p:lambdabracket}) were proposed as a way to interpolate between special holonomy OPE algebras \cite{Odake:1988bh,Shatashvili:1994zw} and the classical symmetries of \cite{Howe:1991im}. A more detailed comprehension of commutator and current algebras of our extended $\mathcal{G}$-structure symmetries would make a useful start about this. This is especially interesting at order $\alpha'$, where the condition $\dd H = 0$ fails, suggesting radical alterations to the algebras. 
Further related works on the Chiral de Rham complex \cite{Witten:2005px,Tan:2006qt} also deserve comparison with our results.
 
It will also be interesting to clarify if our symmetry perhaps can be thought of as the infrared limit of some useful symmetry of gauged linear $\sigma$-model (see e.g. \cite{McOrist:2010ae} and references therein).

More speculatively, since $\mathcal{N}=2$ supersymmetry is a subcase of $\mathcal{G}$-structure symmetries, it is permitted to think that some of the powerful tools following from the former admit a non-linear generalisation to the latter. We might ask for example for a ``$\mathcal{G}$-structure'' analogue of supersymmetric localisation, to name but one, which would encompass $(2,0)$ localisation \cite{Closset:2015ohf}. In any case, whenever they are preserved, these symmetries put strong constraints on the dynamics of the $\sigma$-model and should guide the study of string vacua in the $\alpha'$ expansion from a worldsheet point of view \cite{Gross:1986iv, Grisaru:1986px, Candelas:1986tz}. They might find applications for instance to generalise the results of \cite{Nemeschansky:1986yx} to target spaces other than Calabi--Yau manifolds \cite{Jardine:2018sft,Becker:2014rea}.

Another application of our $\mathcal{G}$-structure symmetry (at tree-level in $\alpha'$) is described in the next chapter: finding marginal deformations of $\sigma$-models used as internal sectors in heterotic string compactifications. By isolating explicitly the symmetry associated with supersymmetric backgrounds, it becomes clear how to impose that it be preserved by deformations.

\chapter{Marginal deformations} \label{chap:MarginalDeformations}

\onehalfspacing
\minitoc
\vspace{25pt}

\doublespacing

In this final chapter, we bring together techniques from various parts of the thesis to study infinitesimal displacements in the space of vacua of heterotic string theory. We do this near a point where a geometric splitting $\mathcal{M}^d \times \mathbb{R}^{1,9-d}$ makes sense and is constrained by minimal supersymmetry in the spacetime factor $\mathbb{R}^{1,9-d}$.

From a worldsheet perspective, this problem can be phrased as follows. Each vacuum corresponds to a choice of internal conformal field theory---for us approximated in the large radius limit by a non-linear $\sigma$-model mapping into $\mathcal{M}$. A priori for heterotic strings, this worldsheet theory has $(1,0)$ supersymmetry. However, as explained in chapter~\ref{chap:G-symmetries}, demanding spacetime supersymmetry enhances the symmetry (super)group of the worldsheet theory. This is correlated with some kind of reduction $SO(d)\rightarrow \mathcal{G}$ of the structure group of $\mathcal{M}$ to a subgroup. Displacements in the space of vacua are then displacements in the ``space of CFTs'', also known as \emph{conformal manifold}. The additional constraint of spacetime supersymmetry intuitively slices a subspace in this manifold. We are interested in local properties of this locus. More precisely, we seek to characterise its tangent vector space at a generic point, whose elements are called infinitesimal \emph{moduli}.\footnote{Obstructions at higher orders in perturbation theory are beyond our scope, but they can hinder the tangent vector interpretation.}

This is a hard problem due to the generality we impose on ourselves: low supersymmetry, no explicit model to perturb about, etc. It is however widely applicable to almost any type of heterotic background and can also easily accommodate the closed-string sector of type II string theories.\footnote{$(1,1)$ supersymmetric worldsheet theories are indeed a particular class of $(1,0)$ theories.} It has relevance to the effective theory in $\mathbb{R}^{1,9-d}$, where moduli appear as massless fields. Mathematical and field theoretical lessons drawn from this problem also make it inherently worthwhile.

We will mainly specialise to the case $d=7$, $\mathcal{G}=G_2$ for which a thorough worldsheet analysis appeared in the published work \cite{Fiset:2017auc}. Work in progress \cite{DeLaOssaFisetToAppear2019} suggests on the other hand that the method of \cite{Fiset:2017auc} is applicable much more generally, in particular to $d=8$, $\mathcal{G}=Spin(7)$. We shall therefore allude to these recent developments and, when possible, keep $d$ unspecified.

The emerging impression from \cite{DeLaOssaFisetToAppear2019, Fiset:2017auc, Melnikov:2011ez}, at least for $d=8, 7,$ and $6$, is the existence of a worldsheet fermionic nilpotent operator $\mathcal{Q}$, the \emph{BRST operator}, whose first cohomology is isomorphic to the infinitesimal moduli space. This fact is well-known in the particular case where $\mathcal{M}$ is Calabi--Yau and ramifies deeply in topological field theory and topological strings. The underlying worldsheet symmetry in that case is $(2,2)$ superconformal invariance \cite{Lerche:1989uy}.
For more general heterotic $d=6$ backgrounds, a worldsheet analysis in $(0,2)$ superspace \cite{Melnikov:2011ez} also supports that a BRST operator should control moduli. 

Meanwhile in \cite{Shatashvili:1994zw, deBoer:2005pt}, partial arguments for a topological twist in $Spin(7)$ and $G_2$ type II string set ups were put forward. The operator algebras $\text{SV}^{Spin(7)}$ and $\text{SV}^{G_2}$---more precisely their hidden minimal model $\langle T_- \rangle$---are crucial in the arguments. The conjectured topological theories are of \emph{cohomological type} and thus come with a BRST operator whose role is analogous to the A and B type BRST operators in $(2,2)$ CFTs \cite{Witten:1988xj, Lerche:1989uy}. The BRST operator in \cite{deBoer:2005pt} is precisely the one we will use to compute $G_2$ heterotic moduli below. Remarkably including NS--NS flux (ignored in the original works) and the heterotic vector bundle $\mathcal{V}\rightarrow \mathcal{M}$, introduce no complication, as we shall illustrate.

\section{Marginal deformations of $(1,0)$ non-linear $\sigma$-models} \label{sec:StartMarginalDeformation}

We start with a Lagrangian theory
\begin{equation}
\text{S}[M,A; X,\Lambda] = \int \frac{\dd^{2|1} \varsigma}{4\pi \alpha'} ~ \text{L}[M,A;X,\Lambda]
\end{equation}
given by the general $(1,0)$ non-linear $\sigma$-model \eqref{eq:S_M}--\eqref{eq:S_V} in chapter~\ref{chap:G-symmetries} with dynamical fields $(X^i, \Lambda^\alpha)$, $i\in\{1,\ldots, d=\text{dim}(\mathcal{M})\}$, $\alpha\in\{1,\ldots, n=\text{rank}(\mathcal{V})\}$. We explicitly keep track of the dependence on the couplings $M_{ij}(X)=G_{ij}(X)+B_{ij}(X)$ and $A_i{}^\alpha{}_\beta(X)$, which correspond (in principle) to a determined target space geometry and to a point in the ``space of theories''. We assume that this point is fixed under the renormalisation group flow.
More precisely, we enforce full conformal symmetry up to a certain order in a suitable perturbative expansion. Here, we work at lowest order in $\alpha'$, where $(1,0)$ superconformal symmetry is automatic for any target space geometry (see section~\ref{sec:sconfsymm}). \detail{(It would be interesting in \cite{DeLaOssaFisetToAppear2019} to extend to $O(\alpha')$ by assuming a solution to the target space equations of motion found by requiring vanishing $\beta$-functions (see section~\ref{sec:Superconformal anomalies}).)}

Next we consider deforming the theory by adding \begin{equation}
\delta \text{S} = \int \frac{\dd^{2|1} \varsigma}{4\pi \alpha'} \, \mathcal{O}\,,
\end{equation}
where $\mathcal{O}$ is a \emph{marginal} operator (at least up to the relevant order). Given the weights of the various superfields and operators, the most general classically marginal operator we can deform the $\sigma$-model by is
\begin{equation} \label{eqn:deformations}
\mathcal{O} = -i \delta M_{ij}(X) DX^i \partial_- X^j + \tr ( \Lambda \delta A_i(X) DX^i \Lambda ) \,.
\end{equation}
Here, $\delta M$ and $\delta A$ are general functions of the superfields $X^i$s allowed on dimensional grounds. The notation is chosen consistently with the $\sigma$-model couplings. For instance, we can identify the symmetric and antisymmetric parts of $\delta M$ with deformations of the metric and B-field: $\delta M_{ij} = \delta M_{(ij)} +\delta M_{[ij]}  = \delta G_{ij} +\delta B_{ij}$.\footnote{We do not consider a deformation of the Fermi fields' metric, as this can be reabsorbed into a deformation of the gauge field.} Effectively, the deformed $\sigma$-model is thus simply
\begin{equation}
\text{S}[M+\delta M,A+\delta A; X, \Lambda] \,.
\end{equation}
This deformation is best thought of as happening inside path integrals. Alternatively it may be translated in the language of effective quantum actions used in section~\ref{sec:Anomalies}. So far the perturbations could be finite, but we will work at linear order in $\delta M$ and $\delta A$. \detail{Necessary?}

Finally we must impose that the deformation preserves supersymmetry in spacetime. One reliable way to do this is to exploit the corresponding $\mathcal{G}$-structure symmetries discussed in chapter~\ref{chap:G-symmetries} by demanding invariance of $\text{S} + \delta \text{S}$. See \cite{Fiset:2017auc} for details. This approach effectively boils down to the following geometric deformation problem.

Assuming a solution to the target space Killing spinor supersymmetry equations, one perturbs it slightly and requires that the deformed system is also a solution to the supersymmetry conditions. This gives differential constraints linear in the deformations. In \cite{delaOssa:2014cia,Anderson:2014xha,Garcia-Fernandez:2015hja,Clarke:2016qtg,delaOssa:2016ivz,delaOssa:2017pqy}, it was shown that the result can, for $d=7$ and $d=6$, be phrased in terms of a target space differential $\mathcal{D}$. The infinitesimal deformations are closed under $\mathcal{D}$ and its exact representatives correspond to trivial deformations (e.g. diffeomorphisms and gauge transformations). Interesting relations with generalised geometry, differential graded Lie algebras and $L_\infty$ algebras were also discovered \cite{Ashmore:2018ybe,Garcia-Fernandez:2018ypt,Garcia-Fernandez:2016azr,Garcia-Fernandez:2015hja,Garcia-Fernandez:2013gja}.

The worldsheet equivalent of this target space differential $\mathcal{D}$ is roughly the BRST operator $\mathcal{Q}$ we are now seeking. It would be satisfying to utilise $\mathcal{G}$-structure worldsheet symmetries to formulate $\mathcal{Q}$ explicitly. At present, the author is not aware of a way to do this. We will instead account for supersymmetry conditions somewhat more indirectly. In the next section, we hark back to chapter~\ref{chap:OA}, to take a closer look at Shatashvili--Vafa operator algebras, in particular in their free field realisations. This will be a long detour, but it will then be clear how to resume the $\sigma$-model derivation of moduli from here, where we now leave it.

\vfill

\section{The BRST operator} \label{sec:BRST}

Recall from chapter~\ref{chap:OA} that the superconformal algebras of interest in superstring compactifications for $d=8, 7$ and $6$, namely $\text{SV}^{Spin(7)}$, $\text{SV}^{G_2}$ and $\text{Od}^{n=3}$, can all be obtained in their free field realisation from the principle that they should contain a hidden \emph{small} Virasoro sector $\langle T_- \rangle$ with $c_- \leq 1$ in addition to the standard superconformal symmetry. The surprising effectiveness of this principle remains a mystery to the author and so does the ubiquitous form taken by the small Virasoro operator:
\begin{equation} \label{eq:TminusBRST}
T_- = \mu (T_f -\widehat{\Psi}) \,.
\end{equation}
Here $\mu$ is a coefficient specific to the case under study; $T_f$ is the fermionic energy-momentum tensor in $(\text{Free})^d$\,:
\begin{equation}
T_f = \sum_{i=1}^d \frac{1}{2}\no{\partial\psi^i \psi^i} \,;
\end{equation}
and $\widehat{\Psi}$ is the operator corresponding to a given constant differential $4$-form:
\begin{equation}
\widehat{\Psi} = \frac{1}{4!} \Psi_{ijkl}\no{\psi^i\no{\psi^j\no{\psi^k\psi^l}}} \,.
\end{equation}

In this section, we exploit this hidden Virasoro symmetry, mainly by considering primary fields or, at the very least, fields with well-defined weight under $T_-$. Primaries are always important in conformal theories, but for $d=8$ and $d=7$, where $T_-$ defines a \emph{Virasoro minimal model}, they are particularly insightful since there are only finitely many of them. The BRST current necessary for our moduli computation ultimately relies on this very fact. Before specialising to the Shatashvili--Vafa algebras, we study this question generically inside $(\text{Free})^d$. The contacts with $\mathcal{G}$-structure geometry we will flesh out provide just enough information about the BRST action on marginal deformations for us to identify moduli in section~\ref{sec:EndMarginalComputation}.

\vfill

\subsection{Form fields and weights under $T_-$}

Let us seek operators $\mathcal{O}\in (\text{Free})^d$ with well-defined weight with respect to the Virasoro operator $T_-$ of the form \eqref{eq:TminusBRST}. Recall that this means that the order-$2$ pole $\boxope{T_- \mathcal{O}}_2$ should be proportional to $\mathcal{O}$.

Since the bosonic and fermionic sectors commute and since $T_-$ itself is expressed solely in terms of fermions, we can focus on the fermionic content by restricting $\mathcal{O}$ to monomials in the $\psi^i$s. As a simplification, we also restrict to operators without any holomorphic derivatives $\partial=\partial/\partial z$ inserted.\footnote{Arguments in favour of focusing on non-derivative combinations are given in \cite{deBoer:2005pt} (sections 4.10 and 5.2).} As mentioned in section~\ref{sec:Free}, there is a correspondence between such operators $\mathcal{O}=\widehat{\Pi}$ and $p$-forms $\Pi\in\Omega^{p}(\mathbb{R}^d,\mathbb{R})$. Since the bosonic content is irrelevant, we restrict without loss of generality to forms with {constant} coefficients.

The fermionic energy-momentum tensor $T_f$ in $T_-$ simply counts conformal weight, so we can concentrate on calculating the OPE between the $4$-form field $\widehat{\Psi}$ and the $p$-form field $\widehat{\Pi}$. More generally, letting $\Psi\in\Omega^{q}(\mathbb{R}^d,\mathbb{R})$, we are interested in the OPE between form fields
\begin{equation}
\widehat{\Psi}=\frac{1}{q!}\Psi_{i_1\ldots i_q} \no{\psi^{i_1\ldots i_q}} \,, \qquad
\widehat{\Pi}=\frac{1}{p!}\Pi_{i_1\ldots i_p} \no{\psi^{i_1\ldots i_p}} \,,
\end{equation}
where we have re-introduced our compact notation for right-nested normal ordered products of fermions from section~\ref{sec:exercise}. We raise and lower indices with a Kronecker delta.

\subsubsection{$\bm{p=0}$:}
There is not much to say for $0$-form fields apart for the fact that the identity $\widehat{\Pi}=\mathds{1}$ is trivially primary with weight $h_-=0$ under $T_-$. Introducing the notation $\ket*{h_-,h-h_-}$, where $h$ is the overall weight under $T=T_f+T_b$, the identity is $\ket*{0,0}$.

To access more interesting cases, we iterate Wick's theorem. It is straightforward to prove
\begin{equation} \label{eq:qFormWith1Form}
\wick{\c \psi^{j_1\ldots j_q}(z) \c \psi_i(w)} = \frac{1}{z-w} \, q \, \delta^{[j_q}_i \psi^{j_1 \ldots j_{q-1}]}(w) \,.
\end{equation}

\detail{
Let us examine this in more detail. We have
\begin{align}
\wick{\c \psi_i (z) \c \psi^{j_1 \ldots j_q}(w)}
&= \frac{1}{2\pi i} \oint \frac{\dd x}{x-w} \left\lbrace \frac{\delta^{j_1}_i}{z-x} \psi^{j_2 \ldots j_q}(w)
-
\psi^{j_1}(x) \wick{\c \psi_i (z) \c \psi^{j_2 \ldots j_q}(w)}
\right\rbrace  \\
&= \frac{1}{z-w}\delta^{j_1}_i\psi^{j_2 \ldots j_q}(w) - \no{\psi^{j_1}(w)\wick{\c \psi_i (z) \c \psi^{j_2 \ldots j_q}(w)}} \label{eq:recurrenceFFOPEsimple}
\end{align}
The first term was easy to evaluate with formulas given in section~\ref{sec:OABasics}. The second term is to be interpreted as a normal ordered product with a singular OPE in $(z-w)$. This form of the second term in Wick's generalised theorem is actually fully general, as can be understood also via the integration formulas, from section~\ref{sec:OABasics}.

Notice that this recurrence relation respects the right-nesting prescription. We can then iterate it until we run out of fermions to contract with. Hence the result is the (expected) alternating sum over all possible contractions. This can be written as
\begin{equation}
\wick{\c \psi_i (z) \c \psi^{j_1 \ldots j_q}(w)}
=
\frac{1}{z-w} \, q \, \delta^{[j_1}_i \psi^{j_2 \ldots j_q]}(w) \,.
\end{equation}
where our antisymmetrisation bracket carries a $1/q!$ normalisation. The formula given above is obtained very easily by permuting the operators and then the indices.
}

\subsubsection{$\bm{p=1}$:}
Since the OPE \eqref{eq:qFormWith1Form} only has a simple pole, most terms in the $T_- \widehat{\Pi}$ OPE are controlled by the fermionic energy-momentum tensor. We find
\begin{equation}
\wick{\c T_-(z) \c {\widehat{\Pi}}(w)} = \frac{\mu/2 ~ \widehat{\Pi}(w)}{(z-w)^2} + \frac{\mu \left( \partial \widehat{\Pi} + \widehat{\Pi \lrcorner \Psi} \right)(w)}{z-w} \,.
\end{equation}
In particular, \emph{all $1$-form operators $\widehat{\Pi}$ have well-defined conformal weight $\mu/2$ with respect to $T_-$} (and $h=1/2$ with respect to $T$). In the notation introduced previously:
\begin{equation}
\widehat{\Pi} \longleftrightarrow \ket*{\frac{\mu}{2},\frac{1-\mu}{2}} \,.
\end{equation}
Recall that the \emph{contraction operator} $\lrcorner$ above is defined in general for a $p$-form $\Pi$ and a $q$-form $\Psi$, $q\geq p$, by
\begin{equation}
\Pi \lrcorner \Psi = \frac{1}{p!(q-p)!} \Pi_{i_1 \ldots i_p}\Psi^{i_1 \ldots i_p}{}_{j_{p+1} \ldots j_q}  \dd x^{j_{p+1} \ldots j_q} \,.
\end{equation}

\detail{Interestingly the order $1$ pole can indeed be interpreted as an exotic ``$\partial$'' since the insertion operator is a derivation! I think the OPE here is how we should define ``primary'' fields under $T_-$; at least for a weight $\mu/2$ primary.}

\sep

In order to study higher order $p$-form fields, we need a generalisation of \eqref{eq:qFormWith1Form}. For $q \geq p$, the result is \cite{Howe:1994tv}
\begin{align} \label{eq:OPEff}
&\wick{\c \psi^{j_1\ldots j_q}(z) \c \psi_{i_1 \ldots i_p}(w)} \\
& = \sum_{m=1}^p \frac{(-1)^{qm+\frac{m(m+1)}{2}}}{(z-w)^m}
\frac{q! p!}{m!(q-m)!(p-m)!}
\delta^{[j_1}_{[i_1} \ldots \delta^{j_m}_{i_m} \no{ \psi^{j_{m+1} \ldots j_q]}(z) \psi_{i_{m+1} \ldots i_p]} (w) } \,, \nonumber
\end{align}
where we have reintroduced $1/k!$ pre-factors in the definition of antisymmetrisation over $k$ indices.
\footnote{The operator on the right hand side needs clarification because it involves normal ordering at \emph{different} points $z$ and $w$. Let $A, B$ be elements of an operator algebra $\mathcal{V}$. Recall that a useful notation for their regular OPE
\begin{equation}
\sum_{n=0}^\infty \frac{(z-w)^n}{n!} \no{\partial^n A B}(w)
\end{equation}
is $\no{A(z)B(w)}$\,. Note that
\begin{equation}
\no{A(w)B(w)} = \no{AB}(w) \,.
\end{equation}
The operator on the right hand side of \eqref{eq:OPEff} should moreover be understood as right-nested. For example,
\begin{equation}
\no{\psi^{ijk}(z)\psi_{lm}(w)} = \no{\psi^i(z)\no{\psi^j(z)\no{\psi^k(z)\no{\psi_l(w)\psi_m(w)}}}} \,.
\end{equation}}


\detail{
In a handwritten note, I could verify \eqref{eq:OPEff} in the limit where derivatives are dropped. More precisely, I verified the first few orders ($m=1$ and $m=2$), from which it is easy to guess the general form. The pre-factor looks simpler and is easier to guess when choosing a different ordering of the indices for the computation; see below. In my derivations, I had first forgotten to take into account possible issues introduced by nested normal ordered products. Corrections arising in this way necessarily involve derivatives, as explained in \cite[p.190]{DiFrancesco:1997nk}. 

Applying the generalised Wick theorem produces a recurrence relation which is a slight generalisation of \eqref{eq:recurrenceFFOPEsimple}:
\begin{equation}
\Big[ \psi^{j_1 \ldots j_q} , \psi_{i_p \ldots i_1} \Big]_m = q \delta^{[j_q}_{i_p} \Big[ \psi^{j_1 \ldots j_{q-1}]} , \psi_{i_{p-1} \ldots i_1} \Big]_{m-1}
+(-1)^q \no{ \psi_{i_p} \Big[\psi^{j_1 \ldots j_q},\psi_{i_{p-1} \ldots i_1} \Big]_m } \nonumber
\end{equation}
(The brackets are just the box maps.) The key is to understand why---if indeed true---it is fine to forget about issues introduced by normal ordering as long as the dependence on $z$ and $w$ is kept. Although I have not explored this in detail, the answer is essentially given by the following rewriting of the recurrence relation:
\begin{equation}
\wick{\c \psi^{j_1\ldots j_q}(z) \c \psi_{i_p \ldots i_1}(w)}
=
\frac{q}{z-w} \delta^{[j_q}_{i_p}\wick{ \c \psi^{j_1 \ldots j_{q-1}]} (z)  \c \psi_{i_{p-1} \ldots i_1} } (w)
+(-1)^q
\no{\psi_{i_p}(w)\wick{\c \psi^{j_1\ldots j_q}(z) \c \psi_{i_{p-1} \ldots i_1}(w)}} \,. \nonumber
\end{equation}
Working in terms of the full OPE directly allows to bypass some issues with operator ordering because OPEs commute up to sign. This is probably enough to solve the problem iteratively in $p$ assuming $q\geq p$. Another nice thing about working with the OPE is that the second term above automatically produces right-nested operators.
}

Contracting \eqref{eq:OPEff} with differential form coefficients, we find (for $q \geq p$)
\begin{equation} \label{eq:PsiPiOPE}
\wick{\c {{\Psi}} (z) \c {{\Pi}} (w)} = \sum_{m=1}^p \frac{(-1)^{qm+\frac{m(m+1)}{2}}}{(z-w)^m} \frac{1}{m!} ~ \widehat{i^{(m)}_{\Psi(z)}(\Pi(w))} \,.
\end{equation}
We have introduced the \emph{order-$m$ insertion operator of the $p$-form $\Pi$ into the $q$-form $\Psi$} defined by
\begin{equation}
i^{(m)}_\Psi (\Pi)  = \Psi^{i_1 \ldots i_m} \wedge \Pi_{i_1 \ldots i_m}
\,,
\end{equation}
where
\begin{align}
\Psi^{i_1 \ldots i_m} = \frac{1}{(q-m)!} \Psi^{i_1 \ldots i_m}{}_{j_{m+1}\ldots j_q} \dd x^{j_{m+1}\ldots j_q} \,, \\
\Pi_{i_1 \ldots i_m} = \frac{1}{(p-m)!} \Pi_{i_1 \ldots i_m i_{m+1} \ldots i_p} \dd x^{i_{m+1} \ldots i_p} \,.
\end{align}
Obviously $m\leq \text{min}(p,q)$. For $m$ saturating this bound (again assuming $q\geq p$ for definitiveness), this reproduces the contraction operator:
\begin{equation}
\Pi \lrcorner \Psi = \frac{1}{p!} i^{(m=p)}_\Psi (\Pi) \,.
\end{equation}
The insertion operator here generalises to $m>1$ the one defined in section~\ref{ssec:GstructureSym}. Only the latter is a derivation.

We care especially about the order $m=2$ and $m=1$ poles in \eqref{eq:PsiPiOPE}. However, unravelling the regular OPEs in this equation yields contributions coming from higher orders $m$ involving products of derivatives of the fermions (e.g. $\no{\partial^{k} \psi^i \psi^j \ldots }$). These contributions also carry insertion operators from higher values of $m$.

\vfill

\subsubsection{$\bm{p=2, 3, 4}$:}

For $p=2$, the most singular pole in \eqref{eq:PsiPiOPE} has order $2$ and the coefficient is proportional to $i^{(2)}_\Psi (\Pi)$. The necessary and sufficient condition for $\widehat{\Pi}$ to have well-defined weight under $T_-$ is therefore the geometric eigenvalue-type equation
\begin{equation} \label{eq:QuasiProjector}
i^{(2)}_\Psi (\Pi) = \lambda \Pi \,, \qquad \text{for some }\lambda \in \mathbb{R} \,.
\end{equation}

For higher values of $p$, \eqref{eq:PsiPiOPE} has poles of order $m\leq p$ proportional to $i^{(m)}_\Psi (\Pi)$. The poles at order $m\geq 3$ contribute derivative corrections at order $2$ proportional to $i^{(m)}_\Psi(\Pi)$. Since they do not involve derivatives, \emph{$p$-form fields for $p=2,3,4$ have well-defined internal weight with respect to $T_-$ if and only if $i^{(m\geq 3)}_\Psi(\Pi) = 0$ and \eqref{eq:QuasiProjector} are satisfied. In this case, the weight is $\mu(\lambda + p)/2$ and thus}
\begin{equation}
\widehat{\Pi} \longleftrightarrow \ket*{\frac{1}{2}\mu\left(\lambda + p\right),\frac{1}{2}\left(p-\mu(\lambda+p)\right)} \,.
\end{equation}



\detail{It would be interesting and very easy to write down the order $1$ pole, but I don't need it. Essentially there will be a $\partial$ contributed from the stress-tensor and an order-1 contraction much like for 1-forms. Then there will be a term from the order-2 pole proportional to $\Pi$ and involving derivatives of the fermions. Presumably it can be interpreted as proportional to $\partial$, thus renormalising the contribution from the stress-tensor.}

It is straightforward to let $p\geq 5$ by permuting $\Psi$ and $\Pi$ in \eqref{eq:PsiPiOPE}, but we will not need to do this. Instead, let us discuss form fields with well-defined weights under $T_-$ in the particular cases $d=8$ and $d=7$.

\vfill

\subsection{Irreducible forms on $G_2$ and $Spin(7)$ manifolds}

We already mentioned in chapter~\ref{chap:G-symmetries} that differential forms on a manifold endowed with a $\mathcal{G}$-structure decompose into irreducible representations (\emph{irreps}) of $\mathcal{G}$:
\begin{align}
\mathcal{G}&=G_2 : \qquad\, \bm{1}, \bm{7}, \bm{14}, \bm{27},\ldots\,, \\
\mathcal{G}&=Spin(7) : \bm{1}, \bm{7}, \bm{8}, \bm{21}, \ldots\,.
\end{align}

In both cases, $p=0$ forms are in the singlet ($\bm{1}$). We write the corresponding state $\ket{0,0}$ found above in the corresponding boxes in tables~\ref{tab:forms-states}. Similarly in both $G_2$ and $Spin(7)$ cases, $p=1$ forms are in the fundamental representation ($\bm{d}$). Hence we put the corresponding state $\ket{\frac{\mu}{2},(1-\mu)/2}$ in the column for the fundamental. The appropriate values of $\mu$ were obtained in sections~\ref{sec:SVSpin(7)}--\ref{sec:SVG2}: $\mu=1/8$ for $Spin(7)$ and $\mu=1/5$ for $G_2$.

\begin{table}
\begin{center}
\renewcommand\arraystretch{1.1}
\begin{tabular}{c c c}
\begin{tabular}{ c | c | c | c | c }
     & \multicolumn{4}{c}{$G_2$ irreps} \\
    $p$ & {\bf 1} & {\bf 7} & {\bf 14} & {\bf 27} \\
    \hline
    $0$ & \cellcolor{gray!30} $\ket{0,0}$ & - & - & - \\
    \hline
    $1$ & - & \cellcolor{gray!30} $\ket{\tfrac{1}{10},\tfrac{2}{5}}$ & - & - \\
    \hline
    $2$ & - & \cellcolor{gray!30} $\ket{\tfrac{3}{5},\tfrac{2}{5}}$ & $\ket{0,1}$ & - \\
    \hline
    $3$ & \cellcolor{gray!30} $\ket{\tfrac{3}{2},0}$ & $\ket{\tfrac{1}{10}+1,\tfrac{2}{5}}$ & - & $\ket{\tfrac{1}{10},\tfrac{7}{5}}$
\end{tabular}
\\~\\~\\
\begin{tabular}{ c | c | c | c | c }
     & \multicolumn{4}{c}{$Spin(7)$ irreps} \\
    $p$ & {\bf 1} & {\bf 7} & {\bf 8} & {\bf 21} \\
    \hline
    $0$ & \cellcolor{gray!30} $\ket{0,0}$ & - & - & - \\
    \hline
    $1$ & - & - & \cellcolor{gray!30} $\ket{\tfrac{1}{16},\tfrac{7}{16}}$ & - \\
    \hline
    $2$ & - & \cellcolor{gray!30} $\ket{\tfrac{1}{2},\tfrac{1}{2}}$ & - & $\ket{0,1}$ \\ \hline
    \textcolor{white}{$\ket{\tfrac{1}{2}}$}
\end{tabular}
\end{tabular}
\caption{Correspondence between irreducible differential forms of low degree $p$ and fields with well-defined weight under $T_-$}
\label{tab:forms-states}
\end{center}
\end{table}

Now let $\Pi$ be a $p=2$ form on a manifold with a $G_2$-structure. It decomposes according to
\begin{equation}
\bm{21} \longrightarrow \bm{7} \oplus \bm{14} \,.
\end{equation}
The projection operators to the irreps on the right hand side can be found in the mathematical literature, e.g. \cite{MR2164593}. We will need later the projection to the $\bm{7}$:
\begin{equation} \label{eq:Pi7_G2}
(\pi^2_{\bm{7}})^{ij}_{mn}=\frac{1}{6}(*\Psi)^{ij}{}_k(*\Psi)^k{}_{mn} \,,
\end{equation}
\detail{Alternatively,
\begin{equation}
\pi^2_7(\Pi) = \frac{1}{3}(\Pi + \Pi \lrcorner \Psi) \,.
\end{equation}
}
where $\Psi$ is the $G_2$-structure $4$-form. Using this projector, one can show that irreducible $2$-forms in the $\bm{7}$ satisfy the eigenvalue-type constraint \eqref{eq:QuasiProjector} and thus have well-defined weight $(3/5)$ with respect to $T_-$. The weight with respect to $T_+$ is $2\cdot 1/2-3/5=2/5$. Similarly irreducible $2$-forms in the $\bm{14}$ have well-defined weight $0$ with respect to $T_-$ and $1$ with respect to $T_+$. We report these results in table~\ref{tab:forms-states}.

The case of $Spin(7)$ has not yet appeared in the literature, so we presently provide more detail. Let $\Pi$ be a $p=2$ form on a manifold with a $Spin(7)$-structure. It decomposes according to
\begin{equation}
\bm{28} \longrightarrow \bm{7} \oplus \bm{21} \,.
\end{equation}
The projectors are \cite{MR2164593}
\begin{align}
\pi^2_{\bm{7}}(\Pi) &= \frac{1}{4}\left(\Pi + *(\Psi \wedge \Pi)\right) \,, 
\label{eq:Pi7_Spin7}
\\
\pi^2_{\bm{21}}(\Pi) &= \frac{1}{4}\left(3\Pi - *(\Psi \wedge \Pi)\right) \,.
\end{align}
We can also write
\begin{equation}
*(\Psi \wedge \Pi) = \Pi \lrcorner \Psi =i^{(2)}_{\Psi}(\Pi)/2\,,
\end{equation}
showing again that irreducible $2$-forms satisfy the eigenvalue constraint \eqref{eq:QuasiProjector} and thus correspond to fields with well-defined weight under $T_-$. The weights are easily obtained from our work in the previous section and they are shown in table~\ref{tab:forms-states}.

We could keep going for higher values of $p$. For $G_2$, this has been done in \cite{deBoer:2005pt}. The result is that \emph{all irreducible $p$-form fields have well-defined weight under $T_-$}. These weights are those of minimal model primaries or their descendants as guaranteed by the representation theory of $T_-$. We can already see this in our results so far: $0$, $1/10$, $3/5$ are weights in the tri-critical Ising minimal model. Moreover we notice that the \emph{weights under $T_+=T-T_-$ keep track of the irreducible representation}: they are constant in the columns of table~\ref{tab:forms-states}. Our results above strongly indicate similar facts for $Spin(7)$: $0$, $1/16$, $1/2$ are weights of primaries in the Ising minimal model.

\subsection{BRST current $G^{\downarrow}$ for $G_2$ and $Spin(7)$ theories}

The occurrence of minimal model primaries in our discussion leads us to recall some of their properties more systematically.

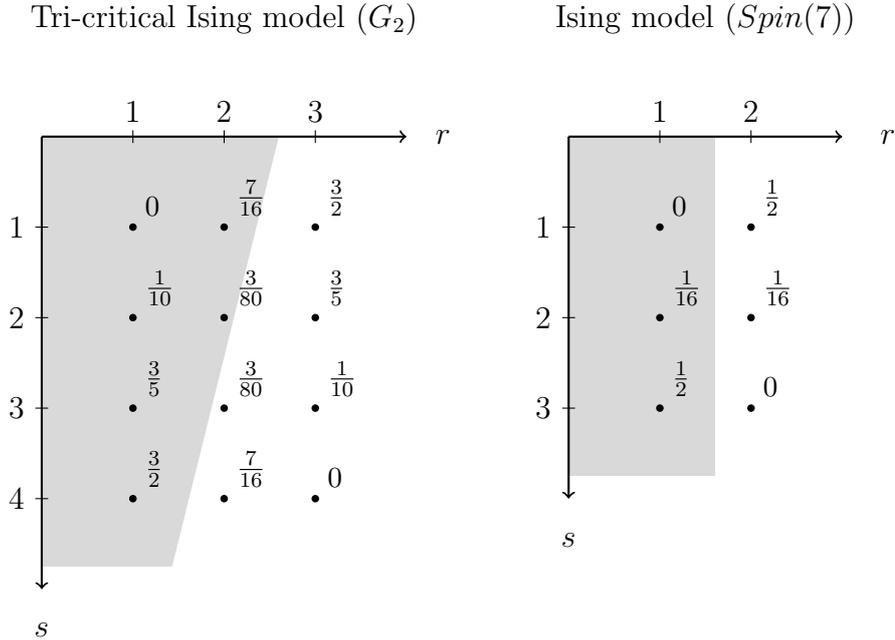
\begin{figure}[tbp]
\begin{center}
\begin{tabular}{c c c}
\begin{tikzpicture}[scale=1.2,label distance=-1pt]
\fill[fill=gray!30] (-2,0) -- (0.6,0) -- (-0.57,-4.75) -- (-2,-4.75);
\draw[line width=0.8pt,->] (-2,0) -- (2,0);
\draw[line width=0.8pt,->] (-2,0) -- (-2,-5);
\foreach \x in {1,2,3}
\draw (\x - 2, 2pt) -- (\x -2 , -2pt) node[above=4pt] {$\x$};
\foreach \y in {1,2,3,4}
\draw (-2cm + 2pt, -\y ) -- (-2cm - 2pt,-\y) node[anchor=east] {$\y$};
\node at (2.4,0) {$r$};
\node at (-2,-5.45) {$s$};
\node at (0,1.3) {Tri-critical Ising model ($G_2$)};

\weight $0$ at (-1,-1)
\weight $\tfrac{7}{16}$ at (0,-1)
\weight $\tfrac{3}{2}$ at (1,-1)
\weight $\tfrac{1}{10}$ at (-1,-2)
\weight $\tfrac{3}{80}$ at (0,-2)
\weight $\tfrac{3}{5}$ at (1,-2)
\weight $\tfrac{3}{5}$ at (-1,-3)
\weight $\tfrac{3}{80}$ at (0,-3)
\weight $\tfrac{1}{10}$ at (1,-3)
\weight $\tfrac{3}{2}$ at (-1,-4)
\weight $\tfrac{7}{16}$ at (0,-4)
\weight $0$ at (1,-4)

\end{tikzpicture}
&
\qquad
&
\begin{tikzpicture}[scale=1.2,label distance=-1pt]
\fill[fill=gray!30] (-1.5,0) -- (0.1,0) -- (0.1,-3.75) -- (-1.5,-3.75);
\draw[line width=0.8pt,->] (-1.5,0) -- (1.5,0);
\draw[line width=0.8pt,->] (-1.5,0) -- (-1.5,-4);
\foreach \x in {1,2}
\draw (\x - 1.5, 2pt) -- (\x - 1.5 , -2pt) node[above=4pt] {$\x$};
\foreach \y in {1,2,3}
\draw (-1.5cm + 2pt, -\y ) -- (-1.5cm - 2pt,-\y) node[anchor=east] {$\y$};
\node at (2,0) {$r$};
\node at (-1.5,-4.45) {$s$};
\node at (2,-5.45) {\textcolor{white}{$s$}};
\node at (0,1.3) {Ising model ($Spin(7)$)};

\weight $0$ at (-0.5,-1)
\weight $\tfrac{1}{2}$ at (0.5,-1)
\weight $\tfrac{1}{16}$ at (-0.5,-2)
\weight $\tfrac{1}{16}$ at (0.5,-2)
\weight $\tfrac{1}{2}$ at (-0.5,-3)
\weight $0$ at (0.5,-3)
\end{tikzpicture}
\end{tabular}
\end{center}
\vspace{-0.5cm}\caption{Kac tables of the Ising and tri-critical Ising minimal models. 
}\label{fig:kac}
\end{figure}

There is a discrete sequence of Virasoro minimal models labelled by the integer $m\geq 3$. $m=3$ and $m=4$ are respectively the Ising and tri-critical Ising models in the conventions of \cite{DiFrancesco:1997nk} used below. Minimal models have finitely many conformal families composed of a primary and its descendants. The conformal families are labelled by a pair of integers $\{r,s\}$, with $1\leq r\leq m-1$, $1\leq s\leq m$ and subject to the identifications $\{r,s\}\sim \{m-r , m+1-s\}$. They may be conveniently organised into a lattice in the $r,s$-plane, the \emph{Kac table} (figure~\ref{fig:kac}). Consistently with the identifications, two copies of each conformal family appear in the Kac table. On each site of this lattice, we record the weight of the primary for the corresponding conformal family given by
\begin{equation}
h_{r,s} = \frac{(r(m+1)-sm)^2-1}{4m(m+1)} \,.
\end{equation}
The reader is invited to compare with tables~\ref{tab:forms-states}.

Conformal families of minimal models obey specific fusion rules. The most relevant for us is
\begin{equation}
\{1,2\}\times\{r,s\}=\{r,s-1\}+\{r,s+1\}\,,
\end{equation}
which states that the OPE between any field in $\{1,2\}$ and any field in $\{r,s\}$ only involves fields in the families immediately above and below $\{r,s\}$ in the Kac table.\footnote{In the case where $\{r,s\}$ is on the boundary, then only the existing neighbour contributes.} Any field in $\{1,2\}$ thus induces two maps obtained by restricting the OPE to either of the neighbouring conformal families. Accordingly $\{1,2\}$ admits a decomposition
\begin{equation}
\{1,2\}=\{1,2\}^{\uparrow}+\{1,2\}^{\downarrow}\,,
\end{equation}
with the following actions:
\begin{align}
\{1,2\}^{\uparrow}&\times\{r,s\}=\{r,s-1\}\,,\\
\{1,2\}^{\downarrow}&\times\{r,s\}=\{r,s+1\}\,.
\end{align}
The arrows convey the idea of moving up or down in the Kac table. For this notation to make sense, we must restrict $\{r,s\}$ to the shaded half of figure~\ref{fig:kac}. 

Returning to the full Shatashvili--Vafa algebras, the $\uparrow\downarrow$ decomposition will apply to any field whose minimal model piece lies fully within the family $\{1,2\}$. This turns out to be the case for the $(1,0)$ supersymmetry current \cite{Shatashvili:1994zw, deBoer:2005pt} and thus:
\begin{equation}
G(z) = G^\uparrow(z) + G^\downarrow (z) \,.
\end{equation}
This can be seen by a direct computation of the $\tilde{X} G$ OPE since $T_- \propto \tilde{X}$ (see sections~\ref{sec:SVSpin(7)}--\ref{sec:SVG2}). In free field realisations, this is even more manifest since $G$ is realised as $\no{{\mathscr{I}}_i \psi^i}$, which is effectively a $p=1$ form field as far as the fermionic content is concerned. Consulting table~\ref{tab:forms-states}, we immediately confirm that $G$ has $T_-$-weight $1/10 = h_{1,2}$ in the tri-critical Ising model and $1/16 = h_{1,2}$ in the Ising model.

The decomposition also carries over to Laurent modes and following \cite{deBoer:2005pt} this finally allows us to define a \emph{BRST operator} for the theory as
\begin{equation} \label{eq:QBRST}
\mathcal{Q} = G_{-1/2}^\downarrow\,,
\end{equation}
where $G_{-1/2}=Q$ is the supersymmetry charge. In \cite{Shatashvili:1994zw, deBoer:2005pt}, evidence was gathered to support the existence of a topological twist for $G_2$ theories, similar to the A and B twists of $(2,2)$ models \cite{Witten:1988xj}. The BRST operator \eqref{eq:QBRST} was proposed in this context, specifically for $(1,1)$ models with $G_2$ holonomy targets. As we will soon see, more general $G_2$-structure targets of $(1,0)$ $\sigma$-models are also concerned by this BRST operator, even including the heterotic vector bundle. The generalisation to $Spin(7)$ is also immediate.

\vfill

\section{Marginal couplings in BRST cohomology} \label{sec:EndMarginalComputation}

We are at last equipped to complete the moduli computation started in section~\ref{sec:StartMarginalDeformation} using the general $(1,0)$ non-linear $\sigma$-model.

Notice first that the marginal deformation $\mathcal{O}$ in \eqref{eqn:deformations},
\begin{equation} \label{eqn:deformationsBis}
\mathcal{O} = -i (\delta M_{ij}DX^i) \partial_- X^j -( \delta A_iDX^i )_{\alpha\beta}\Lambda^\alpha \Lambda^\beta \,,
\end{equation}
is only defined up to total derivatives since it is used in an integral. This includes superspace derivatives since $\int \dd \theta \partial_\theta (\ldots) = 0$. Hence the deformation is really by an equivalence class
\begin{equation}
[\mathcal{O}] = \{ \mathcal{O} + \text{total derivatives}\} \,.
\end{equation}
To identify the moduli of the theory, we seek marginal deformations $[\mathcal{O}]$ that are closed under the BRST operator $Q^\downarrow$. Let us worry about projection later and start by acting with the supercharge. To simplify this calculation, we notice that
\begin{equation}
Q [\mathcal{O}] = D [\mathcal{O}] \,.
\end{equation}
Indeed $D$ acting on the representative $\mathcal{O}$ of $[\mathcal{O}]$ is identical to $Q$ on a different representative $\mathcal{O}'=\mathcal{O}-D\left(2\theta \mathcal{O}\right)$:
\begin{align}
Q\left(\mathcal{O}-D\left(2\theta \mathcal{O}\right)\right) = Q\mathcal{O}+2i\theta \partial_+ \mathcal{O} = D\mathcal{O} \,.
\end{align}
We used the explicit superspace definitions \eqref{eq:DQSuperspace}.

We can thus evaluate $D\mathcal{O}$, for $\mathcal{O}$ in \eqref{eqn:deformationsBis}. This straightforward calculation produces terms with $D \partial_- X^i$ and $D\Lambda^\alpha$ amongst other things. We get rid of these factors by making use of the equations of motion of the unperturbed $(1,0)$ $\sigma$-model calculated in section~\ref{sec:Vars}. (Working on-shell in this way is not uncommon, as evidenced by similar studies on moduli of $(0,2)$ \cite{Melnikov:2011ez} and 4d $\mathcal{N}=1$ \cite{Green:2010da} $\sigma$-models.) The result is

\begin{align} \label{eq:DO}
D\mathcal{O} &=
\delta M_{ij} \partial_+ X^i \partial_-X^j  \\
&\quad+i\left(\partial_j \delta M_{ik}-\Gamma^{-l}_{jk}\delta M_{il}\right)DX^{ij} \partial_- X^k
+\frac{1}{2}\tr(F^k{}_j\delta M_{ik} DX^{ij} \Lambda \Lambda) \nonumber \\
&\quad
-\left( \partial_i \delta A_{j\alpha\beta} - 2A_i{}^{\gamma}{}_\alpha \delta A_{j \gamma \beta}\right) DX^{ij} \Lambda^\alpha \Lambda^\beta
-i \, \tr \,\delta A_i \partial_+ X^i \Lambda \Lambda
\quad \text{on-shell}.\nonumber
\end{align}

Now notice that $\mathcal{O}$ involves exactly one factor of $DX^i = \psi^i + \ldots$, so its leading component is a left-moving $p=1$ form field in the language of section~\ref{sec:BRST}. We shall interpret $\mathcal{O}$ as a normal ordered composite of free fields in order to apply the technology we developed above, mildly generalised to superspace. This approximation is subject to sub-leading curvature corrections.

Both $G$ and $\mathcal{O}$ are in the conformal family $\{1,2\}$ of the minimal model sector $\langle T_- \rangle$. According to fusion, the leading component of \eqref{eq:DO} should be composed of fields in the families $\{1,1\}$ and/or $\{1,3\}$ but nothing else. Indeed this is the case since this expression is effectively composed of $p=0$ form fields and $p=2$ form fields which, as seen in table~\ref{tab:forms-states}, are in these families. The BRST action is thus simply found by projecting to the $\{1,3\}$ family. The first and last terms of \eqref{eq:DO} lie in the image of $G^\uparrow_{-1/2}$ and can be dropped. To isolate further the image of $\mathcal{Q}=G^\downarrow_{-1/2}$, we retain only components in the ${\bf 7}$ representation either of $G_2$ or $Spin(7)$. Explicitly in the $G_2$ case, using $\pi^2_{\bm{7}}$ in \eqref{eq:Pi7_G2}, we find the constraint
\begin{align}
\label{eq:MClosure}
&(*\Psi)_m{}^{ij}\left(\partial_j\delta M_{ik} -\Gamma^{-l}_{jk}\delta M_{il}\right) = 0 \,, \\
\label{eq:G2Atiyah}
& (*\Psi)_m{}^{ij}\left({D_A}_{i}\delta A_{j}^{\alpha\beta} + \tfrac12 F^{\alpha\beta}_{ki}\delta M_{j}{}^k \right) =0 \,,
\end{align}
where ${D_A}_i$ is the gauge covariant derivative, acting as
\begin{equation}
{D_A}_i (\delta A)_{j\alpha\beta} = \partial_i (\delta A)_{j\alpha\beta} + [A_i,\delta A_j]_{\alpha} \,.
\end{equation}
These are precisely the (tree-level) relations derived from spacetime considerations in~\cite{delaOssa:2016ivz,delaOssa:2017pqy}. The first condition was derived in \cite{deBoer:2005pt} for type II models in the special case $H=0$. The second condition is of course unique to the heterotic string.

Since our interest is in the cohomology of $\mathcal{Q}=G^\downarrow_{-1/2}$, we must also identify
\begin{equation}
\mathcal{O} \sim \mathcal{O} + \mathcal{Q}\mathcal{O}'\,,
\end{equation}
where $\mathcal{O}'$ must have weights $(0,1)$. The most general $\mathcal{O}'$ is therefore of the form
\begin{equation}
\mathcal{O}' = -iC_i(X)\partial_- X^i +\Xi_{\alpha\beta}(X)\Lambda^\alpha \Lambda^\beta\,.
\end{equation}

Working through the action of $\mathcal{Q}$ on-shell as above leads us to the following identifications:
\begin{align}
\delta M_{ij} &\sim \delta M_{ij} + \nabla^-_i C_j \,, \label{eqn:Midentify}\\
\delta A_i^{\alpha\beta} &\sim \delta A_i^{\alpha\beta} - {D_A}_i\Xi^{\alpha\beta}- \tfrac{1}{2}  F_i{}^j{}^{\alpha\beta}C_j \,.\label{eqn:Aidentify}
\end{align}
It is somewhat illuminating to unpack the above relations for $\delta G$ and $\delta B$ separately:
\begin{align}
\delta G_{ij} &\sim \delta G_{ij} + \nabla_{(i}C_{j)} \,, \\
\delta B_{ij} &\sim \delta B_{ij} + \partial_{[i}C_{j]} - \tfrac12 H_{ij}{}^kC_k \,,\\
\delta A_i^{\alpha\beta} &\sim \delta A_i^{\alpha\beta} - {D_A}_i\Xi^{\alpha\beta} - \tfrac12 F_i{}^j{}^{\alpha\beta}C_j\,.
\end{align}
We can clearly identify the redundancies associated with diffeomorphisms, B-field transformations and gauge transformations. The final terms on the second and third lines are associated with the effects of diffeomorphisms on the curvatures $H$ and $F$.

\vfill

\section{Conclusion}

We have just reproduced from the worldsheet point of view the results of \cite{delaOssa:2016ivz,delaOssa:2017pqy} on the infinitesimal moduli of the heterotic string compactified on geometries with $G_2$-structure. It should be clear from our presentation that a completely analogous approach would yield moduli in the $Spin(7)$ case: simply use $\pi^2_{\bm{7}}$ for $Spin(7)$ in \eqref{eq:Pi7_Spin7}. While this is a natural guess from the worldsheet approach, it remains to compare with heterotic supergravity. This will be addressed in \cite{DeLaOssaFisetToAppear2019}.

Another obvious question currently under investigation concerns $\alpha'$ corrections. Including such corrections to the $(1,0)$ worldsheet theory tends to be a rather complicated endeavour, in spite of the grasp on symmetries at one-loop described in chapter~\ref{chap:G-symmetries}. As a guiding principle, one can however use the first order supergravity analysis performed in \cite{delaOssa:2017pqy}. In the supergravity analysis, the infinitesimal moduli were identified with first order cohomology of a bundle-valued complex on target space, suggesting that a BRST derivation is possible from the worldsheet even when including $\alpha'$ corrections.

Next it might be interesting to go beyond infinitesimal deformations and consider the exactly marginal deformations, corresponding to integrable deformations of the geometry from the supergravity point of view. In doing so, we could compare with results on the deformation algebra described in \cite{Ashmore:2018ybe}.

Finally it is hard to avoid speculating on possible quasi-topological sectors for heterotic $G_2$ systems. Similar sectors have been found in the $(0,2)$ setting \cite{Adams:2005tc} and been used to compute exact results in the worldsheet theory \cite{McOrist:2007kp,McOrist:2008ji,Donagi:2011uz,Donagi:2011va}. The fact that we reproduce expected results on heterotic moduli gives further credence to the original topological twist and BRST proposals in \cite{Shatashvili:1994zw, deBoer:2005pt}. It also suggests that the BRST operator $\mathcal{Q}$ continues to apply beyond the realm of $(1,1)$ models and vanishing torsion. Such topological sectors would also be relevant for understanding the nature of any topological theory that might govern the heterotic $G_2$ deformation algebra and might also help to shed some light on open mathematical problems concerning $G_2$ structure manifolds with instantons bundles.

%


\newpage


\bibliographystyle{alpha}

\onehalfspacing

\bibliography{Bibliography_Fiset_Jun2019}

\end{document}